\documentclass[dvipdfm,epsf]{iopart}

\eqnobysec
\usepackage{iopams,amsfonts,rotating,graphicx,fancybox,fancyhdr}
\usepackage{eucal}
\usepackage{appendix}
\newcommand{\imag}{\dot{\iota}}
\newcommand{\bm}{\boldsymbol}

\newcommand{\mkbox}[3]{\hbox{\vrule
      \vbox to  #1{\hrule \vss
                  \hbox to #2{\hss#3\hss}\vss
                  \hrule}\vrule}}

\begin{document}

\title[Correlations of RMT Characteristic Polynomials and Integrability: Hermitian Matrices]
        {Correlations of RMT Characteristic Polynomials and Integrability: Hermitean Matrices}

\author{Vladimir Al. Osipov${}^{1}$ and Eugene Kanzieper${}^{2,\,3}$}

\address{
    ${}^{1}$ Fakult\"at f\"ur Physik, Universit\"at Duisburg-Essen, D-47057 Duisburg, Germany \\
    ${}^{2}$ Department of Applied Mathematics, H.I.T. -- Holon Institute of
    Technology, Holon 58102,
    Israel\\
    ${}^{3}$ Department of Physics of Complex Systems, Weizmann Institute of Science, Rehovot 76100,
    Israel}
    \eads{\mailto{Vladimir.Osipov@uni-due.de},
    \mailto{Eugene.Kanzieper@hit.ac.il}}

\begin{abstract}
Integrable theory is formulated for correlation functions of characteristic polynomials associated with invariant
non-Gaussian ensembles of Hermitean random matrices. By embedding the correlation functions of interest into a
more general theory of $\tau$ functions, we (i) identify a zoo of hierarchical relations satisfied by $\tau$ functions
in an abstract infinite-dimensional space, and (ii) present a technology to translate these relations into hierarchically
structured nonlinear differential equations describing the correlation functions of characteristic polynomials in
the physical, spectral space. Implications of this formalism for fermionic, bosonic, and supersymmetric variations of
zero-dimensional replica field theories are discussed at length. A particular emphasis is placed on the phenomenon
of fermionic-bosonic factorisation of random-matrix-theory correlation functions.\newline\newline\noindent
\texttt{Annals of Physics {\bf 325} (2010) 2251-2306 [arXiv:~1003.0757]}
\end{abstract}

%\pacs{02.10.Yn, 02.50.-r, 05.40.-a, 75.10.Nr}

\tableofcontents

\newpage
\section{Introduction} \label{Sec1}

\subsection{Motivation and definitions}

Correlation functions of characteristic polynomials (CFCP) appear in various fields of mathematical and theoretical physics. (i) In quantum chaology, CFCP (i.a) provide a convenient way to describe the universal features of spectral statistics of a particle confined in a finite system exhibiting chaotic classical dynamics (Bohigas, Giannoni and Schmit 1984; Andreev, Agam, Simons and Altshuler 1996; M\"uller, Heusler, Braun, Haake and Altland 2004) and (i.b) facilitate calculations of a variety of important distribution functions whose generating functions may often be expressed in terms of CFCP (see, e.g., Andreev and Simons 1995). (ii) In the random matrix theory approach to quantum chromodynamics, CFCP allow to probe various QCD partition functions (see, e.g., Verbaarschot 2010). (iii) In the number theory, CFCP have been successfully used to model behaviour of the Riemann zeta function along the critical line (Keating and Snaith 2000a, 2000b; Hughes, Keating and O'Connell 2000). (iv) Recently, CFCP surfaced in the studies of random energy landscapes (Fyodorov 2004). (v) For the r\^ole played by CFCP in the algebraic geometry, the reader is referred to the paper by Br\'ezin and Hikami (2008) and references therein.

In what follows, we adopt a formal setup which turns an $n\times n$ Hermitian matrix ${\boldsymbol{\cal H}}={\boldsymbol{\cal H}^\dagger}$ into a central object of our study. For a {\it fixed} matrix ${\boldsymbol{\cal H}}$, the characteristic polynomial ${\rm det}_n(\varsigma-{\boldsymbol{\cal H}})$ contains complete information about the matrix spectrum. To study the {\it statistics} of spectral fluctuations in an {\it ensemble} of random matrices, it is convenient to introduce the correlation function $\Pi_{n|p} ({\boldsymbol \varsigma}; {\boldsymbol \kappa})$ of characteristic polynomials
\begin{eqnarray}
\label{def-cf}
    \Pi_{n|p} ({\boldsymbol \varsigma}; {\boldsymbol \kappa}) =
    \left<
        \prod_{\alpha=1}^{p} {\rm det}_n^{\kappa_\alpha}(\varsigma_\alpha-{\boldsymbol{\cal H}})
    \right>_{\boldsymbol {\cal H}}.
\end{eqnarray}
Here, the vectors ${\boldsymbol \varsigma} = (\varsigma_1,\cdots,\varsigma_p)$ and ${\boldsymbol \kappa}=(\kappa_1,\cdots,\kappa_p)$ accommodate the energy and the ``replica'' parameters, respectively. The angular brackets $\left< f({\boldsymbol {\cal H}}) \right>_{\boldsymbol {\cal H}}$ stand for the ensemble average
\begin{eqnarray}
\label{def}
    \left< f({\boldsymbol {\cal H}}) \right>_{\boldsymbol {\cal H}}
     = \int d\mu_n({\boldsymbol {\cal H}})  \, f(\boldsymbol {\cal H})
\end{eqnarray}
with respect to a proper probability measure
\begin{eqnarray}
    d\mu_n({\boldsymbol {\cal H}})&= P_n({\boldsymbol {\cal H}})\,({\cal D}_n {\boldsymbol {\cal H}}), \\
    ({\cal D}_n {\boldsymbol {\cal H}}) &=
     \prod_{j=1}^n d{\cal H}_{jj} \,\prod_{j<k}^n d{\rm Re}{\cal H}_{jk}
    \, d{\rm Im}{\cal H}_{jk}
\end{eqnarray}
normalised to unity. Throughout the paper, the probability density
function $P_n({\boldsymbol {\cal H}})$ is assumed to follow the
trace-like law
\begin{eqnarray}
    P_n({\boldsymbol {\cal H}}) = {\cal C}_n^{-1}
    \exp\left[-{\rm tr}_n\, V({\boldsymbol {\cal H}})
    \right]
\end{eqnarray}
with $V({\boldsymbol {\cal H}})$ to be referred to as the
confinement potential.

There exist two canonical ways to relate the spectral statistics of
${\boldsymbol {\cal H}}$ encoded into the average $p$-point Green
function
\begin{eqnarray}
    G_{n|p}({\boldsymbol \varsigma}) = \left<
        \prod_{\alpha=1}^p {\rm tr}_n\left( \varsigma_\alpha - {\boldsymbol{\cal H}}
        \right)^{-1}
    \right>_{\boldsymbol{\cal H}}
\end{eqnarray}
to the correlation function $\Pi_{n|p}({\boldsymbol \varsigma};
{\boldsymbol \kappa})$ of characteristic polynomials.
\newline
\begin{itemize}
  \item The supersymmetry-like prescription (Efetov 1983, Verbaarschot, Weidenm\"uller and Zirnbauer 1985, Guhr 1991),
        \begin{eqnarray} \label{susy-r}
    G_{n|p}({\boldsymbol \varsigma}) =
    \left( \prod_{\alpha=1}^p
    \lim_{\varsigma_\alpha^\prime\rightarrow \varsigma_\alpha} \frac{\partial}{\partial \varsigma_\alpha}
    \right) \, \Pi^{\rm{(susy)}}_{n|p+p}({\boldsymbol \varsigma},{\boldsymbol \varsigma}^\prime),
\end{eqnarray}
makes use of the correlation function
\begin{eqnarray}
    \Pi^{\rm{(susy)}}_{n|q+q^\prime}({\boldsymbol \varsigma},{\boldsymbol \varsigma}^\prime)
    =\left<
        \prod_{\alpha=1}^{q} {\rm det}_n(\varsigma_\alpha-{\boldsymbol{\cal H}})
        \prod_{\beta=1}^{q^\prime} {\rm det}_n^{-1}(\varsigma_\beta^\prime-{\boldsymbol{\cal H}})
    \right>_{\boldsymbol {\cal H}}
\end{eqnarray}
obtainable from $\Pi_{n|q+q^\prime}({\boldsymbol
\varsigma},{\boldsymbol \varsigma}^\prime; {\boldsymbol
\kappa},{\boldsymbol \kappa}^\prime)$ by setting the replica
parameters ${\boldsymbol \kappa}$ and ${\boldsymbol
\kappa}^\prime$ to the {\it integers} $\pm 1$.
\newline
  \item On the contrary, the replica-like prescription (Hardy, Littlewood and P\'olya 1934, Edwards and Anderson 1975),
  \begin{eqnarray}
  \label{RL}
    G_{n|p}({\boldsymbol \varsigma}) = \left( \prod_{\alpha=1}^p
    \lim_{\kappa_\alpha\rightarrow 0} \kappa_\alpha^{-1} \frac{\partial}{\partial \varsigma_\alpha}
    \right) \, \Pi_{n|p}({\boldsymbol \varsigma}; {\boldsymbol \kappa}),
\end{eqnarray}
entirely relies on the behaviour of the correlation function
$\Pi_{n|p}({\boldsymbol \varsigma}; {\boldsymbol \kappa})$ for
{\it real-valued} replica parameters, ${\boldsymbol \kappa} \in
{\mathbb R}^p$, as suggested by the limiting procedure in Eq.
(\ref{RL}). In this case, the notation $\Pi_{n|p} ({\boldsymbol \varsigma}; {\boldsymbol \kappa})$ should be understood as the principal value of the r.h.s. in Eq.~(\ref{def-cf}). Existence of the CFCP is guaranteed by a proper choice of imaginary parts of $\bm \varsigma$.
\newline
\end{itemize}
A nonperturbative calculation of the correlation function
$\Pi_{n|p}({\boldsymbol \varsigma}; {\boldsymbol \kappa})$ of
characteristic polynomials is a nontrivial problem. So far, the solutions reported by several groups have always reduced $\Pi_{n|p}({\boldsymbol \varsigma}; {\boldsymbol \kappa})$ to a {\it determinant form}. Its simplest -- Hankel determinant -- version follows from the eigenvalue representation \footnote{See Eq.~(\ref{rpf-1}) and a brief discussion around it.} of Eq.~(\ref{def-cf}) by virtue of the Andr\'eief--de Bruijn formula [Eq.~(\ref{adB}) below]
\begin{eqnarray}
\label{han-det}\fl \qquad
    \Pi_{n|p}({\boldsymbol \varsigma}; {\boldsymbol \kappa}) = n!\,\frac{{\cal V}_n}{{\cal C}_n}\,
    {\rm det}_n \left[
        \int_{\mathbb R} d\lambda \, \lambda^{j+k} e^{-V(\lambda)} \prod_{\alpha=1}^p (\varsigma_\alpha -\lambda)^{\kappa_\alpha}
    \right]_{0 \le j,k \le n-1}.
\end{eqnarray}
Here, ${\mathcal V}_n$ denotes a volume of the unitary group ${\bm {\cal U}}(n)$ as defined by Eq.~(\ref{un-vol}). Unfortunately, the Hankel determinant Eq.~(\ref{han-det}) is
difficult to handle in the physically interesting thermodynamic limit: finding its asymptotics in the domain $n\gg 1$ remains to a large extent an open problem (Basor, Chen and Widom 2001, Garoni 2005, Krasovsky 2007, Its and Krasovsky 2008) especially as the integral in Eq.~(\ref{han-det}) has unbounded support.

For ${\boldsymbol \kappa}$ {\it integers}, ${\boldsymbol \kappa} \in {\mathbb Z}^p$, so-called duality relations (see, e.g., Br\'ezin and Hikami 2000, Mehta and Normand 2001, Desrosiers 2009 and references therein) make it possible to identify a more convenient determinant representation of $\Pi_{n|p}({\boldsymbol \varsigma}; {\boldsymbol \kappa})$: Apart from being expressed through a determinant of a reduced size (see below), such an alternative representation of CFCP displays an explicit $n$-dependence hereby making an asymptotic large-$n$ analysis more viable. For instance, the
correlation function
\begin{eqnarray} \label{cfg} \fl
    \Pi_{n|q+q^\prime}({\boldsymbol \varsigma},{\boldsymbol \varsigma}^\prime;{\boldsymbol m}, {\boldsymbol m}^\prime)
    =\left<
        \prod_{\alpha=1}^{q} {\rm det}_n^{m_\alpha}(\varsigma_\alpha-{\boldsymbol{\cal H}})
        \prod_{\beta=1}^{q^\prime} {\rm det}_n^{-m^\prime_\beta}(\varsigma_\beta^\prime-{\boldsymbol{\cal H}})
    \right>_{\boldsymbol {\cal H}}
\end{eqnarray}
with ${\boldsymbol m} \in {\mathbb Z}_+^q$ and ${\boldsymbol
m^\prime} \in {\mathbb Z}_+^{q^\prime}$ can be deduced
from the result \footnote[1]{See also much earlier works by Uvarov (1959, 1969). Alternative representations for
$\Pi^{\rm{(susy)}}_{n|q+q^\prime}({\boldsymbol
\varsigma},{\boldsymbol \varsigma}^\prime)$ have been obtained by
Strahov and Fyodorov (2003), Baik, Deift and Strahov (2003),
Borodin and Strahov (2005), Borodin, Olshanski and Strahov (2006),
and Guhr (2006).} by Fyodorov and Strahov (2003)
\begin{eqnarray}\label{pink-det}\fl
    \Pi^{\rm{(susy)}}_{n|q+q^\prime}({\boldsymbol \varsigma},{\boldsymbol \varsigma}^\prime)
    =
    \frac{ c_{n,q^\prime}}{\Delta_{q}({\boldsymbol \varsigma})\Delta_{q^\prime}({\boldsymbol \varsigma}^\prime)}
    \, {\rm det}_{q+q^\prime}
    \left[
      \begin{array}{c}
        \left[ h_{n-q^\prime+k}(\varsigma^\prime_j)\right]_{j=1,\cdots,q;\; k=0,\cdots,q+q^\prime-1}  \\
        \left[ \pi_{n-q^\prime+k}(\varsigma_j)\right]_{j=1,\cdots,q^\prime;\; k=0,\cdots,q+q^\prime-1} \\
        \end{array}
    \right] \label{fs}
\end{eqnarray}
by inducing a proper degeneracy of energy variables. The validity of this alternative representation (which still possesses a {\it determinant form}) is restricted to $q^\prime \le n$ (Baik, Deift and
Strahov 2003). Here,
\begin{eqnarray}
    \Delta_q({\boldsymbol \varsigma}) = {\rm det}_q \left[\varsigma_\alpha^{\beta-1}
    \right] =  \prod_{\alpha<\beta}^q (\varsigma_\beta-\varsigma_\alpha)
\end{eqnarray}
is the Vandermonde determinant; the two sets of functions,
$\pi_k(\varsigma)$ and $h_k(\varsigma)$, are the average
characteristic polynomial
\begin{eqnarray}
    \pi_k(\varsigma) = \left<
         {\rm det}_k(\varsigma-{\boldsymbol{\cal H}})
    \right>_{\boldsymbol {\cal H}}
\end{eqnarray}
and, up to a prefactor, the average {\it inverse} characteristic
polynomial \footnote[2]{Making use of the Heine formula (Heine 1878,
Szeg\"o 1939), it can be shown that $\pi_k(\varsigma)$ is a monic
polynomial orthogonal on ${\mathbb R}$ with respect to the measure
$d\tilde{\mu}(\varsigma)=\exp[-V(\varsigma)]\,d\varsigma$. The
function $h_k(\varsigma)$ is its Cauchy-Hilbert transform (see,
e.g., Fyodorov and Strahov 2003):
\begin{eqnarray}
    h_k(\varsigma) =\frac{1}{2\pi \imag} \int_{\mathbb R} \frac{d{\tilde \mu}(\varsigma^\prime)}{\varsigma^\prime-\varsigma}
    \, \pi_k(\varsigma^\prime), \;\;\; {\rm Im\,} \varsigma \neq 0. \nonumber
\end{eqnarray}}
\begin{eqnarray}
    h_{k-1}(\varsigma) = c_{k,1}\,
    \left<
         {\rm det}_k^{-1}(\varsigma-{\boldsymbol{\cal H}})
    \right>_{\boldsymbol {\cal H}}.
\end{eqnarray}
Finally, the constant $c_{n,q^\prime}$ is
\begin{eqnarray}
    c_{n,q^\prime} =
    \frac{(2\pi)^{q^\prime}}{\imag^{\lceil q^\prime/2\rceil - \lfloor q^\prime/2 \rfloor}}
    \frac{n!}{(n-q^\prime)!}\frac{{\cal V}_n}{{\cal V}_{n-q^\prime}}\frac{{\cal N}_{n-q^\prime}}{{\cal N}_{n}},
\end{eqnarray}
where ${\cal V}_n$ is a volume of the unitary group ${\bm {\cal U}}(n)$,
\begin{eqnarray}
\label{un-vol}
    {\cal V}_n = \frac{\pi^{n(n-1)/2}}{\prod_{j=1}^n j!}.
\end{eqnarray}
The result Eq.~(\ref{fs}) is quite surprising since it expresses the
higher-order spectral correlation functions $G_{n|p}({\boldsymbol
\varsigma})$ in terms of one-point averages (Gr\"onqvist, Guhr and Kohler 2004).

For ${\boldsymbol \kappa}$ {\it reals}, ${\boldsymbol \kappa} \in {\mathbb R}^p$, the duality relations are sadly unavailable; consequently, determinant representation  Eq.~(\ref{pink-det}) and determinant representations of the same ilk
(see, e.g., Strahov and Fyodorov 2003, Baik, Deift and Strahov 2003,
Borodin and Strahov 2005, Borodin, Olshanski and Strahov 2006,
and Guhr 2006) no longer exist.

{\it The natural question to ask is what structures come instead of determinants?} This question is the core issue of the present paper in which we develop a completely different way of treating of CFCP. Heavily influenced by a series of remarkable works by Adler, van Moerbeke and collaborators
(Adler, Shiota and van Moerbeke 1995, Adler and van Moerbeke 2001, and reference therein), we make use of the ideas of integrability \footnote{For a review on integrability and matrix models, the reader is referred to Morozov (1994).} to develop an {\it integrable theory of CFCP} whose main outcome is an {\it implicit} characterisation of CFCP in terms of solutions to certain nonlinear differential equations.

As will be argued later, such a theory is of utmost importance for
advancing the idea of exact integrability of zero-dimensional
replica field theories (Kanzieper 2002, Splittorff and Verbaarschot
2003, Kanzieper 2005, Osipov and Kanzieper 2007, Kanzieper 2010). In fact, it is
this particular application of our integrable theory of CFCP that motivated the present study.

\subsection{Main results at a glance}
\label{Sec-1-2}
This work, consisting of two parts, puts both the ideology and technology in the first place. Consequently, its {\it main outcome} is not a single explicit formula (or a set of them) for the correlation function $\Pi_{n|p} ({\boldsymbol \varsigma}; {\boldsymbol \kappa})$ of characteristic polynomials but
\begin{itemize}
  \item a {\it regular formalism} tailor-made for a nonperturbative description of $\Pi_{n|p} ({\boldsymbol \varsigma}; {\boldsymbol \kappa})$ considered at {\it real valued} replica parameters ${\boldsymbol \kappa} \in {\mathbb R}^p$, and
  \item a {\it comparative analysis} of three alternative versions of the replica method (fermionic, bosonic, and supersymmetric) which sheds new light on the phenomenon of fermionic-bosonic factorisation \footnote{A quantum correlation function is said to possess the factorisation property if it can be expressed in terms of a single fermionic and a single bosonic partition function (Splittorff and Verbaarschot 2003, Splittorff and Verbaarschot 2004).} of quantum correlation functions.
\end{itemize}
\noindent
More specifically, in the first part of the paper (comprised of Sections \ref{Sec-2}, \ref{Sec-3} and \ref{Sec-4} written in a tutorial manner) we show that
the correlation function $\Pi_{n|p} ({\boldsymbol \varsigma}; {\boldsymbol \kappa})$ of characteristic polynomials satisfies an infinite set of {\it nonlinear differential hierarchically structured} relations. Although these hierarchical relations do not supply {\it explicit} (determinant) expressions for $\Pi_{n|p} ({\boldsymbol \varsigma}; {\boldsymbol \kappa})$ as predicted by classical theories (which routinely assume ${\bm \kappa} \in {\mathbb Z}^p$), they do provide an {\it implicit} nonperturbative characterisation of $\Pi_{n|p} ({\boldsymbol \varsigma}; {\boldsymbol \kappa})$ which turns out to be much beneficial for an in-depth analysis of the mathematical foundations of zero-dimensional replica field theories arising in the random-matrix-theory context (Verbaarschot and Zirnbauer 1985).

Such an analysis is performed in the second part of the paper (Section \ref{Sec-5}) which turns the fermionic-bosonic factorisation of spectral correlation functions into its central motif. In brief, focussing on the finite-$N$ average density of eigenlevels in the paradigmatic Gaussian Unitary Ensemble (GUE), we have used the integrable theory of CFCP (developed in the first part of the paper) in conjunction with the Hamiltonian theory of Painlev\'e transcendents (Noumi 2004) to associate fictitious Hamiltonian systems ${H}_{\rm f}\left\{P(t),Q(t),t\right\}$ and $H_{\rm b}\left\{ P(t),Q(t),t\right\}$ with fermionic and bosonic replica field theories, respectively. Using this language, we demonstrate that a proper replica limit yields the average density of eigenlevels in an anticipated factorised form. Depending on the nature (fermionic or bosonic) of the replica limit, the compact and noncompact contributions can be assigned to a derivative of the canonical ``coordinate'' and canonical ``momentum'' of the corresponding Hamiltonian system. Hence, the appearance of a noncompact (bosonic) contribution in the fermionic replica limit is no longer a ``mystery'' (Splittorff and Verbaarschot 2003).

\subsection{Outline of the paper}

To help a physics oriented reader navigate through an involved integrable theory of CFCP, in Section \ref{Sec-2} we outline a general structure of the theory. Along with introducing the notation to be used throughout the paper, we list three major ingredients of the theory -- the $\tau$ function $\tau_n^{(s)} ({\boldsymbol \varsigma}, {\boldsymbol \kappa}; {\bm t})$ assigned to the correlation function $\Pi_{n|p} ({\boldsymbol \varsigma}; {\boldsymbol \kappa})$, the bilinear identity in an integral form, and the Virasoro constraints -- and further discuss an interrelation between them and the original correlation function $\Pi_{n|p} ({\boldsymbol \varsigma}; {\boldsymbol \kappa})$. Two integrable hierarchies playing a central r\^ole in our theory -- the Kadomtsev-Petviashvili and the Toda Lattice hierarchies for the $\tau$ function -- are presented in the so-called Hirota form. Both the Hirota derivative \footnote{The properties of Hirota differential operators are reviewed in Appendix \ref{App-hi}.} and Schur functions appearing in the above integrable hierarchies are defined.

Having explained a general structure of the theory, we start its detailed exposition in Section \ref{Sec-3}. In Section \ref{Sec-3-1}, a determinant structure of the $\tau$ function is established and associated matrix of moments and a symmetry dictated scalar product are introduced. The bilinear identity in an integral form which governs the behavior of $\tau$ function is derived in Section \ref{Sec-3-2}. The bilinear identity in Hirota form is derived \footnote{An alternative derivation can be found in Appendix \ref{App-bi}.} in Section \ref{Sec-3-3}. In Section \ref{Sec-3-4}, the bilinear identity is ``deciphered'' to produce a zoo of bilinear integrable hierarchies satisfied by the $\tau$ function; their complete classification is given by Eqs.~(\ref{TL}) -- (\ref{346}). The two most important integrable hierarchies -- the Kadomtsev-Petviashvili (KP) and the Toda Lattice (TL) hierarchies -- are discussed in Section \ref{Sec-3-5}, where explicit formulae are given for the two first nontrivial members of KP and TL hierachies. Section \ref{Sec-3-6} contains a detailed derivation of the Virasoro constraints, the last ingredient of the integrable theory of characteristic polynomials.

Section \ref{Sec-4} shows how the properties of the $\tau$ function studied in Section \ref{Sec-3} can be translated to those of the correlation function $\Pi_{n|p} ({\boldsymbol \varsigma}; {\boldsymbol \kappa})$ of characteristic polynomials. This is done for Gaussian Unitary Ensemble (GUE) and Laguerre Unitary Ensemble (LUE) whose treatment is very detailed. Correlation functions for two more matrix models -- Jacobi Unitary Ensemble (JUE) and Cauchy Unitary Ensemble (CyUE) -- are addressed in the Appendices \ref{App-JUE} and \ref{App-CyUE}.

Finally, in Section \ref{Sec-5}, we apply the integrable theory of CFCP to a comparative analysis of three alternative formulations of the replica method, with a special emphasis placed on the phenomenon of fermionic-bosonic factorisation of spectral correlation functions; some technical calculations involving functions of parabolic cylinder are collected in Appendix \ref{App-D-int}. To make the paper self-sufficient, we have included Appendix \ref{App-chazy} containing an overview of very basic facts on the six Painlev\'e transcendents and a closely related differential equation belonging to the Chazy I class.

The conclusions are presented in Section \ref{Sec-6}.

\section{Structure of the Theory}
\label{Sec-2}
The correlation function
\begin{eqnarray}\fl
\label{rpf-1}
    \Pi_{n|p}({\bm {\varsigma}};{\bm \kappa}) =\frac{1}{{\cal N}_n} \int_{{\cal D}^{n}} \prod_{j=1}^{n}
        \left( d\lambda_j \, e^{-V_n(\lambda_j)}\prod_{\alpha=1}^p (\varsigma_\alpha - \lambda_j)^{\kappa_\alpha} \right)
    \cdot
    \Delta_{n}^2({\boldsymbol \lambda})
\end{eqnarray}
to be considered in this section can be viewed as a natural
extension of its primary definition Eq.~(\ref{def-cf}). Written in
the eigenvalue representation (induced by the unitary rotation ${\bm
{\cal H}}={\bm {\cal U}}^\dagger {\bm \Lambda} {\bm {\cal U}}$ such
that ${\bm {\cal U}}\in{\bm {\cal U}}(n)$ and ${\bm \Lambda}={\rm
diag}(\lambda_1,\cdots,\lambda_n)$) it accommodates an
{\it $n$-dependent confinement potential} \footnote{Matrix integrals with $n$-dependent weights are known to appear in the bosonic formulations of replica field theories, see Osipov and Kanzieper (2007). This was the motivation behind our choice of the definition Eq.~(\ref{rpf-1}).} $V_n(\lambda)$ and also allows
for a generic eigenvalue integration domain \footnote[1]{In
applications to be considered in Section \ref{Sec-5}, the integration domain
${\mathcal D}$ will be set to ${\mathcal D}=[-1,+1]$ for (compact)
fermionic replicas, and to ${\mathcal D}=[0,+\infty]$ for
(noncompact) bosonic replicas. A more general setting Eq.
(\ref{D-dom}) does not complicate
    the theory we present.
}
\begin{eqnarray}
\label{D-dom}
    {\cal D} = \bigcup_{j=1}^r \left[ c_{2j-1}, c_{2j} \right],\;\;\; c_1 < \cdots < c_{2r}.
\end{eqnarray}
The normalisation constant ${\mathcal N}_n$ is
\begin{eqnarray}\label{norm}
    {\mathcal N}_{n} = \int_{{\cal D}^{n}} \prod_{j=1}^{n}
         d\lambda_j \, e^{-V_n(\lambda_j)}    \cdot
    \Delta_{n}^2({\boldsymbol \lambda}).
\end{eqnarray}
While for ${\bm \kappa}\in {\mathbb Z}_\pm^p$, the correlation
function $\Pi_{n|p}({\bm {\varsigma}};{\bm \kappa})$ can readily be
calculated by utilising the formalism due to Fyodorov and Strahov
(2003), there seems to be no simple extension of their method to
${\bm \kappa}\in {\mathbb R}^p$. It is this latter domain that will
be covered by our theory.

Contrary to the existing approaches which represent the CFCP {\it explicitly} in a determinant form
(akin to Eq.~(\ref{fs})), our formalism does not yield any closed
expression for the correlation function $\Pi_{n|p}({\bm
{\varsigma}};{\bm \kappa})$. Instead, it describes $\Pi_{n|p}({\bm
{\varsigma}};{\bm \kappa})$ {\it implicitly} in terms of a solution
to a nonlinear (partial) differential equation which -- along with an infinite
set of nonlinear (partial) differential hierarchies satisfied by
$\Pi_{n|p}({\bm {\varsigma}};{\bm \kappa})$ -- can be generated in a
regular way starting with Eq.~(\ref{rpf-1}). Let us stress that a
lack of explicitness is by no means a weak point of our theory: the
representations emerging from it save the day when a replica limit
is implemented in Eq.~(\ref{RL}).

Before plunging into the technicalities of the integrable theory of
CFCP, we wish to outline its general
structure.
\newline\newline\noindent
{\it Deformation.}---To determine the correlation function
$\Pi_{n|p}({\bm {\varsigma}};{\bm \kappa})$ nonperturbatively, we
adopt the ``deform-and-study'' approach, a standard string theory
method of revealing hidden structures. Its main idea consists of
``embedding'' $\Pi_{n|p}({\bm {\varsigma}};{\bm \kappa})$ into a
more general theory of the $\tau$ function
\begin{eqnarray}
\label{tau-f}
    \tau_{n}^{(s)}({\boldsymbol \varsigma},{\bm \kappa}; {\boldsymbol t}) = \frac{1}{n!}
    \int_{{\mathcal D}^{n}} \prod_{j=1}^{n}
    \left(
        d\lambda_j\,  \Gamma_{n-s}({\boldsymbol \varsigma},{\bm \kappa};\lambda_j)\,
        e^{v({\boldsymbol t};\lambda_j)}\right) \cdot
     \Delta_{n}^2({\boldsymbol \lambda})
\end{eqnarray}
which posseses an infinite-dimensional parameter space $(s;{\bm
t})=(s;t_1, t_2,\cdots)$ arising as the result of the
$(s;{\boldsymbol t})$-deformation of the weight function
\begin{eqnarray}
\label{Gamma-n}
    \Gamma_n({\boldsymbol \varsigma},{\bm \kappa};\lambda) =
     e^{-V_{n}(\lambda)}\prod_{\alpha=1}^p (\varsigma_\alpha - \lambda)^{\kappa_\alpha}
\end{eqnarray}
appearing in the original definition Eq.~(\ref{rpf-1}). The
parameter $s$ is assumed to be an integer, $s\in {\mathbb Z}$, and
$v({\bm t};\lambda)$ is defined as an infinite series
\begin{eqnarray}
\label{vt-def}
v({\boldsymbol t};\lambda) = \sum_{k=1}^\infty t_k\, \lambda^k,\;\;\;
{\bm t} = (t_1,t_2,\cdots).
\end{eqnarray}
Notice that a somewhat unusual $(s\in {\mathbb Z})$-deformation of
$\Gamma_n({\bm \varsigma}, {\bm \kappa};\lambda)$ is needed to
account for the $n$-dependent confinement potential $V_n(\lambda)$
in Eq.~(\ref{rpf-1}).
\newline\newline\noindent
{\it Bilinear identity and integrable hierarchies.}---Having
embedded the correlation function $\Pi_{n|p}({\bm \varsigma};{\bm
\kappa})$ into a set of $\tau$ functions $\tau_n^{(s)}({\bm \varsigma},{\bm
\kappa}; {\bm t})$, one studies the evolution of $\tau$ functions in
the extended parameter space $(n,s,{\bm t})$ in order to identify
nontrivial nonlinear differential hierarchical relations between
them. It turns out that an infinite set of hierarchically structured
nonlinear differential equations in the variables ${\bm
t}=(t_1,t_2,\cdots)$ can be encoded into a single {\it bilinear
identity}
\begin{eqnarray} \label{bi-id} \fl
    \oint_{{\cal C}_\infty} dz\,
    e^{(a-1)v({\bm t}-{\bm t}^\prime;z)} \, \tau_m^{(s)}({\bm t^\prime}-[\bm{z}^{-1}])
    \frac{\tau_{\ell+1}^{(\ell+1+s-m)}({\bm t}+[\bm{z}^{-1}])}{z^{\ell+1-m}} \nonumber \\
    =
    \nonumber \\
    \oint_{{\cal C}_\infty} dz\,
    e^{a\,v({\bm t}-{\bm t}^\prime;z)}\, \tau_\ell^{(\ell+s-m)} ({\bm t} - [\bm{z}^{-1}])
    \frac{\tau_{m+1}^{(s+1)}({\bm t}^\prime + [\bm{z}^{-1}])}{z^{m+1-\ell}},
\end{eqnarray}
where the integration contour ${\cal C}_\infty$ encompasses the
point $z=\infty$. Here, $a\in {\mathbb R}$ is a free parameter;  the
notation ${\bm t} \pm [{\bm z}^{-1}]$ stands for the infinite set of
parameters $\{t_k\pm z^{-k}/k\}_{k\in{\mathbb Z}_+}$; for brevity,
the physical parameters ${\bm \varsigma}$ and ${\bm \kappa}$ were
dropped from the arguments of $\tau$ functions.

To the best of our knowledge, this is the most general form of bilinear identity
that has ever appeared in the literature for Hermitean matrix models: not only it accounts for the $n$-dependent probability measure (``confinement potential'') but also
it generates, in a unified way, a whole zoo of integrable hierarchies satisfied by the $\tau$ function Eq.~(\ref{tau-f}). The latter was made possible by the daedal introduction of the free parameter $a$ in Eq.~(\ref{bi-id}) prompted by the study by Tu, Shaw and Yen (1996).

The bilinear
identity generates various integrable hierarchies in the $(n,s,{\bm
t})$ space. The {\it Kadomtsev-Petviashvili (KP) hierarchy},
  \begin{equation}
   \label{kph} \fl \qquad
    \frac{1}{2}\,D_1 D_k\, \tau_n^{(s)}(\bm{t})
    \circ \tau_n^{(s)}(\bm{t}) =  s_{k+1}([\bm{D}]) \, \tau_n^{(s)}(\bm{t})
    \circ \tau_n^{(s)}(\bm{t}),
\end{equation}
and the {\it Toda Lattice (TL) hierarchy},
    \begin{equation} \fl \qquad
    \label{tlh}
    \frac{1}{2}\, D_1 D_k  \,\tau_n^{(s)}(\bm{t})\circ
    \tau_{n}^{(s)}(\bm{t})= s_{k-1}([\bm{D}])\,
    \tau_{n+1}^{(s+1)}({\bm t}) \circ
    \tau_{n-1}^{(s-1)}({\bm t}),
\end{equation}
are central to our approach. In the above formulae, the vector ${\bm D}$ stands for ${\bm D} = (D_1, D_2,\cdots, D_k,\cdots)$ whilst
the $k$-th component of the vector $[{\bm D}]$ equals $k^{-1} D_k$. The operator symbol $D_k \, f({\bm t})\circ
g({\bm t})$ denotes the Hirota derivative \footnote[7]{The properties of Hirota differential
operators are briefly reviewed in Appendix \ref{App-hi}; see also the book by Hirota (2004).}
\begin{eqnarray}
   D_k\,f({\bm t})\circ g({\bm t}) = \frac{\partial}{\partial x_k}
   f({\bm t}+{\bm x})\, g({\bm t}-{\bm x})\Big|_{{\bm x}={\bm 0}}.
\end{eqnarray}
The functions $s_k({\bm t})$ are the Schur polynomials (Macdonald
1998) defined by the expansion
\begin{eqnarray} \label{SCHUR}
   \exp\left( \sum_{j=1}^\infty t_j x^j \right) = \sum_{\ell=0}^\infty
x^\ell s_\ell({\bm t}),
\end{eqnarray}
see also Table~\ref{schur-table}.  A complete list of emerging
hierarchies will be presented in Section \ref{Sec-3-4}.

\begin{table}
\caption{\label{schur-table} Explicit formulae for the lowest-order
Schur polynomials $s_\ell({\bm t})$ defined by the relation
$\exp\left( \sum_{j=1}^\infty t_j x^j \right) = \sum_{\ell=0}^\infty
x^\ell s_\ell({\bm t})$.}
\begin{indented}
\lineup
\item[]
\begin{tabular}{@{}*{2}{l}} \br
$\0\0 \ell$&$\ell!\,s_\ell({\bm t})$ \cr \mr

$\0\0 0$&$1$ \vspace{0.1cm} \cr

$\0\0 1$&$t_1$ \vspace{0.1cm} \cr

$\0\0 2$&$t_1^2+2t_2$ \vspace{0.1cm}\cr

$\0\0 3$&$t_1^3+6 t_1 t_2+6 t_3$\vspace{0.1cm} \cr

$\0\0 4$&$t_1^4+24 t_1 t_3+ 12 t_1^2 t_2+ 12 t_2^2+ 24 t_4$ \vspace{0.1cm}\cr

$\0\0 5$&$t_1^5+20 t_1^3 t_2 + 60 t_1^2 t_3 + 60 t_1 t_2^2+120 t_1 t_4+120 t_2 t_3+120 t_5$\vspace{0.1cm}\cr

\br
\end{tabular}\newline
\footnotesize{The Schur polynomials admit the representation (Macdonald 1998)
\begin{eqnarray}
    s_{\ell}({\bm t}) =
    \sum_{|\bm{\lambda}|=\ell}\prod_{j=1}^g\frac{t_{\ell_j}^{\sigma_j}}{\sigma_j!},
    \nonumber
\end{eqnarray}
where the summation runs over all partitions
$\blambda = (\ell_1^{\sigma_1},\cdots, \ell_g^{\sigma_g})$ of the
size $|\blambda|=\ell$. The notation
$\blambda = (\ell_1^{\sigma_1},\cdots,
    \ell_g^{\sigma_g})$, known as the frequency representation of
    the partition $\blambda$ of the size $|\blambda|=\ell$, implies
    that the part $\ell_j$ appears
    $\sigma_j$ times so that $\ell = \sum_{j=1}^g \ell_j\, \sigma_j$,
    where $g$ is the number of inequivalent parts of
the partition. Another way to compute $s_\ell({\bm t})$ is based on
the recursion equation
\begin{eqnarray}
s_\ell({\bm t})=\frac{1}{\ell}\sum_{j=1}^{\ell}
j\, t_j s_{\ell-j}({\bm t}),\;\;\;\ell \ge 1, \nonumber
\end{eqnarray}
supplemented by the condition $s_0({\bm t})=1$.}
\end{indented}
\end{table}
\noindent\newline{\it Projection.}---The projection formula
\begin{eqnarray}
\label{pf}
    \Pi_{n|p}({\bm \varsigma};{\bm \kappa}) = \frac{n!}{{\cal N}_n}
    \, \tau_n^{(s)}({\bm \varsigma},{\bm \kappa};{\bm t})\Big|_{s=0,\,{\bm t}={\bm 0}}
\end{eqnarray}
makes it tempting to assume that nonlinear integrable hierarchies
satisfied by $\tau$ functions in the $(n, s, {\bm t})$-space should
induce similar, hierarchically structured, nonlinear differential
equations for the correlation function $\Pi_{n|p}({\bm
\varsigma};{\bm \kappa})$. To identify them, one has to seek an
additional block of the theory that would make a link between the partial
$\{t_k\}_{k\in{\mathbb Z}_+}$ derivatives of $\tau$ functions taken
at ${\bm t}={\bm 0}$ and the partial derivatives of $\Pi_{n|p}({\bm
\varsigma};{\bm \kappa})$ over the {\it physical parameters}
$\{\varsigma_\alpha\}_{\alpha\in{\mathbb Z}_+}$. The study by Adler,
Shiota and van Moerbeke (1995) suggests that the missing block is
the {\it Virasoro constraints} for $\tau$
functions.
\newline\newline\noindent {\it Virasoro
constraints.}---The Virasoro constraints reflect the invariance of
$\tau$ functions [Eq.~(\ref{tau-f})] under a change of
integration variables. In the context of CFCP, it is useful to demand the invariance under an infinite
set of transformations \footnote[7]{The specific choice
Eq.~(\ref{vt}) will be advocated in Section \ref{Sec-3-6}.}
\begin{eqnarray}\label{vt}
    \lambda_j \rightarrow \mu_j + \epsilon \mu_j^{q+1} f(\mu_j)
    \prod_{k=1}^{\varrho} (\mu_j - c_k^\prime),\;\;\;
    q \ge -1,
\end{eqnarray}
labeled by integers $q$. Here, $\epsilon>0$, the vector $\bm{c^\prime}$ is ${\bm
c}^\prime=\{c_1,\cdots,c_{2r}\}\setminus\{\pm \infty, {\cal Z}_0\}$ with ${\cal Z}_0$ denoting a set of zeros of $f(\lambda)$, and $\varrho = {\rm dim\,}(\bm{c^\prime})$. The function $f(\lambda)$ is, in turn, related to the confinement potential
$V_{n-s}(\lambda)$ through the parameterisation
\begin{eqnarray}
\label{vns} \frac{dV_{n-s}}{d\lambda} =
\frac{g(\lambda)}{f(\lambda)},\;\;\;g(\lambda)=\sum_{k=0}^\infty b_k
\lambda^k,\;\;\; f(\lambda)=\sum_{k=0}^\infty a_k \lambda^k
\end{eqnarray}
in which both $g(\lambda)$ and $f(\lambda)$ depend on $n-s$ as do
the coefficients $b_k$ and $a_k$ in the above expansions. The
transformation Eq.~(\ref{vt}) induces the Virasoro-like constraints~\footnote[8]{The very notation $\hat{\cal L}_q^V$ suggests that this
operator originates from the confinement-potential-part $e^{-V_n}$
in Eqs.~(\ref{Gamma-n}) and (\ref{tau-f}). On the contrary, the
operator ${\hat {\cal L}}_q^{\rm det}$ is due to the
determinant-like product $\prod_{\alpha}(\varsigma_\alpha -
    \lambda)^{\kappa_\alpha}$ in Eq.~(\ref{Gamma-n}). Indeed,
    setting $\kappa_\alpha=0$ nullifies the operator ${\hat {\cal L}}_q^{\rm
    det}$. See Section \ref{Sec-3-6} for a detailed derivation.}
\begin{equation}
\label{2-Vir}
    \left[ \hat{{\cal L}}_{q}^V({\bm t}) + \hat{{\cal L}}_q^{\rm det}({\bm \varsigma};{\bm t})
     \right] \tau_n^{(s)}({\bm \varsigma};{\bm t})
    ={\hat {\cal B}}_q^V ({\bm \varsigma})\,\tau_n^{(s)}({\bm \varsigma};{\bm t}),
\end{equation}
where the differential operators
\begin{eqnarray} \fl
\label{vLv}
     \hat{{\cal L}}_{q}^V({\bm t}) = \sum_{\ell = 0}^\infty
    \sum_{k=0}^{\varrho}  s_{\varrho-k}(-{\bm p}_{\varrho} (\bm{c^\prime}))
       \left(
        a_\ell \hat{\cal L}_{q+k+\ell}({\bm t}) - b_\ell \frac{\partial}{\partial t_{q+k+\ell+1}}
    \right)
\end{eqnarray}
and
\begin{eqnarray} \fl
\label{vLG}
     \hat{{\cal L}}_{q}^{\rm det}({\bm t}) = \sum_{\ell = 0}^\infty
       a_\ell
    \sum_{k=0}^{\varrho} s_{\varrho-k}(-{\bm p}_{\varrho} (\bm{c^\prime}))
       \sum_{m=0}^{q+k+\ell} \left(\sum_{\alpha=1}^p \kappa_\alpha\,\varsigma_\alpha^m\right)
       \frac{\partial}{\partial t_{q+k+\ell-m}}
\end{eqnarray}
act in the ${\bm t}$-space whilst the differential operator
\begin{eqnarray}
\label{bq}
    {\hat {\cal B}}_q^V ({\bm \varsigma}) = \sum_{\alpha=1}^p
    \left( \prod_{k=1}^{\varrho} (\varsigma_\alpha - c_k^\prime) \right)
     f(\varsigma_\alpha) \,
    \varsigma_{\alpha}^{q+1} \frac{\partial}{\partial \varsigma_\alpha}
\end{eqnarray}
acts in the space of {\it physical parameters}
$\{\varsigma_\alpha\}_{\alpha\in{\mathbb Z}_+}$. The notation $s_k(-{\bm p}_{\varrho} (\bm{c^\prime}))$ stands for the Schur
polynomial and ${\bm p}_{\varrho}(\bm{c^\prime})$ is an infinite dimensional vector
\begin{eqnarray} \label{b9090}
    {\bm p}_\varrho(\bm{c^\prime})=\left(
    {\rm tr}_\varrho(\bm{c^\prime}), \frac{1}{2} {\rm tr}_\varrho(\bm{c^\prime})^2,\cdots,
    \frac{1}{k} {\rm tr}_\varrho(\bm{c^\prime})^k,\cdots
    \right)
\end{eqnarray}
with
\begin{eqnarray}
    {\rm tr}_\varrho(\bm{c^\prime})^k =
    \sum_{j=1}^{\varrho} (c_j^\prime)^k.
\end{eqnarray}
Notice that the operator $\hat{{\cal L}}_{q}^V({\bm t})$ is
expressed in terms of the Virasoro operators \footnote{~For $q=-1$, the second sum in Eq.~(\ref{vo}) is interpreted as zero.}
\begin{eqnarray}
    \label{vo}
    \hat{{\cal L}}_q({\bm t}) = \sum_{j=1}^\infty jt_j \,\frac{\partial}{\partial t_{q+j}}
    +
    \sum_{j=0}^q \frac{\partial^2}{\partial {t_j}\partial {t_{q-j}}},
\end{eqnarray}
obeying the Virasoro algebra
\begin{eqnarray}
\label{va}
[\hat{{\cal L}}_p,\hat{{\cal L}}_q] = (p-q)\hat{{\cal L}}_{p+q}, \;\;\;
p,q\ge -1.
\end{eqnarray}
\newline\noindent
{\it Projection (continued).}---Equations (\ref{pf}) and
(\ref{2-Vir}) suggest that there exists an infinite set of equations
which express various combinations of the derivatives
\begin{eqnarray} \nonumber
    \frac{\partial}{\partial t_j} \,\tau_n^{(s)}({\bm \varsigma};{\bm t})\Big|_{s=0,{\bm t}={\bm 0}}\;\;\; {\rm and}\;\;\;
    \frac{\partial^2}{\partial t_j \partial t_{k}} \,\tau_n^{(s)}({\bm \varsigma};{\bm t})\Big|_{
    s=0,{\bm t}={\bm 0}}
\end{eqnarray}
in terms of $
    {\hat {\cal B}}_q^V ({\bm \varsigma})\, \Pi_{n|p}({\bm \varsigma};{\bm \kappa})
$. This observation makes it tempting to project the hierarchical
relations Eqs. (\ref{kph}) and (\ref{tlh}) onto the hyperplane
$(s=0,{\bm t}={\bm 0})$ in an attempt to generate their analogues in
the space of {\it physical parameters}. In particular, such a
projection of the first equation of the KP hierarchy,
\begin{eqnarray} \fl
\left(
    \frac{\partial^4}{\partial t_1^4} + 3 \frac{\partial^2}{\partial t_2^2}
    - 4 \frac{\partial^2}{\partial t_1 \partial t_3}
\right)\log\, \tau_n^{(s)}({\bm t}) + 6 \left(
    \frac{\partial^2}{\partial t_1^2} \log\, \tau_n^{(s)}({\bm t})
\right)^2 = 0,
\end{eqnarray}
is expected \footnote[3]{Whether or not the projected Virasoro
constraints and the hierarchical equations always form a closed
system is a separate question that lies beyond the scope of the
present paper.} to bring a closed nonlinear differential equation
for the correlation function $\Pi_{n|p}({\boldsymbol
{\varsigma}};{\bm \kappa})$ of characteristic polynomials. It is
this equation which, being supplemented by appropriate boundary
conditions, provides an exact, nonperturbative description of the
averages of characteristic polynomials. Similarly, projections of
other equations from the hierarchies Eqs. (\ref{kph}) and
(\ref{tlh}) will reveal additional nontrivial nonlinear differential
relations that would involve not only $\Pi_{n|p}({\boldsymbol
{\varsigma}};{\bm \kappa})$ but its ``neighbours'' $\Pi_{n\pm
q|p}({\boldsymbol {\varsigma}};{\bm \kappa})$, as explained in Section \ref{Sec-4}.

Having exhibited the general structure of the theory, let us turn to
the detailed exposition of its main ingredients.

\section{From Characteristic Polynomials to $\tau$ Functions}
\label{Sec-3}
\subsection{The $\tau$ function, symmetry and associated scalar
product}
\label{Sec-3-1}

Integrability derives from the symmetry. In the context of $\tau$
functions Eq.~(\ref{tau-f}), the symmetry is encoded into
$\Delta_n^2({\bm \lambda})$, the squared Vandermonde determinant, as
it appears in the integrand below \footnote[4]{For the sake of
brevity, the physical parameters ${\bm \varsigma}$ and ${\bm
\kappa}$ were dropped from the arguments of $\tau_n^{(s)}$ and
$\Gamma_{n-s}$.}:
\begin{eqnarray}
\label{tau-f-c1}
    \tau_{n}^{(s)}({\boldsymbol t}) = \frac{1}{n!}
    \int_{{\mathcal D}^{n}} \prod_{j=1}^{n}
    \left(
        d\lambda_j\,  \Gamma_{n-s}(\lambda_j)\,
        e^{v({\boldsymbol t};\lambda_j)}\right) \cdot
     \Delta_{n}^2({\boldsymbol \lambda}).
\end{eqnarray}
In the random matrix theory language, the $\tau$ function Eq.~(\ref{tau-f-c1}) is said to posses the $\beta=2$ symmetry. Using the identity
\begin{eqnarray}
\Delta_n({\bm \lambda}) = {\rm det}[\lambda_j^{k-1}]_{1\le j,k \le
n}
    = {\rm det}[P_{k-1}(\lambda_j)]_{1\le j,k \le n}
\end{eqnarray}
with $P_k(\lambda)$ being an {\it arbitrary} set of monic
polynomials and the integration formula
(Andr\'eief 1883, de Bruijn 1955)
\begin{eqnarray} \fl
\label{adB}
    \int_{{\mathcal D}^n} \prod_{j=1}^n d\lambda_j \, {\rm det}_n[\varphi_j(\lambda_k)]\,
    {\rm det}_n[\psi_j(\lambda_k)] = n!\, {\rm det}_n\left[
        \int_{\mathcal D} d\lambda\, \varphi_j(\lambda)\,\psi_k(\lambda)
    \right],
\end{eqnarray}
the $\tau$ function Eq.~(\ref{tau-f-c1}) can be written as the
determinant
\begin{eqnarray}
\label{mm-det}
       \tau_{n}^{(s)}({\boldsymbol t})
        ={\rm det}\, \left[ \mu_{jk}^{(n-s)}({\bm t})
    \right]_{0\le j,k \le n-1}
\end{eqnarray}
of the matrix of moments
\begin{eqnarray}
    \mu_{jk}^{(m)}({\bm t}) =
    \left<
        P_j | P_k
    \right>_{\Gamma_{m}\, e^v} = \int_{\mathcal D} d\lambda\, \Gamma_{m}(\lambda)\, e^{v({\bm t};\lambda)}
        P_j(\lambda)\, P_k(\lambda)
\end{eqnarray}
Both the determinant representation and the scalar product
\begin{eqnarray}
\label{sc-prod}
    \left< f | g \right>_w = \int_{\mathcal D} d\lambda\, w(\lambda)\, f(\lambda)\, g(\lambda)
\end{eqnarray}
are dictated by the $\beta=2$ symmetry \footnote[3]{
The $\tau$ function Eq.~(\ref{tau-f-c1}) is a particular case of a more general $\tau$ function
\begin{eqnarray}
\label{tau-f-beta}
    \tau_{n}^{(s)}({\boldsymbol t};\beta) = \frac{1}{n!}
    \int_{{\mathcal D}^{n}} \prod_{j=1}^{n}
    \left(
        d\lambda_j\,  \Gamma_{n-s}(\lambda_j)\,
        e^{v({\boldsymbol t};\lambda_j)}\right) \cdot
     |\Delta_{n}({\boldsymbol \lambda})|^\beta. \nonumber
\end{eqnarray}
In accordance with the Dyson ``three-fold'' way (Dyson 1962), the symmetry parameter $\beta$ may also take the values $\beta=1$ and $\beta=4$. For these cases, the $\tau$ function Eq. (\ref{tau-f-beta}) admits the {\it Pfaffian} rather than determinant representation (Adler and van Moerbeke 2001):
\begin{eqnarray}
    \tau_n^{(s)} ({\bm t};\beta) = {\rm pf\,} \left[
        \mu_{jk}^{(n-s)}({\bm t};\beta)
    \right]_{0\le j,k \le n-1}, \nonumber
\end{eqnarray}
where the matrix of moments $\mu_{jk}^{(m)}({\bm t};\beta) =
    \left<
        P_j | P_k
    \right>_{\Gamma_{m}\, e^v}^{(\beta)}$ is defined through the scalar product
\begin{eqnarray}
    \left< f | g \right>_w^{(\beta)} = \cases{\int_{{\cal D}^2} d\lambda d\lambda^\prime
     w(\lambda) \, f(\lambda) \,{\rm sgn\,} (\lambda^\prime-\lambda) w(\lambda^\prime)g(\lambda^\prime), & $\beta=1$;\\
     \int_{{\cal D}} d\lambda
     w(\lambda) \left[ f(\lambda) g^\prime(\lambda)-g(\lambda) f^\prime(\lambda)\right], & $\beta=4$.}\nonumber
\end{eqnarray}

}
of the $\tau$ function.

\subsection{Bilinear identity in integral form}\label{Sec-3-2}\noindent
In this subsection, the bilinear identity Eq.~(\ref{bi-id}) will be proven.\newline\newline\noindent
{\it The $\tau$ function and orthogonal polynomials}.---The
representation Eq.~(\ref{mm-det}) reveals a special r\^ole played by
the monic polynomials $P_{k}^{(m)}({\bm t}; \lambda)$ {\it
orthogonal} on ${\mathcal D}$ with respect to the measure $\Gamma_m(\lambda)\, e^{v({\bm t};\lambda)}d\lambda$. Indeed, the
orthogonality relation
\begin{eqnarray} \fl \label{or}
  \left<
        P_k | P_j
  \right>_{\Gamma_m\,e^v}= \int_{\mathcal D} d\lambda\, \Gamma_{m}(\lambda)\, e^{v({\bm t};\lambda)}
  P_{k}^{(m)}({\bm t}; \lambda)\, P_{j}^{(m)}({\bm t}; \lambda) =  h_{k}^{(m)}({\bm t})\, \delta_{jk},
\end{eqnarray}
shows that the choice $P_j(\lambda)\mapsto P_{j}^{(n-s)}({\bm t};
\lambda)$ diagonalises the matrix of moments in Eq.~(\ref{mm-det})
resulting in the fairly compact representation
\begin{eqnarray}
        \tau_{n}^{(s)}({\boldsymbol t}) = \prod_{j=0}^{n-1} h_{j}^{(n-s)}({\bm t}).
\end{eqnarray}

Remarkably, the monic orthogonal polynomials $P_k^{(m)}({\bm
t};\lambda)$, that were introduced as a most economic tool for the
calculation of $\tau_n^{(s)}$, can themselves be expressed in terms
of $\tau$ functions:
\begin{eqnarray}\label{p-tau}
    P_k^{(m)}({\bm t};\lambda) = \lambda^k \,\frac{\tau_k^{(k-m)} ({\bm t} - [{\bm \lambda}^{-1}])}{\tau_k^{(k-m)}({\bm t})}.
\end{eqnarray}
Here, the notation ${\bm t} - [{\bm \lambda}^{-1}]$ stands for an
infinite-dimensional vector with the components
\begin{eqnarray}
    \label{t-shift}
    {\bm t} \pm [{\bm \lambda}^{-1}] = \left(
        t_1 \pm \frac{1}{\lambda},\, t_2 \pm \frac{1}{2\lambda^2},\,\cdots, t_k \pm \frac{1}{k\,\lambda^k},
        \cdots
    \right).
\end{eqnarray}
The statement Eq.~(\ref{p-tau}) readily follows from the definitions
Eqs.~(\ref{tau-f-c1}) and (\ref{vt-def}), the formal relation
\begin{eqnarray}
\label{v-shifted}
    e^{v({\bm t} \pm [{\bm \lambda}^{-1}];\,\lambda_j)} = e^{v({\bm t};\,\lambda_j)}
    \left(
            1  - \frac{\lambda_j}{\lambda}
    \right)^{\mp 1},
\end{eqnarray}
and the Heine formula (Heine 1878, Szeg\"o 1939)
\begin{eqnarray} \fl
\label{szego}
    P_k^{(m)}({\bm t};\lambda) = \frac{1}{k!\,\tau_k^{(k-m)}({\bm t})}
    \int_{{\mathcal D}^{k}} \prod_{j=1}^{k}
    \left(
        d\lambda_j\,  (\lambda-\lambda_j)\, \Gamma_{m}(\lambda_j)\,
        e^{v({\boldsymbol t};\lambda_j)}\right) \cdot
     \Delta_{k}^2({\boldsymbol \lambda}).
\end{eqnarray}
\newline\newline\noindent
{\it The $\tau$ function and Cauchy transform of orthogonal
polynomials}.---As will be seen later, the Cauchy transform of
orthogonal polynomials is an important ingredient of our proof of
the bilinear identity. Viewed as the scalar product,
\begin{eqnarray} \fl
\label{q-cauchy}
    Q_k^{(m)}({\bm t};z) = \left<
        P_k^{(m)}({\bm t};\lambda)\Big| \frac{1}{z-\lambda}
    \right>_{\Gamma_m \,e^v} = \int_{{\mathcal D}} d\lambda
    \,  \Gamma_{m}(\lambda)\,
        e^{v({\boldsymbol t};\lambda)} \, \frac{P_k^{(m)}({\bm t};\lambda)}{z-\lambda},
\end{eqnarray}
it can also be expressed in terms of $\tau$ function:
\begin{eqnarray}
\label{q-tau}
    Q_{k}^{(m)}({\bm t};z) = z^{-k-1} \frac{\tau_{k+1}^{(k+1-m)} ({\bm t}+[{\bm z}^{-1}])}{\tau_k^{(k-m)}({\bm t})}.
\end{eqnarray}
To prove Eq.~(\ref{q-tau}), we  substitute Eq.~(\ref{szego}) into
Eq.~(\ref{q-cauchy}) to derive:
\begin{eqnarray} \fl
\label{qc-st-1}
    Q_{k}^{(m)}({\bm t};z) &= \frac{1}{k!\,\tau_k^{(k-m)}({\bm t})}
    \int_{{\mathcal D}^{k+1}} \prod_{j=1}^{k+1}
    \left(
        d\lambda_j\,  \Gamma_{m}(\lambda_j)\,
        e^{v({\boldsymbol t};\lambda_j)}\right) \cdot \Delta_{k+1}^2({\boldsymbol \lambda})
        \nonumber \\ \fl
        &\times
     \frac{1}{(z-\lambda_{k+1})}\prod_{j=1}^{k} \frac{1}{\lambda_{k+1}-\lambda_j}.
\end{eqnarray}
Owing to the identity
\begin{eqnarray}
\label{id-2}
    \prod_{j=1}^n \frac{1}{z-\lambda_j} = \sum_{\alpha=1}^n\left(
        \frac{1}{z-\lambda_\alpha} \prod_{j=1,
        \;j\neq\alpha}^n\frac{1}{\lambda_\alpha-\lambda_j}
    \right)
\end{eqnarray}
taken at $n=k+1$, the factor
\begin{eqnarray}
    \frac{1}{(z-\lambda_{k+1})}\prod_{j=1}^{k} \frac{1}{\lambda_{k+1}-\lambda_j} \nonumber
\end{eqnarray}
in the integrand of Eq.~(\ref{qc-st-1}) can be symmetrised
\begin{eqnarray} \fl
    \frac{1}{(z-\lambda_{k+1})}\prod_{j=1}^{k} \frac{1}{\lambda_{k+1}-\lambda_j} \mapsto
    \frac{1}{k+1} \sum_{\alpha=1}^{k+1}\left(
        \frac{1}{z-\lambda_\alpha} \prod_{j=1,
        \;j\neq\alpha}^{k+1}\frac{1}{\lambda_\alpha-\lambda_j}
    \right) \nonumber\\
    \qquad \qquad= \frac{1}{k+1} \prod_{j=1}^{k+1} \frac{1}{z-\lambda_j} \nonumber
\end{eqnarray}
to yield the representation
\begin{eqnarray}
\label{qc-st-2} \fl
    Q_{k}^{(m)}({\bm t};z) &= \frac{1}{(k+1)!\,\tau_k^{(k-m)}({\bm t})}
    \int_{{\mathcal D}^{k+1}} \prod_{j=1}^{k+1}
    \left(
        \frac{d\lambda_j}{z-\lambda_j}\,  \Gamma_{m}(\lambda_j)\,
        e^{v({\boldsymbol t};\lambda_j)}\right) \cdot \Delta_{k+1}^2({\boldsymbol \lambda}).
\end{eqnarray}
In view of Eq.~(\ref{v-shifted}), this is seen to coincide with
\begin{eqnarray}
    \frac{z^{-k-1} }{(k+1)!\,\tau_k^{(k-m)}({\bm t})}
    \int_{{\mathcal D}^{k+1}} \prod_{j=1}^{k+1} \left(
    d\lambda_j\, \Gamma_{m}(\lambda_j)\,
        e^{v({\boldsymbol t}+\bm{[z^{-1}]};\lambda_j)}\right) \cdot \Delta_{k+1}^2({\boldsymbol \lambda}). \nonumber
\end{eqnarray}
Comparison with the definition Eq.~(\ref{tau-f-c1}) completes the
proof of Eq.~(\ref{q-tau}).
\newline\newline\noindent
{\it Proof of the bilinear identity.}---Now we are ready to prove
the bilinear identity
\begin{eqnarray} \label{bi-id-rep} \fl
\oint_{{\cal C}_\infty} dz\,
    e^{(a-1)v({\bm t}-{\bm t}^\prime;z)} \, \tau_m^{(s)}({\bm t^\prime}-[\bm{z}^{-1}])
    \frac{\tau_{\ell+1}^{(\ell+1+s-m)}({\bm t}+[\bm{z}^{-1}])}{z^{\ell+1-m}} \nonumber \\
    =
    \nonumber \\
    \oint_{{\cal C}_\infty} dz\,
    e^{a\,v({\bm t}-{\bm t}^\prime;z)}\, \tau_\ell^{(\ell+s-m)} ({\bm t} - [\bm{z}^{-1}])
    \frac{\tau_{m+1}^{(s+1)}({\bm t}^\prime + [\bm{z}^{-1}])}{z^{m+1-\ell}},
\end{eqnarray}
where the integration contour ${\cal C}_\infty$ encompasses the
point $z=\infty$, and $a\in {\mathbb R}$ is a free parameter.

We start with the needlessly fancy identity
\begin{eqnarray} \fl \label{fancy}
    \int_{\cal D} d\lambda \, \Gamma_n(\lambda)\, e^{v({\bm t};\lambda)}
    e^{(a-1)v({\bm t}-{\bm t}^\prime;\lambda)} P_\ell^{(n)}({\bm t};\lambda)
    P_m^{(n)}({\bm t}^\prime;\lambda) \nonumber \\=
    \int_{\cal D} d\lambda \, \Gamma_n(\lambda)\, e^{v({\bm t}^\prime;\lambda)}
    e^{a\,v({\bm t}-{\bm t}^\prime;\lambda)} P_\ell^{(n)}({\bm t};\lambda)
    P_m^{(n)}({\bm t}^\prime;\lambda)
\end{eqnarray}
whose structure is prompted by the scalar product Eq.~(\ref{or}) and
which trivially holds due to a linearity of the ${\bm
t}$-deformation
\begin{eqnarray}
    v({\bm t};\lambda) + (a-1)\,v({\bm t}-{\bm t}^\prime;\lambda)
    =
    v({\bm t}^\prime;\lambda) + a\,v({\bm t}-{\bm t}^\prime;\lambda),
\end{eqnarray}
see Eq.~(\ref{vt-def}).

The formulae relating the orthogonal
polynomials and their Cauchy transforms to $\tau$ functions
[Eqs.~(\ref{p-tau}) and (\ref{q-tau})] make it possible to express
both sides of Eq.~(\ref{fancy}) in terms of $\tau$ functions with
shifted arguments. \newline\newline\noindent
(i) Due to the Cauchy integral representation
\begin{eqnarray} \label{lhs-cauchy}\fl
    e^{(a-1)v({\bm t}-{\bm t}^\prime;\lambda)} P_m^{(n)}({\bm t}^\prime;\lambda) =
    \frac{1}{2\pi i} \oint_{{\cal C}_\infty} dz\,
    e^{(a-1)v({\bm t}-{\bm t}^\prime;z)} \frac{P_m^{(n)}({\bm t}^\prime;z)}{z-\lambda},
\end{eqnarray}
the l.h.s. of Eq.~(\ref{fancy}) can be transformed as follows:
\begin{eqnarray} \fl
    {\rm l.h.s.} = \frac{1}{2\pi i}
    \oint_{{\cal C}_\infty} dz\,
    e^{(a-1)v({\bm t}-{\bm t}^\prime;z)} \, P_m^{(n)}({\bm t}^\prime;z)
    \underbrace{\int_{\cal D} d\lambda \, \Gamma_n(\lambda)\, e^{v({\bm t};\lambda)}
    \frac{P_\ell^{(n)}({\bm t};\lambda)}{z-\lambda}}_{Q_\ell^{(n)}({\bm t};z)\;\;\; [{\rm Eq.~(\ref{q-cauchy})}]} \nonumber\\
    =
    \frac{1}{2\pi i}
    \oint_{{\cal C}_\infty} dz\,
    e^{(a-1)v({\bm t}-{\bm t}^\prime;z)} \, P_m^{(n)}({\bm t}^\prime;z)
    Q_\ell^{(n)}({\bm t};z).
\end{eqnarray}
Taking into account Eqs.~(\ref{p-tau}) and (\ref{q-tau}), this is further
reduced to
\begin{eqnarray} \fl
    \label{lhs-1}
    {\rm l.h.s.} = \frac{1}{2\pi i} \frac{1}{\tau_\ell^{(\ell-n)}({\bm t})\, \tau_{m}^{(m-n)}({\bm t^\prime})} \nonumber \\
    \times
    \oint_{{\cal C}_\infty} dz\,
    e^{(a-1)v({\bm t}-{\bm t}^\prime;z)} \, \tau_m^{(m-n)}({\bm t^\prime}-[\bm{z}^{-1}])
    \frac{\tau_{\ell+1}^{(\ell+1-n)}({\bm t}+[\bm{z}^{-1}])}{z^{\ell+1-m}}.
\end{eqnarray}
\newline\newline\noindent (ii) To transform the r.h.s. of
Eq.~(\ref{fancy}), we make use of the Cauchy theorem in the form
\begin{eqnarray} \label{rhs-cauchy}
    e^{a\,v({\bm t}-{\bm t}^\prime;\lambda)} P_\ell^{(n)}({\bm t};\lambda)=
    \frac{1}{2\pi i}
    \oint_{{\cal C}_\infty} dz\,
    e^{a\,v({\bm t}-{\bm t}^\prime;z)} \frac{P_\ell^{(n)}({\bm t};z)}{z-\lambda},
\end{eqnarray}
to get:
\begin{eqnarray} \fl
    {\rm r.h.s.} = \frac{1}{2\pi i}
        \oint_{{\cal C}_\infty} dz\,
    e^{a\,v({\bm t}-{\bm t}^\prime;z)} P_\ell^{(n)}({\bm t};z)
        \underbrace{\int_{\cal D} d\lambda \, \Gamma_n(\lambda)\, e^{v({\bm t}^\prime;\lambda)}
    \frac{
    P_m^{(n)}({\bm t};\lambda)}{z-\lambda}}_{Q_m^{(n)}({\bm t^\prime};z)\;\;\; [{\rm Eq.~(\ref{q-cauchy})}]}
    \nonumber \\
    =
    \frac{1}{2\pi i}
        \oint_{{\cal C}_\infty} dz\,
    e^{a\,v({\bm t}-{\bm t}^\prime;z)} P_\ell^{(n)}({\bm t};z)\, Q_m^{(n)}({\bm t^\prime};z).
\end{eqnarray}
Taking into account Eqs.~(\ref{p-tau}) and (\ref{q-tau}), this is further
reduced to
\begin{eqnarray} \fl \label{rhs-1}
    {\rm r.h.s.} = \frac{1}{2\pi i} \frac{1}{\tau_\ell^{(\ell-n)}({\bm t})\, \tau_{m}^{(m-n)}({\bm t^\prime})}
    \nonumber \\
        \times \oint_{{\cal C}_\infty} dz\,
    e^{a\,v({\bm t}-{\bm t}^\prime;z)}\, \tau_\ell^{(\ell-n)} ({\bm t} - [\bm{z}^{-1}])
    \frac{\tau_{m+1}^{(m+1-n)}({\bm t}^\prime + [\bm{z}^{-1}])}{z^{m+1-\ell}}.
\end{eqnarray}
The bilinear identity Eq.~(\ref{bi-id-rep}) follows from
Eqs.~(\ref{lhs-1}) and (\ref{rhs-1}) after setting $n=m-s$. End of proof.

\subsection{Bilinear identity in Hirota form}\label{Sec-3-3}
\label{bi-hirota}
The bilinear identity Eq.~(\ref{bi-id-rep}) can alternatively be written in the Hirota form:
\begin{eqnarray}
\label{bi-hf} \fl
    e^{\beta
    \bm{(x}\cdot\bm{D)}}
    \sum_{k=0}^\infty s_k\left( (2a-1-\beta){\bm x}\right)
    s_{k+q+1}\left(
        [\bm{D}]
    \right)\, \tau_{p}^{(s)}({\bm t})\,\circ
    \tau_{p+q}^{(s+q)}({\bm t}) \nonumber \\ \fl \qquad
    = e^{-\beta
    \bm{(x}\cdot\bm{D)}}
    \sum_{k=q+1}^\infty s_k\left((2a-1+\beta){\bm x}\right)
    s_{k-q-1}\left(
        [\bm{D}]
    \right)\,\tau_{p+q+1}^{(s+q+1)}({\bm t})\circ
    \tau_{p-1}^{(s-1)} ({\bm t})
\end{eqnarray}
where $\beta=\pm 1$~(not to be confused with the Dyson symmetry index!), $p \ge 1$ and $q \ge -1$. Let us remind that the vector ${\bm D}$ appearing in the scalar product $({\bm x}\cdot {\bm D})=\sum_{k} x_k D_k$ stands for ${\bm D} = (D_1, D_2,\cdots, D_k,\cdots)$; the $k$-th component of the vector $[{\bm D}]$ equals $k^{-1} D_k$ (compare with Eq.~(\ref{t-shift})). The generic Hirota differential
operator ${\cal P}({\bm D})\, f({\bm t})
\circ g({\bm t})$ is defined in Appendix \ref{App-hi}. Also, $s_k({\bm x})$ are the Schur polynomials defined in Eq.~(\ref{SCHUR}).
\noindent\newline\newline
To prove Eq.~(\ref{bi-hf}), we proceed in two steps.

(i) First, we set the vectors ${\bm t}$ and ${\bm
t}^\prime$ in Eq.~(\ref{bi-id-rep}) to be
\begin{eqnarray} \label{ttprime}
    ({\bm t},{\bm t}^\prime) \mapsto ({\bm t}+{\bm x},{\bm t}-{\bm x}).
\end{eqnarray}
The parameterisation Eq.~(\ref{ttprime}) allows us to rewrite the $({\bm t},{\bm t}^\prime)$ dependent part of the
integrand in the l.h.s. of Eq.~(\ref{bi-id-rep})
\begin{eqnarray}
e^{(a-1)v({\bm t}-{\bm t}^\prime;z)} \, \tau_m^{(s)}({\bm t^\prime}-[\bm{z}^{-1}])
    \tau_{\ell+1}^{(\ell+1+s-m)}({\bm t}+[\bm{z}^{-1}]) \nonumber
\end{eqnarray}
as follows:
\begin{eqnarray} \label{is-01} \fl
    \exp\left[ v\big(2(a-1){\bm x};z\big)\right] \, \tau_m^{(s)}({\bm t}-{\bm x}-[\bm{z}^{-1}])
    \tau_{\ell+1}^{(\ell+1+s-m)}({\bm t}+{\bm
    x}+[\bm{z}^{-1}])\nonumber\\ \fl
    =
    \exp\left[ v\big(2(a-1){\bm x};z\big)\right] \nonumber\\
    \fl \qquad \times \exp\left[ \bm{(x}\cdot\bm{\partial_\xi)} + \bm{(}[\bm{z}^{-1}]\cdot\bm{\partial_\xi)}\right]
    \tau_{\ell+1}^{(\ell+1+s-m)}({\bm t}+{\bm \xi})\,
    \tau_m^{(s)}({\bm t}-{\bm \xi})\Big|_{{\bm \xi}=0}.
\end{eqnarray}
Here, we have used the linearity of the $t$-deformation, $\alpha \,
v({\bm t};z) = v(\alpha{\bm t};z)$. Further, we spot the identity
$\bm{(}[\bm{z}^{-1}]\cdot\bm{\partial_\xi)}=v \left(
[\bm{\partial_\xi}]; z^{-1}\right)$ to reduce Eq.~(\ref{is-01}) to \footnote[5]{Here,
\begin{eqnarray}
    [\bm{\partial_\xi}] = \left(
        \frac{\partial}{\partial \xi_1}, \frac{1}{2}\frac{\partial}{\partial \xi_2},\cdots,
        \frac{1}{k}\frac{\partial}{\partial \xi_k},\cdots
    \right). \nonumber
\end{eqnarray}}
\begin{eqnarray}    \fl
    \exp\left[ v\big(2(a-1){\bm x};z\big)\right]\nonumber \\ \fl
    \qquad \times \exp\left[
    \bm{(x}\cdot\bm{\partial_\xi)}\right]\,
    \exp\left[
        v \left( [\bm{\partial_\xi}]; z^{-1}\right)
    \right]
    \tau_{\ell+1}^{(\ell+1+s-m)}({\bm t}+{\bm \xi})\,
    \tau_m^{(s)}({\bm t}-{\bm \xi})\Big|_{{\bm \xi}=0}. \nonumber
\end{eqnarray}
The latter can be rewritten in
terms of Hirota differential operators [see Eq.~(\ref{b2})] with the
result being
\begin{eqnarray} \fl
    \exp\left[ v\big(2(a-1){\bm x};z\big)\right]
    \exp\left[
    \bm{(x}\cdot\bm{D)}\right]\,
    \exp\left[
        v \left( [\bm{D}]; z^{-1}\right)
    \right]
    \tau_{\ell+1}^{(\ell+1+s-m)}({\bm t})\,\circ
    \tau_m^{(s)}({\bm t}).
\end{eqnarray}
By the same token, the $({\bm t},{\bm t}^\prime)$ dependent part of
the integrand in the r.h.s. of Eq.~(\ref{bi-id-rep}),
\begin{eqnarray}
e^{a\,v({\bm t}-{\bm t}^\prime;z)}\, \tau_\ell^{(\ell+s-m)} ({\bm t}
- [\bm{z}^{-1}])\,\tau_{m+1}^{(s+1)}({\bm t}^\prime + [\bm{z}^{-1}])
\nonumber
\end{eqnarray}
can be reduced to
\begin{eqnarray} \fl
    \exp\left[ v\big(2a{\bm x};z\big)\right]
    \exp\left[-
    \bm{(x}\cdot\bm{D)}\right]\,
    \exp\left[
         v \left( [\bm{D}]; z^{-1}\right)
    \right]
    \,\tau_{m+1}^{(s+1)}({\bm t})\circ
    \tau_\ell^{(\ell+s-m)} ({\bm t}).
\end{eqnarray}
Thus, we end up with the alternative representation for the bilinear
identity Eq.~(\ref{bi-id-rep}):
\begin{eqnarray} \label{316-rew}\fl
    e^{
    \bm{(x}\cdot\bm{D)}}
 \oint_{{\cal C}_\infty} \frac{dz}{z^{\ell-m+1}} \,
 \exp\left[ v\big(2(a-1){\bm x};z\big)\right]
    \,
    \exp\left[
        v \left( [\bm{D}]; z^{-1}\right)
    \right]
    \tau_{\ell+1}^{(\ell+1+s-m)}({\bm t})\,\circ
    \tau_m^{(s)}({\bm t}) \nonumber \\ \fl \quad
    = e^{-
    \bm{(x}\cdot\bm{D)}} \oint_{{\cal C}_\infty} \frac{dz}{z^{m-\ell+1}} \,
    \exp\left[ v\big(2a{\bm x};z\big)\right]
    \,
    \exp\left[
         v \left( [\bm{D}]; z^{-1}\right)
    \right]
    \,\tau_{m+1}^{(s+1)}({\bm t})\circ
    \tau_\ell^{(\ell+s-m)} ({\bm t}). \nonumber\\{}
\end{eqnarray}

(ii) Second, to facilitate the integration in Eq.~(\ref{316-rew}), we rewrite the integrands therein in the form of Laurent series in $z$ by employing the identity
\begin{eqnarray}
    e^{v({\bm t};z)} = \exp\left(
        \sum_{j=1}^\infty t_j z^j
    \right) = \sum_{k=0}^\infty s_k({\bm t})\, z^k.
\end{eqnarray}
Now, the integrals in Eq.~(\ref{316-rew}) are easily performed to yield
\begin{eqnarray} \label{bid-2}\fl
    e^{
    \bm{(x}\cdot\bm{D)}}
    \sum_{k=\max(0,\ell-m)}^\infty s_k\left(2(a-1){\bm x}\right)
    s_{k+m-\ell}\left(
        [\bm{D}]
    \right)\, \tau_{\ell+1}^{(\ell+1+s-m)}({\bm t})\,\circ
    \tau_m^{(s)}({\bm t}) \nonumber \\ \fl \quad
    = e^{-
    \bm{(x}\cdot\bm{D)}}
    \sum_{k=\max(0,m-\ell)}^\infty s_k\left(2a{\bm x}\right)
    s_{k+\ell-m}\left(
        [\bm{D}]
    \right)\,\tau_{m+1}^{(s+1)}({\bm t})\circ
    \tau_\ell^{(\ell+s-m)} ({\bm t}).
\end{eqnarray}

It remains to verify that Eq.~(\ref{bid-2}) is equivalent to the announced result Eq.~(\ref{bi-hf}). To this end we
distinguish between two different cases. (i) If $\ell \le m$, we set $\ell=p-1$, $m=p+q$ and $s \mapsto s+q$ in Eq.~(\ref{bid-2}) to find out that it reduces to Eq.~(\ref{bi-hf}) taken at $\beta=+1$; (ii) If $\ell > m$, we set $\ell=p+q$, $m=p-1$ and $s \mapsto s-1$ in Eq.~(\ref{bid-2}) to find out that it reduces to Eq.~(\ref{bi-hf}) taken at $\beta=-1$. This ends the proof.

For an
alternative derivation of Eq.~(\ref{bi-hf}) the reader is referred to Appendix \ref{App-bi}.

\subsection{Zoo of integrable
hierarchies}\label{Sec-3-4}
The bilinear identity, in either form, encodes an infinite set of
hierarchically structured nonlinear differential equations in the
variables ${\bm t}$. Two of these hierarchies -- the KP and the TL hierarchies --
were mentioned in Section \ref{Sec-2}. Below, we provide a complete list of integrable hierarchies
associated with the $\tau$ function Eq.~(\ref{tau-f-c1}).

To identify them, we expand the bilinear identity in Hirota form [Eq.~(\ref{bi-hf})] around
${\bm x}={\bm 0}$ and $a=0$, keeping only linear in ${\bm x}$ terms. Since $s_0({\bm t})=1$ and
\begin{eqnarray}
    s_k({\bm t})\Big|_{{\bm t}\rightarrow {\bm 0}} =   t_k + {\cal O}({\bm
    t}^2),\quad k = 1,\, 2,\,\dots
\end{eqnarray}
we obtain:
\begin{eqnarray} \label{bi-hf-exp} \fl
 (1-\delta_{q,-1})\, s_{q+1}([{\bm D}]) \, \tau_{p}^{(s)}({\bm t})\,\circ
    \tau_{p+q}^{(s+q)}({\bm t}) \nonumber\\ \fl
    + \sum_{k=1}^\infty x_k\left[
    (2a-1-\beta)s_{k+q+1}\left([\bm{D}]\right) + \beta D_k \Big(
    s_{q+1}\left([\bm{D}]\right) + \delta_{q,-1}\Big)
    \right]
    \, \tau_{p}^{(s)}({\bm t})\,\circ\tau_{p+q}^{(s+q)}({\bm t})
    \nonumber \\ \fl
    -
    (2a-1+\beta)\sum_{k=\max(1,q+1)}^\infty x_k
    s_{k-q-1}\left(
        [\bm{D}]
    \right)\,\tau_{p+q+1}^{(s+q+1)}({\bm t})\circ
    \tau_{p-1}^{(s-1)} ({\bm t})+ {\cal O}({\bm x}^2)=0.
\end{eqnarray}
As soon as Eq.~(\ref{bi-hf}) holds for arbitrary $a$ and ${\bm
x}$, Eq.~(\ref{bi-hf-exp}) generates four
identities.\newline\newline\noindent
(i) The first identity
\begin{eqnarray} \label{i-3}
\quad
    s_{k+q+1}\left([\bm{D}]\right) \, \tau_{p}^{(s)}({\bm t})\,\circ\tau_{p+q}^{(s+q)}({\bm
    t}) = 0
\end{eqnarray}
holds for $q\ge 1$ and $k=0,\,1,\,\dots\,,\,q$.
\newline\newline\noindent
(ii) The second identity
\begin{eqnarray} \label{i-2} \fl
    \qquad \big[
    (1+\beta)\,s_{k+q+1}\left([\bm{D}]\right) - \beta D_k
    s_{q+1}\left([\bm{D}]\right)
    \big]
    \, \tau_{p}^{(s)}({\bm t})\,\circ\tau_{p+q}^{(s+q)}({\bm t})=0
\end{eqnarray}
holds for $q\ge 1$ and $k=1,\,2,\,\dots\,,\,q$.
\newline\newline\noindent
(iii) The third identity
\begin{eqnarray} \label{i-4} \fl
    \big[
    (1+\beta)\,s_{k+q+1}\left([\bm{D}]\right) - \beta D_k
    \big( s_{q+1}\left([\bm{D}]\right) + \delta_{q,-1} \big)
    \big]
    \, \tau_{p}^{(s)}({\bm t})\,\circ\tau_{p+q}^{(s+q)}({\bm t}) \nonumber \\
    = (1-\beta)\, s_{k-q-1}\left(
        [\bm{D}]
    \right)\,\tau_{p+q+1}^{(s+q+1)}({\bm t})\circ
    \tau_{p-1}^{(s-1)} ({\bm t})
\end{eqnarray}
holds for $q\ge -1$ and $k\ge\max(1,q+1)$.
\newline\newline\noindent
(iv) The last, fourth identity
\begin{eqnarray} \label{i-5} \fl
    s_{k+q+1}\left([\bm{D}]\right)
    \, \tau_{p}^{(s)}({\bm t})\,\circ\tau_{p+q}^{(s+q)}({\bm t})
    =  s_{k-q-1}\left(
        [\bm{D}]
    \right)\,\tau_{p+q+1}^{(s+q+1)}({\bm t})\circ
    \tau_{p-1}^{(s-1)} ({\bm t})
\end{eqnarray}
holds for $q\ge 0$ and $k\ge q+1$.
\newline\newline
Equations (\ref{i-3}) -- (\ref{i-5}) can further be classified to yield the following bilinear hierarchies:
\newline
\begin{itemize}
  \item {\it Toda Lattice (TL) hierarchy:}
    \begin{eqnarray} \fl\label{TL}
    \frac{1}{2} D_1 D_k
    \, \tau_{p}^{(s)}({\bm t})\,\circ\tau_{p}^{(s)}({\bm t}) =
    s_{k-1}\left([\bm{D}]\right)
    \,\tau_{p+1}^{(s+1)}({\bm t})\circ
    \tau_{p-1}^{(s-1)} ({\bm t})
    \end{eqnarray}
with $k\ge 1$.
\newline
  \item {\it q-modified Toda Lattice hierarchy:}
  \begin{eqnarray} \fl \label{qTL}
    \frac{1}{2} D_k s_{q+1}\left([\bm{D}]\right)
    \, \tau_{p}^{(s)}({\bm t})\,\circ\tau_{p+q}^{(s+q)}({\bm t}) =
    s_{k-q-1}\left([\bm{D}]\right)
    \,\tau_{p+q+1}^{(s+q+1)}({\bm t})\circ
    \tau_{p-1}^{(s-1)} ({\bm t})
\end{eqnarray}
with $q\ge 0$ and $k\ge q+1$. (For $q=0$, it reduces to the above Toda
Lattice hierarchy.)
\newline
  \item {\it Kadomtsev-Petviashvili (KP) hierarchy:}
\begin{eqnarray} \fl \label{KP}
    \left[
    \frac{1}{2} D_1 D_k - s_{k+1}\left([\bm{D}]\right) \right]
    \, \tau_{p}^{(s)}({\bm t})\,\circ\tau_{p}^{(s)}({\bm t}) = 0
\end{eqnarray}
with \footnote[8]{Both $k=1$ and $k=2$ bring trivial statements, see Appendix
\ref{App-hi}.} $k\ge 3$.
\newline
  \item {\it q-modified Kadomtsev-Petviashvili hierarchy:}
  \begin{eqnarray} \fl \label{qKP}
    \frac{1}{2} D_k s_{q+1}\left([\bm{D}]\right)
    \, \tau_{p}^{(s)}({\bm t})\,\circ\tau_{p+q}^{(s+q)}({\bm t}) =
    s_{k+q+1}\left([\bm{D}]\right)
    \, \tau_{p}^{(s)}({\bm t})\,\circ\tau_{p+q}^{(s+q)}({\bm
    t})
\end{eqnarray}
    with $q\ge 0$ and $k\ge q+1$. (For $q=0$, it reduces to the above KP
    hierarchy.)\newline

   \item {\it Left q-modified Kadomtsev-Petviashvili hierarchy:}
\begin{eqnarray} \fl
    D_k s_{q+1}\left([\bm{D}]\right)
    \, \tau_{p}^{(s)}({\bm t})\,\circ\tau_{p+q}^{(s+q)}({\bm t}) =
    0
\end{eqnarray}
with $q\ge 1$ and $1 \le k \le q$. \newline

   \item {\it Right q-modified Kadomtsev-Petviashvili hierarchy:}
       \begin{eqnarray} \fl
    s_{k+q+1}\left([\bm{D}]\right)
    \, \tau_{p}^{(s)}({\bm t})\,\circ\tau_{p+q}^{(s+q)}({\bm t}) =
    0
\end{eqnarray}
with $q\ge 1$ and $0 \le k \le q$.
\newline
\item {\it $(-1)$-modified Kadomtsev-Petviashvili hierarchy:}
\begin{eqnarray} \label{346}\fl
    \big[
        s_{k}\left([\bm{D}]\right) - D_k
    \big]\, \tau_{p}^{(s)}({\bm t})\,\circ\tau_{p-1}^{(s-1)}({\bm
    t})=0
\end{eqnarray}
with $k\ge 2$.
\newline
\end{itemize}
Notice, that the modified hierarchies will play no r\^ole in further
consideration.

\subsection{KP and Toda Lattice hierarchies}\label{Sec-3-5}
As was pointed out in Section \ref{Sec-2}, the KP and Toda Lattice
hierarchies are of primary importance for our formalism. In this
subsection, we explicitly present a few first members of these
hierarchies.
\newline\newline\noindent
{\it KP hierarchy.}---Due to the properties of Hirota symbol
reviewed in Appendix \ref{App-hi}, the first nontrivial equation of the KP
hierarchy corresponds to $k=3$ in Eq.~(\ref{KP}). Consulting Table
\ref{schur-table} and having in mind that $[{\bm D}]_k = k^{-1}
D_k$, we derive the first two members, ${\rm KP}_1$ and ${\rm
KP}_2$, of the KP hierarchy in Hirota form
\begin{eqnarray}
\label{kp-1}
   {\rm KP}_1:\quad (D_1^4 - 4 D_1 D_3 + 3 D_2^2) \, \tau_{p}^{(s)}({\bm t})\,\circ\tau_{p}^{(s)}({\bm
   t}) = 0, \\
   \label{kp-2}
    {\rm KP}_2:\quad (D_1^3D_2 + 2 D_2 D_3 - 3 D_1 D_4) \, \tau_{p}^{(s)}({\bm t})\,\circ\tau_{p}^{(s)}({\bm
   t}) = 0.
\end{eqnarray}
In deriving Eqs.~(\ref{kp-1}) and (\ref{kp-2}), we have used the Property 2a from Appendix \ref{App-hi}.

Making use of the Property 2b from Appendix \ref{App-hi}, the two
equations can be written explicitly:
\begin{eqnarray} \fl \label{kp1-exp}
{\rm KP}_1:\quad \left(
    \frac{\partial^4}{\partial t_1^4} + 3 \frac{\partial^2}{\partial t_2^2}
    - 4 \frac{\partial^2}{\partial t_1 \partial t_3}
\right)\log\, \tau_p^{(s)}({\bm t}) + 6 \left(
    \frac{\partial^2}{\partial t_1^2} \log\, \tau_p^{(s)}({\bm t})
\right)^2 = 0,\\ \fl \label{kp2-exp}
{\rm KP}_2:\quad \left(
    \frac{\partial^4}{\partial t_1^3 \partial t_2} - 3 \frac{\partial^2}{\partial t_1 \partial t_4}
    + 2 \frac{\partial^2}{\partial t_2 \partial t_3}
\right)\log\, \tau_p^{(s)}({\bm t}) \nonumber \\
\quad \quad \quad \quad + 6 \left(
    \frac{\partial^2}{\partial t_1^2} \log\, \tau_p^{(s)}({\bm t})
\right) \left(
    \frac{\partial^2}{\partial t_1 \partial t_2} \log\, \tau_p^{(s)}({\bm t})
\right)= 0.
\end{eqnarray}
Only ${\rm KP}_1$ will further be used.
\newline\newline\noindent
{\it Toda Lattice hierarchy.}---The first nontrivial equations of
the Toda Lattice hierarchy can be derived along the same lines from
Eq.~(\ref{TL}).
\begin{eqnarray}
\label{tl-1}
   {\rm TL}_1:\quad \frac{1}{2} D_1^2 \, \tau_{p}^{(s)}({\bm t})\,\circ\tau_{p}^{(s)}({\bm
   t}) =  \tau_{p+1}^{(s+1)}({\bm t})\circ
    \tau_{p-1}^{(s-1)} ({\bm t}), \\
   \label{tl-2}
    {\rm TL}_2: \quad \frac{1}{2} D_1 D_2
    \, \tau_{p}^{(s)}({\bm t})\,\circ\tau_{p}^{(s)}({\bm t}) =
     D_1
    \,\tau_{p+1}^{(s+1)}({\bm t})\circ
    \tau_{p-1}^{(s-1)} ({\bm t}).
\end{eqnarray}
Explicitly, one has:
\begin{eqnarray} \fl\label{tl-1-expl}
    {\rm TL}_1:\quad \tau_p^{(s)}({\bm t}) \frac{\partial^2
    \tau_p^{(s)}({\bm t})}{\partial t_1^2} - \left(
        \frac{\partial
    \tau_p^{(s)}({\bm t})}{\partial t_1}
    \right)^2 = \tau_{p+1}^{(s+1)}({\bm t})
    \tau_{p-1}^{(s-1)} ({\bm t}),
    \\ \fl\label{tl-2-expl} {\rm TL}_2:\quad \tau_p^{(s)}({\bm t}) \frac{\partial^2
    \tau_p^{(s)}({\bm t})}{\partial t_1 \partial t_2} -
        \frac{\partial
    \tau_p^{(s)}({\bm t})}{\partial t_1}
    \frac{\partial
    \tau_p^{(s)}({\bm t})}{\partial t_2} \nonumber \\ =
     \tau_{p-1}^{(s-1)}({\bm t}) \frac{\partial
    \tau_{p+1}^{(s+1)}({\bm t})}{\partial t_1 } -
     \tau_{p+1}^{(s+1)}({\bm t}) \frac{\partial
    \tau_{p-1}^{(s-1)}({\bm t})}{\partial t_1 }.
\end{eqnarray}

Higher order members of the KP and Toda Lattice hierarchies can
readily be generated from Eqs.~(\ref{KP}) and~(\ref{TL}),
respectively.

\subsection{Virasoro constraints}\label{Sec-3-6}

Virasoro constraints satisfied by the $\tau$ function
Eq.~(\ref{tau-fff}) below is yet another important ingredient of the
``deform-and-study'' approach to the correlation functions of characteristic
polynomials $\Pi_{n|p} ({\boldsymbol \varsigma}; {\boldsymbol
\kappa})$. In accordance with the discussion in Section \ref{Sec-2},
Virasoro constraints are needed to translate nonlinear integrable
hierarchies Eqs.~(\ref{TL}) -- (\ref{346}), satisfied by the $\tau$ function
\begin{eqnarray} \fl
\label{tau-fff}
    \tau_{n}^{(s)}({\boldsymbol \varsigma},{\bm \kappa}; {\boldsymbol t}) = \frac{1}{n!}
    \int_{{\mathcal D}^{n}} \prod_{j=1}^{n}
    \left(
        d\lambda_j\,  e^{-V_{n-s}(\lambda_j)} \, \prod_{\alpha=1}^p
        (\varsigma_\alpha-\lambda_j)^{\kappa_\alpha}\,  e^{v({\boldsymbol t};\lambda_j)}\right) \cdot
     \Delta_{n}^2({\boldsymbol \lambda}),
\end{eqnarray}
into nonlinear, hierarchically structured differential equations for
the correlation function
\begin{eqnarray}\fl
\label{rpf-111}
    \Pi_{n|p}({\bm {\varsigma}};{\bm \kappa}) =\frac{1}{{\cal N}_n} \int_{{\cal D}^{n}} \prod_{j=1}^{n}
        \left( d\lambda_j \, e^{-V_n(\lambda_j)}\prod_{\alpha=1}^p (\varsigma_\alpha - \lambda_j)^{\kappa_\alpha} \right)
    \cdot
    \Delta_{n}^2({\boldsymbol \lambda})
\end{eqnarray}
obtained from Eq.~(\ref{tau-fff}) by setting ${\bm t}=0$ and $s=0$.
The normalisation constant ${\cal N}_n$ is defined in
Eq.~(\ref{norm}).

The Virasoro constraints reflect invariance of the $\tau$
function Eq.~(\ref{tau-fff}) under the change of integration
variables
\begin{eqnarray}
\label{vc-var}
    \lambda_j \rightarrow \mu_j + \epsilon \mu_j^{q+1}
    R(\mu_j),\;\;\; q \ge -1,
\end{eqnarray}
labeled by the integer $q$; here $\epsilon>0$ is an infinitesimally small parameter, and
$R(\mu)$ is a suitable benign function (e.g., a polynomial). The
function $f(\lambda)$ is related to the confinement potential
$V_{n-s}(\lambda)$ through the parameterisation (Adler, Shiota and van Moerbeke 1995)
\begin{eqnarray}
\label{fg} \frac{dV_{n-s}}{d\lambda} =
\frac{g(\lambda)}{f(\lambda)},\;\;\;g(\lambda)=\sum_{k=0}^\infty b_k
\lambda^k,\;\;\; f(\lambda)=\sum_{k=0}^\infty a_k \lambda^k
\end{eqnarray}
in which both $g(\lambda)$ and $f(\lambda)$ depend on $n-s$ as do
the coefficients $b_k$ and $a_k$ in the above expansions. We also
assume that
\begin{eqnarray}
 \lim_{\lambda\rightarrow \pm \infty} f(\lambda)\,\lambda^k \,
 e^{-V_{n-s}(\lambda)} = 0, \quad k\ge 0.
\end{eqnarray}

To derive the Virasoro constraints announced in
Eqs.~(\ref{2-Vir})--(\ref{bq}), we transform the integration
variables in Eq.~(\ref{tau-fff}) as specified in Eq.~(\ref{vc-var})
and further expand Eq.~(\ref{tau-fff}) in $\epsilon$. Invariance of
the integral under this transformation implies that the linear in
$\epsilon$ terms must vanish:
\begin{eqnarray}\fl
\label{vc-1}
    \int_{{\cal D}^n} (d{\bm \mu})\,
    \Bigg(
    \beta \,\sum_{i>j} \frac{\mu_i^{q+1} R(\mu_i) - \mu_j^{q+1}
    R(\mu_j)}{\mu_i-\mu_j} + \sum_{\ell=1}^n
  \mu_\ell^{q+1} R^\prime (\mu_\ell) \nonumber \\ \fl
   \qquad +
    \sum_{\ell=1}^n R(\mu_\ell) \left[ (q+1)\mu_\ell^q +  v^\prime({\bm t};\mu_\ell)\, \mu_\ell^{q+1}
           - \frac{g(\mu_\ell)}{f(\mu_\ell)} \,\mu_\ell^{q+1} \right]
        \Bigg)
    I_n^{(s)}({\bm t}; {\bm \mu}, {\bm \varsigma})
    \nonumber \\ \fl
    \qquad
    -\int_{{\cal D}^n} (d{\bm \mu})\, \left( \sum_{\ell=1}^n  \mu_\ell^{q+1} R(\mu_\ell)  \sum_{\alpha=1}^p \frac{\kappa_\alpha}{\varsigma_\alpha - \mu_\ell}
    \right) \, I_n^{(s)}({\bm t}; {\bm \mu}, {\bm \varsigma}) \nonumber \\ \fl
    \qquad -
    \left( \sum_{i=1}^{{\rm dim}({\bm c^\prime})}
        R(c_i^\prime)\, c_i^{\prime\,{q+1}} \frac{\partial}{\partial c_i^\prime}
    \right)\, \int_{{\cal D}^n} (d{\bm \mu})\, I_n^{(s)}({\bm t}; {\bm \mu}, {\bm \varsigma})=0.
\end{eqnarray}
Here, ${\bm c}^\prime=\{c_1,\cdots,c_{2r}\}\setminus\{\pm \infty\}$,
\begin{eqnarray} \fl
    I_n^{(s)}({\bm t}; {\bm \mu}, {\bm \varsigma}) =  | \Delta_{n}({\boldsymbol \mu})|^\beta\,\prod_{k=1}^n \left[
      e^{-V_{n-s}(\mu_j)} \, \prod_{\alpha=1}^p
        (\varsigma_\alpha-\mu_j)^{\kappa_\alpha}\,  e^{v({\boldsymbol t};\mu_j)}
        \right],
\end{eqnarray}
and
\begin{eqnarray}
    (d{\bm \mu})\, = \prod_{k=1}^n d\mu_k.
\end{eqnarray}
In the above formulae, we reinstated $\beta>0$; it will be set to
$\beta=2$ when needed.

The choice of $R(\mu)$ is dictated by problem in question and,
hence, is not unique. If one is interested in studying matrix
integrals as functions of the parameters $\{c_1,\cdots,c_{2r}\}$
defining the integration domain ${\cal D}$, the suitable choice of
$R(\mu)$ is
\begin{eqnarray}
    R(\mu) = f(\mu).
\end{eqnarray}
In this case, the differential operator (Adler, Shiota and van Moerbeke 1995)
\begin{eqnarray} \label{c-op}
    \sum_{i=1}^{2r}
        R(c_i^\prime)\, c_i^{\prime\, {q+1}} \frac{\partial}{\partial c_i^\prime}
\end{eqnarray}
becomes an essential part of the Virasoro constraints. In the
context of characteristic polynomials, the integration domain ${\cal
D}$ is normally fixed whilst the {\it physical parameters}
$\{\varsigma_\alpha\}$ are allowed to vary. This prompts the choice
\begin{eqnarray}\label{R}
    R(\mu) =  f(\mu)
    \prod_{k=1}^{\varrho} (\mu - c_k^\prime),\;\;\;
    \varrho ={\rm dim}({\bm c}^\prime)
\end{eqnarray}
that nullifies the differential operator Eq.~(\ref{c-op}).
Equivalently,
\begin{eqnarray}
\label{rmu}
    R(\mu) = f(\mu)\,\sum_{k=0}^{\varrho} \mu^k
    s_{\varrho - k}(-{\bm p}_{\varrho}
    (\bm{c^\prime})
    ).
\end{eqnarray}
Here, the notation $s_k(-{\bm p}_{\varrho} (\bm{c^\prime}))$ stands for the Schur
polynomial and ${\bm p}_{\varrho}(\bm{c^\prime})$ is an infinite dimensional vector
\begin{eqnarray} \label{b909}
    {\bm p}_\varrho(\bm{c^\prime})=\left(
    {\rm tr}_\varrho(\bm{c^\prime}), \frac{1}{2} {\rm tr}_\varrho(\bm{c^\prime})^2,\cdots,
    \frac{1}{k} {\rm tr}_\varrho(\bm{c^\prime})^k,\cdots
    \right)
\end{eqnarray}
with
\begin{eqnarray}
     {\rm tr}_\varrho(\bm{c^\prime})^k =
    \sum_{j=1}^{\varrho} (c_j^\prime)^k.
\end{eqnarray}
{\it Remark.}  Equation (\ref{R}) assumes that none of $c_k^\prime$'s are zeros of
$f(\mu)$. If this is not the case, the set ${\bm c^\prime}$ must be redefined:
\begin{eqnarray} \label{c-redef}
{\bm c^\prime} \rightarrow {\bm c^\prime} \setminus \{{\cal Z}_0\},
\end{eqnarray}
where ${\cal Z}_0$ is comprised of zeros
of $f(\mu)$.
\newline\newline
Substituting Eqs.~(\ref{rmu}), (\ref{fg}) and (\ref{vt-def}) into
Eq.~(\ref{vc-1}), we derive:
\begin{eqnarray} \fl
    \label{vc-2}
    \int_{{\cal D}^n} (d{\bm \mu})\,
    \sum_{k=0}^{\varrho}
    s_{\varrho-k}(-{\bm p}_{\varrho} (\bm{c^\prime})) \Bigg[
    \sum_{i=0}^\infty a_i \, \Bigg( \frac{\beta}{2}
    \sum_{j=0}^{q+k+i} {\rm tr}_n ({\bm \mu}^j) \, {\rm tr}_n ({\bm
    \mu}^{q+k+i-j}) \nonumber \\
    \fl
    \qquad +\left( 1- \frac{\beta}{2} \right) \, (i+k+q+1) \, {\rm tr}_n ({\bm \mu}^{q+k+i})
    + \sum_{j=0}^{\infty} jt_j {\rm tr}_n ({\bm
    \mu}^{q+k+i+j}) \nonumber\\ \fl
    \qquad + \sum_{\alpha=1}^p \kappa_\alpha \sum_{m=0}^{q+k+i}
    \varsigma_{\alpha}^m \,{\rm tr}_n ({\bm \mu}^{q+k+i-m})
    - \sum_{\alpha=1}^p  \varsigma_\alpha^{q+k+i+1} \frac{\partial}{\partial \varsigma_\alpha}
    \Bigg)  \Bigg]\, I_n^{(s)}({\bm t}; {\bm \mu}, {\bm \varsigma}) \nonumber \\
    \fl \qquad = \int_{{\cal D}^n} (d{\bm \mu})\,
    \sum_{k=0}^{\varrho}
    s_{\varrho-k}(-{\bm p}_{\varrho} (\bm{c^\prime}))
    \sum_{i=0}^\infty b_i \, {\rm tr}_n ({\bm
    \mu}^{q+k+i+1})\, I_n^{(s)}({\bm t}; {\bm \mu}, {\bm \varsigma}).
\end{eqnarray}
The ${\bm \varsigma}$-dependent part in Eq.~(\ref{vc-2}),
\begin{eqnarray} \fl \label{vc-3}
\int_{{\cal D}^n} (d{\bm \mu})\,
    \sum_{k=0}^{\varrho}
    s_{\varrho-k}(-{\bm p}_{\varrho} (\bm{c^\prime}))
    \sum_{i=0}^\infty a_i \, \nonumber \\
    \fl \qquad  \times
   \left( \sum_{\alpha=1}^p \kappa_\alpha \sum_{m=0}^{q+k+i}
    \varsigma_{\alpha}^m \,{\rm tr}_n ({\bm \mu}^{q+k+i-m})
    - \sum_{\alpha=1}^p  \varsigma_\alpha^{q+k+i+1} \frac{\partial}{\partial
    \varsigma_\alpha} \right) \, I_n^{(s)}({\bm t}; {\bm \mu}, {\bm \varsigma}),
\end{eqnarray}
originates from the term
\begin{eqnarray}
\label{vc-piece}
    - \int_{{\cal D}^n} (d{\bm \mu})\, \left( \sum_{\ell=1}^n \mu_\ell^{q+1} R(\mu_\ell)  \sum_{\alpha=1}^p \frac{\kappa_\alpha}{\varsigma_\alpha - \mu_\ell}
    \right) \, I_n^{(s)}({\bm t}; {\bm \mu}, {\bm \varsigma})
\end{eqnarray}
in Eq.~(\ref{vc-1}). Indeed, substituting Eqs.~(\ref{rmu}) and
(\ref{fg}) into Eq.~(\ref{vc-piece}), the latter reduces to
\begin{eqnarray} \fl
    \int_{{\cal D}^n} (d{\bm \mu})\, \sum_{k=0}^{\varrho}
    s_{\varrho-k}(-{\bm p}_{\varrho} (\bm{c^\prime}))
    \sum_{i=0}^\infty a_i \, \left( \sum_{\alpha=1}^p \kappa_\alpha
     \sum_{\ell=1}^n \frac{ \mu_\ell^{q+k+i+1}}{ \mu_\ell - \varsigma_\alpha }
    \right) \, I_n^{(s)}({\bm t}; {\bm \mu}, {\bm \varsigma}).
\end{eqnarray}
The double sum in parentheses can conveniently be divided into two
pieces,
\begin{eqnarray}
\label{c-1}
    \sum_{\alpha=1}^p \kappa_\alpha
     \sum_{\ell=1}^n \frac{ \mu_\ell^{q+k+i+1}- \varsigma_\alpha^{q+k+i+1}}{ \mu_\ell - \varsigma_\alpha }
\end{eqnarray}
and
\begin{eqnarray}
\label{c-2}
    \sum_{\alpha=1}^p \kappa_\alpha
     \sum_{\ell=1}^n \frac{ \varsigma_\alpha^{q+k+i+1}}{ \mu_\ell - \varsigma_\alpha
     }.
\end{eqnarray}
Due to the identities
\begin{eqnarray}
    \sum_{\ell=1}^n \frac{ \mu_\ell^{q+k+i+1}- \varsigma_\alpha^{q+k+i+1}}{ \mu_\ell - \varsigma_\alpha
    } = \sum_{m=0}^{q+k+i}
    \varsigma_{\alpha}^m \,{\rm tr}_n ({\bm \mu}^{q+k+i-m})
\end{eqnarray}
and
\begin{eqnarray}
    \kappa_\alpha
     \sum_{\ell=1}^n \frac{ 1}{ \mu_\ell - \varsigma_\alpha
     }\, I_n^{(s)}({\bm t}; {\bm \mu}, {\bm \varsigma}) = - \frac{\partial}{\partial \varsigma_\alpha} I_n^{(s)}({\bm t}; {\bm \mu}, {\bm
     \varsigma}),
\end{eqnarray}
we conclude that Eq.~(\ref{vc-piece}) reduces to the sought
Eq.~(\ref{vc-3}). We found it more convenient to rewrite the
$\partial/\partial \varsigma_\alpha$-term in Eq.~(\ref{vc-3}) in a
more compact way,
\begin{eqnarray} \fl \label{vc-4}
\int_{{\cal D}^n} (d{\bm \mu})\,
    \sum_{k=0}^{\varrho}
    s_{\varrho-k}(-{\bm p}_{\varrho} (\bm{c^\prime}))
    \sum_{i=0}^\infty a_i \left( \sum_{\alpha=1}^p \varsigma_\alpha^{q+k+i+1} \frac{\partial}{\partial
    \varsigma_\alpha}\right) \, I_n^{(s)}({\bm t}; {\bm \mu}, {\bm \varsigma})\nonumber\\
    \qquad \qquad \qquad \qquad \qquad= {\hat {\cal B}}_q^V({\bm \varsigma})  \int_{{\cal D}^n} (d{\bm
    \mu})\,  I_n^{(s)}({\bm t}; {\bm \mu}, {\bm \varsigma})
\end{eqnarray}
with the differential operator ${\hat {\cal B}}_q^V({\bm \varsigma})$
being
\begin{eqnarray}
\label{bq-rep}
    {\hat {\cal B}}_q^V ({\bm \varsigma}) = \sum_{\alpha=1}^p
    \left( \prod_{k=1}^{\varrho} (\varsigma_\alpha - c_k^\prime) \right)
     f(\varsigma_\alpha) \,
    \varsigma_{\alpha}^{q+1} \frac{\partial}{\partial
    \varsigma_\alpha}.
\end{eqnarray}
Equation (\ref{vc-4}) follows from the expansions Eqs.~(\ref{fg}),
(\ref{R}) and (\ref{rmu}).

To complete the derivation of Virasoro constraints, we further
notice that terms ${\rm \tr}_n ({\bm \mu}^j)$ in Eq.~(\ref{vc-2})
can be generated by differentiating $I_n^{(s)}$ over $t_j$. Since
${\rm \tr}_n ({\bm \mu}^0)=n$, the derivative $\partial/\partial
t_0$ should formally be understood as $\partial/\partial t_0 \equiv
n$. This observation yields Virasoro constraints for the $\tau$
function
\begin{eqnarray}
    \tau_{n}^{(s)}({\boldsymbol \varsigma},{\bm \kappa}; {\boldsymbol t}) = \frac{1}{n!}
    \int_{{\mathcal D}^{n}} (d{\bm \mu})\, I_n^{(s)}({\bm t}; {\bm \mu}, {\bm \varsigma})
\end{eqnarray}
in the form ($q\ge -1$)
\begin{equation}
\label{2-Vir-repp}
    \left[ \hat{{\cal L}}_{q}^V({\bm t}) + \hat{{\cal L}}_q^{\rm det}({\bm \varsigma};{\bm t})
     \right] \tau_n^{(s)}({\bm \varsigma};{\bm t})
    ={\hat {\cal B}}_q^V ({\bm \varsigma})\,\tau_n^{(s)}({\bm \varsigma};{\bm
    t}).
\end{equation}
Here, the differential operators
\begin{eqnarray} \fl
\label{vLv-repp}
     \hat{{\cal L}}_{q}^V({\bm t}) = \sum_{\ell = 0}^\infty
    \sum_{k=0}^{\varrho}  s_{\varrho-k}(-{\bm p}_{\varrho} (\bm{c^\prime}))
       \left(
        a_\ell \hat{\cal L}_{q+k+\ell}^{(\beta)}({\bm t}) - b_\ell \frac{\partial}{\partial t_{q+k+\ell+1}}
    \right)
\end{eqnarray}
and
\begin{eqnarray} \fl
\label{vLG-repp}
     \hat{{\cal L}}_{q}^{\rm det}({\bm t}) = \sum_{\ell = 0}^\infty
       a_\ell
    \sum_{k=0}^{\varrho} s_{\varrho-k}(-{\bm p}_{\varrho} (\bm{c^\prime}))
       \sum_{m=0}^{q+k+\ell} \left(\sum_{\alpha=1}^p \kappa_\alpha\,\varsigma_\alpha^m\right)
       \frac{\partial}{\partial t_{q+k+\ell-m}}
\end{eqnarray}
act in the ${\bm t}$-space whilst the differential operator ${\hat
{\cal B}}_q^V ({\bm \varsigma})$ acts in the space of {\it physical
parameters} $\{\varsigma_\alpha\}_{\alpha\in{\mathbb Z}_+}$. Notice
that the operator $\hat{{\cal L}}_{q}^V({\bm t})$ is expressed in
terms of the Virasoro operators
\begin{eqnarray} \fl
    \label{vo-repp}
    \hat{{\cal L}}_q^{(\beta)}({\bm t}) = \sum_{j=1}^\infty jt_j \,\frac{\partial}{\partial t_{q+j}}
    + \frac{\beta}{2}
    \sum_{j=0}^q \frac{\partial^2}{\partial {t_j}\partial {t_{q-j}}} + \left(
    1 -\frac{\beta}{2}
    \right)(q+1)\frac{\partial}{\partial t_q},
\end{eqnarray}
obeying the Virasoro algebra
\begin{eqnarray}
\label{va-repp} [\hat{{\cal L}}_p^{(\beta)},\hat{{\cal
L}}_q^{(\beta)}] = (p-q)\hat{{\cal L}}_{p+q}^{(\beta)}, \;\;\;
p,q\ge -1.
\end{eqnarray}
The Virasoro constraints derived in this section stay valid for
arbitrary $\beta>0$; for $\beta=2$, they are reduced to
Eqs.~(\ref{2-Vir}) -- (\ref{va}) announced in Sec. \ref{Sec-2}.
\newline\newline\noindent
This concludes the derivation of three main ingredients of
integrable theory of average characteristic polynomials -- the
bilinear identity, integrable hierarchies emanating from it, and the Virasoro constraints.

\section{From $\tau$ Functions to Characteristic Polynomials}
\label{Sec-4}
The general calculational scheme formulated in Section \ref{Sec-2} and detailed in Section \ref{Sec-3} applies to a variety of random matrix ensembles. In this Section we deal with CFCP for the Gaussian Unitary Ensemble (GUE) and Laguerre Unitary Ensemble (LUE). A detailed treatment of the GUE case is needed to lay the basis for further comparative analysis of the three variations of the replica approach that will be presented in Section \ref{Sec-5}. The study of the LUE relevant to the QCD physics (Verbaarschot 2010) is included for didactic purposes. A sketchy exposition of the theory for Jacobi Unitary Ensemble (JUE) and Cauchy Unitary Ensemble (CyUE) appearing in the context of universal aspects of quantum transport through chaotic cavities (Beenakker 1997) can be found in Appendices~\ref{App-JUE}~and~\ref{App-CyUE}.

\subsection{Gaussian Unitary Ensemble (GUE)}
The correlation function of characteristic polynomials in GUE is defined by the $n$-fold integral
\begin{eqnarray}\fl
\label{rpf-gue}
    \Pi_{n|p}^{\rm G}({\bm {\varsigma}};{\bm \kappa}) =\frac{1}{{\cal N}_n^{\rm G}} \int_{{\mathbb R}^{n}} \prod_{j=1}^{n}
        \left( d\lambda_j \, e^{-\lambda_j^2}\prod_{\alpha=1}^p (\varsigma_\alpha - \lambda_j)^{\kappa_\alpha} \right)
    \cdot
    \Delta_{n}^2({\boldsymbol \lambda})
\end{eqnarray}
where
\begin{eqnarray}
\label{N-gue} \fl
    {\cal N}_n^{\rm G} =  \int_{{\mathbb R}^{n}} \prod_{j=1}^{n}
        \left( d\lambda_j \, e^{-\lambda_j^2} \right)
    \cdot
    \Delta_{n}^2({\boldsymbol \lambda}) =  \pi^{n/2} 2^{-n(n-1)/2}\prod_{j=1}^n \Gamma(j+1)
\end{eqnarray}
is the normalisation constant. The associated $\tau$ function equals
\begin{eqnarray}\fl
\label{tau-gue}
    \tau_{n}^{\rm G}({\bm {\varsigma}},{\bm \kappa};{\bm t}) =\frac{1}{n!} \int_{{\mathbb R}^{n}} \prod_{j=1}^{n}
        \left( d\lambda_j \, e^{-\lambda_j^2 + v({\bm t};\lambda_j)}
        \prod_{\alpha=1}^p (\varsigma_\alpha - \lambda_j)^{\kappa_\alpha} \right)
    \cdot
    \Delta_{n}^2({\boldsymbol \lambda}),
\end{eqnarray}
see Section \ref{Sec-2}. (In the above definitions, the
superscript ${\rm G}$ stands for GUE but it will further be omitted
when notational confusion is unlikely to arise.)

\subsubsection{Virasoro constraints}
\noindent\newline\newline In the notation of Section \ref{Sec-3},
the definition Eq.~(\ref{rpf-gue}) implies that
\begin{eqnarray}
    f(\lambda)=1 \;\; &\mapsto& \;\;\; a_k=\delta_{k,0},\\
    g(\lambda) = 2\lambda\;\;&\mapsto& \;\;\;b_k = 2 \delta_{k,1},\\
    {\cal D} = {\mathbb R} \;\; &\mapsto&\;\;\; {\rm dim}(\bm{c^\prime}) =0.
\end{eqnarray}
This brings the Virasoro constraints Eqs.~(\ref{2-Vir}) --
(\ref{vo}) for the $\tau$ function Eq.~(\ref{tau-gue}):
\begin{equation}
\label{2-Vir-G} \fl
    \left[ \hat{\cal L}_{q}({\bm t}) - 2
        \frac{\partial}{\partial t_{q+2}}
 + \sum_{m=0}^{q}
    \left(\sum_{\alpha=1}^p \kappa_\alpha\,\varsigma_\alpha^m\right)
       \frac{\partial}{\partial t_{q-m}}
     \right] \tau_n({\bm \varsigma},{\bm \kappa};{\bm t})
    ={\hat {\mathcal B}}_{q} \,\tau_n({\bm \varsigma},{\bm \kappa};{\bm t}),
\end{equation}
where
\begin{eqnarray} \label{Bq}
    {\hat {\mathcal B}}_q  = \sum_{\alpha=1}^p
    \varsigma_{\alpha}^{q+1} \frac{\partial}{\partial \varsigma_\alpha},
\end{eqnarray}
the short-hand notation $\vartheta_m({\bm \varsigma},{\bm \kappa})$ is defined as
\begin{eqnarray}
\label{theta-m}
    \vartheta_m({\bm \varsigma},{\bm \kappa}) = \sum_{\alpha=1}^p \kappa_\alpha \varsigma_\alpha^m,
\end{eqnarray}
so that
\begin{eqnarray}
\label{theta-0}
        \vartheta_0({\bm \varsigma},{\bm \kappa})
            ={\rm tr}_p\,\bm\kappa=\sum_{\alpha=1}^p \kappa_\alpha \equiv \kappa.
\end{eqnarray}
Also, $\hat{\cal L}_{q}({\bm t})$ is the Virasoro operator given by
Eq.~(\ref{vo}). Notice that the $\tau$ function in
Eq.~(\ref{2-Vir-G}) does not bear the superscript $(s)$ since the
GUE confinement potential $V(\lambda)= \lambda^2$ does not depend on
$n$.

In what follows, we need the three lowest Virasoro constraints
labeled by $q=-1$, $q=0$ and~$q=+1$. Written for the logarithm of $\tau$
function, they
read:
\begin{equation}
\label{2-Vir-q=-1} \fl
    \left(
        \sum_{j=2}^\infty jt_j\frac{\partial}{\partial t_{j-1}}
     - 2
        \frac{\partial}{\partial t_{1}}
     \right) \log \tau_n ({\bm \varsigma},{\bm \kappa};{\bm t}) + nt_1
    ={\hat {\mathcal B}}_{-1}\,\log \tau_n ({\bm \varsigma},{\bm \kappa};{\bm t}),
\end{equation}
\begin{equation}
\label{2-Vir-q=0} \fl
    \left(
        \sum_{j=1}^\infty jt_j\frac{\partial}{\partial t_{j}}
         - 2
        \frac{\partial}{\partial t_{2}} \right)\,\log \tau_n ({\bm \varsigma},{\bm \kappa};{\bm t})
     + n \left( n+ \kappa\right)
    ={\hat {\mathcal B}}_{0}\,\log\tau_n ({\bm \varsigma},{\bm \kappa};{\bm t}),
\end{equation}
\begin{eqnarray}
\label{2-Vir-q=+1} \fl
    \left( \sum_{j=1}^\infty jt_j\frac{\partial}{\partial t_{j+1}} - 2
        \frac{\partial}{\partial t_{3}}
    +
    \left(2n+\kappa\right)
       \frac{\partial}{\partial t_{1}}
     \right) \log\tau_n ({\bm \varsigma},{\bm \kappa};{\bm t}) \nonumber \\
     \qquad \qquad\qquad\quad+
       n \,\vartheta_1({\bm \varsigma},{\bm \kappa})
    ={\hat {\mathcal B}}_{1} \,\log\tau_n({\bm \varsigma},{\bm \kappa};{\bm t}).
\end{eqnarray}

\subsubsection{Toda Lattice hierarchy}\noindent
\newline\newline
Projection of the Toda Lattice hierarchy Eq.~(\ref{tlh}) for the
${\bm t}$-dependent $\tau$ function Eq.~(\ref{tau-gue}) onto the
hyperplane ${\bm t} = {\bm 0}$ generates the Toda Lattice hierarchy
for the correlation function $\Pi^{\rm G}_{n|p}({\bm
{\varsigma}};{\bm \kappa})$ [Eq.~(\ref{rpf-gue})] of the GUE characteristic
polynomials,
\begin{eqnarray}
\label{tau-pi}
    \Pi^{\rm G}_{n|p}({\bm {\varsigma}};{\bm \kappa}) = \frac{n!}{{\cal N}_n^{\rm G}}\,
    \tau^{\rm G}_n({\bm \varsigma},{\bm \kappa};{\bm t})\Big|_{{\bm t}={\rm 0}}.
\end{eqnarray}
Below, the first [Eq.~(\ref{tl-1-expl})] and second
[Eq.~(\ref{tl-2-expl})] equation of the TL hierarchy will be
considered:
\begin{eqnarray} \fl\label{tl-1-equiv}
    {\rm TL}_1:\quad \frac{\partial^2}{\partial t_1^2}\log \tau_n({\bm t})= \frac{\tau_{n+1}({\bm t}) \,
    \tau_{n-1} ({\bm t})}{\tau_{n}^2({\bm t})},
    \\ \fl
    \label{tl-2-equiv} {\rm TL}_2:\quad \frac{\partial^2}{\partial t_1 \partial t_2}
    \log \tau_n({\bm t})
     =
     \frac{\tau_{n+1}({\bm t}) \,
    \tau_{n-1} ({\bm t})}{\tau_{n}^2({\bm t})}
    \,
        \frac{\partial}{\partial t_1} \log \left(\frac{\tau_{n+1}({\bm t})}{\tau_{n-1}({\bm t})}\right).
\end{eqnarray}
The equivalence of Eqs.~(\ref{tl-1-equiv}) and (\ref{tl-2-equiv}) to
Eqs.~(\ref{tl-1-expl}) and (\ref{tl-2-expl}) is easily established.
\noindent\newline\newline (i) To derive the first equation of the TL
hierarchy for $\Pi^{\rm G}_{n|p}({\bm {\varsigma}};{\bm \kappa})$
from Eqs.~(\ref{tau-pi}) and (\ref{tl-1-equiv}), we have to
determine
$$
    \frac{\partial^2}{\partial t_1^2}\log\tau_n({\bm t})\Big|_{\bm{t}=0}
$$
with the help of Virasoro constraints. This is achieved in two
steps. First, we differentiate Eq.~(\ref{2-Vir-q=-1}) over $t_1$ and
set ${\bm t}={\bm 0}$ afterwards, to derive:
\begin{eqnarray}
\label{d-11}
    2     \frac{\partial^2}{\partial t_1^2}\log\tau_n({\bm t})\Big|_{\bm{t}=0}
    = n -
    {\hat {\mathcal B}}_{-1} \,\frac{\partial}{\partial t_1}
    \log\tau_n({\bm t}) \Big|_{\bm{t}=0}.
\end{eqnarray}
Second, we set ${\bm t}={\bm 0}$ in Eq.~(\ref{2-Vir-q=-1}) to
identify the relation
\begin{eqnarray}
\label{d-1}
    2  \frac{\partial}{\partial t_1}\log\tau_n({\bm t})\Big|_{\bm{t}=0}
    = -
    {\hat {\mathcal B}}_{-1} \,
    \log\tau_n({\bm 0}).
\end{eqnarray}
Combining Eqs.~(\ref{d-11}) and (\ref{d-1}), we conclude that
\begin{eqnarray}
        4 \frac{\partial^2}{\partial t_1^2}\log\tau_n({\bm t})\Big|_{\bm{t}=0}
        =2 n + {\hat {\mathcal B}}_{-1}^2 \,
        \log\tau_n({\bm 0}).
\end{eqnarray}
Finally, substituting this result back to Eq.~(\ref{tl-1-equiv}),
and taking into account Eqs.~(\ref{tau-pi}) and (\ref{N-gue}), we
end up with the first TL equation
\begin{eqnarray}
\label{gue-TL-1}\fl
\widetilde{{\rm TL}}_1^{\rm G}: \qquad
    {\hat {\mathcal B}}_{-1}^2 \,
        \log \Pi_{n|p}({\bm {\varsigma}};{\bm \kappa})  = 2 n\, \left(
            \frac{\Pi_{n+1|p}({\bm {\varsigma}};{\bm \kappa}) \,
            \Pi_{n-1|p}({\bm {\varsigma}};{\bm \kappa})}{\Pi_{n|p}^2({\bm {\varsigma}};{\bm \kappa})} -1
        \right)
\end{eqnarray}
written in the space of physical parameters ${\bm \varsigma}$.
\noindent\newline\newline (ii) The second equation of the TL
hierarchy for $\Pi_{n|p}({\bm {\varsigma}};{\bm \kappa})$ can be
derived along the same lines. Equation (\ref{tl-2-equiv}) suggests
that, in addition to the derivative $\partial/\partial t_1
\log\tau_n$ at ${\bm t}={\bm 0}$ given by Eq.~(\ref{d-1}), one needs
to know the mixed derivative
$$
    \frac{\partial^2}{\partial t_1\partial t_2}\log\tau_n({\bm t})\Big|_{\bm{t}=0}.
$$
It can be calculated by combining Eq.~(\ref{2-Vir-q=-1})
differentiated over $t_2$ with Eqs.~(\ref{2-Vir-q=0}) and
(\ref{d-1}). The result reads:
\begin{eqnarray}\label{gue-TL-2} \fl
\widetilde{{\rm TL}}_2^{\rm G}: \qquad
     (1 - {\hat {\mathcal B}}_{0}) {\hat {\mathcal B}}_{-1} \,
        \log \Pi_{n|p}({\bm {\varsigma}};{\bm \kappa}) \nonumber \\
         =
        n\,
            \frac{\Pi_{n+1|p}({\bm {\varsigma}};{\bm \kappa}) \,
            \Pi_{n-1|p}({\bm {\varsigma}};{\bm \kappa})}{\Pi_{n|p}^2({\bm {\varsigma}};{\bm \kappa})}
            \,{\hat {\mathcal B}}_{-1}
         \log \left(
            \frac{\Pi_{n+1|p}({\bm {\varsigma}};{\bm \kappa})}{\Pi_{n-1|p}({\bm {\varsigma}};{\bm \kappa})}
        \right).
\end{eqnarray}
\newline\newline\noindent
Higher order equations of the TL hierarchy for the correlation
functions $\Pi_{n|p}$ can be derived in a similar fashion.
\newline\newline
{\it Remark.}---For $p=1$, the equations $\widetilde{{\rm
TL}}_1^{\rm G}$ and $\widetilde{{\rm TL}}_2^{\rm G}$ become
particularly simple:
\begin{eqnarray}
\label{TL-1-cc} \fl
\widetilde{{\rm TL}}_1^{\rm G}: \quad
    \frac{\partial^2}{\partial \varsigma^2}\,
        \log {\Pi}_{n}(\varsigma;\kappa)  = 2 n\,\left(
            \frac{{\Pi}_{n+1}(\varsigma;\kappa) \,
            { \Pi}_{n-1}(\varsigma;\kappa)}{{ \Pi}_{n}^2(\varsigma;\kappa)}-1\right), \\
            \fl \label{TL-2-cc}
\widetilde{{\rm TL}}_2^{\rm G}: \quad
      \left(1- \varsigma \frac{\partial}{\partial \varsigma} \right) \frac{\partial}{\partial \varsigma}\,
        \log {\Pi}_{n}(\varsigma; \kappa)
         =
        n\,
            \frac{{\Pi}_{n+1}(\varsigma;\kappa) \,
            {\Pi}_{n-1}(\varsigma;\kappa)}{{\Pi}_{n}^2(\varsigma;\kappa)}
            \,\frac{\partial}{\partial \varsigma}
         \log \left(
            \frac{{\Pi}_{n+1}(\varsigma;\kappa)}{{\Pi}_{n-1}(\varsigma;\kappa)}
        \right). \nonumber \\
        {}
\end{eqnarray}

\subsubsection{KP hierarchy and Painlev\'e IV equation}
\noindent
\newline\newline
The technology used in the previous subsection can equally be
employed to project the KP hirerachy Eq.~(\ref{kph}) onto the
hyperplane ${\bm t}={\bm 0}$. Below, only the first KP equation
\begin{eqnarray} \fl \label{kp1-exp-rep}
{\rm KP}_1:\quad \left(
    \frac{\partial^4}{\partial t_1^4} + 3 \frac{\partial^2}{\partial t_2^2}
    - 4 \frac{\partial^2}{\partial t_1 \partial t_3}
\right)\log\, \tau_n({\bm t}) + 6 \left(
    \frac{\partial^2}{\partial t_1^2} \log\, \tau_n({\bm t})
\right)^2 = 0
\end{eqnarray}
will be treated. Notice that no superscript $(s)$ appears in
Eq.~(\ref{kp1-exp-rep}) as the GUE confinement potential does not
depend on the matrix size $n$. Proceeding along the lines of the
previous subsection, we make use of the three Virasoro constraints
Eqs.~(\ref{2-Vir-q=-1}) -- (\ref{2-Vir-q=0}) to derive:
\begin{eqnarray} \fl \label{kp-e1}
        16 \frac{\partial^4}{\partial t_1^4}\log\tau_n({\bm t})\Big|_{\bm{t}=0}
        ={\hat {\mathcal B}}_{-1}^4 \,
        \log\tau_n({\bm 0}), \\ \fl  \label{kp-e2}
        4 \frac{\partial^2}{\partial t_2^2}\log\tau_n({\bm t})\Big|_{\bm{t}=0}
        =2n\left( n+ \kappa \right)- ( 2 - {\hat {\mathcal B}}_{0} ){\hat {\mathcal B}}_{0} \,
        \log\tau_n({\bm 0}), \\ \fl  \label{kp-e3}
        4 \frac{\partial^2}{\partial t_1 \partial t_3}\log\tau_n({\bm t})\Big|_{\bm{t}=0}
        =n\left( 3 n+ 2 \kappa \right) \nonumber\\ - \left(
        {\hat {\mathcal B}}_{0} - {\hat {\mathcal B}}_{1} {\hat {\mathcal B}}_{-1} -
        \frac{1}{2} \left(
            2 n+  \kappa
        \right) {\hat {\mathcal B}}_{-1}^2
        \right) \,
        \log\tau_n({\bm 0}).
\end{eqnarray}
Substitution of Eqs.~(\ref{kp-e1}) -- (\ref{kp-e3}) and (\ref{d-11})
into Eq.~(\ref{kp1-exp-rep}) generates a closed nonlinear
differential equation for $\log \Pi_{n|p}({\bm \varsigma};{\bm
\kappa})$ in the form
\begin{eqnarray} \fl \label{ch-gue}
\widetilde{{\rm KP}}_1^{\rm G}: \quad
    \left[
        {\hat {\mathcal B}}_{-1}^4 + 8 (n-\kappa) {\hat {\mathcal B}}_{-1}^2
        -4 ( 2{\hat {\mathcal B}}_{0} - 3 {\hat {\mathcal B}}_0^2 + 4 {\hat {\mathcal B}}_1 {\hat {\mathcal B}}_{-1}  )
    \right] \, \log \Pi_{n|p}({\bm \varsigma};{\bm \kappa}) \nonumber\\
    \qquad \qquad\qquad\qquad+ 6 \left(
        {\hat {\mathcal B}}_{-1}^2 \log \Pi_{n|p}({\bm \varsigma};{\bm \kappa})
    \right)^2 = 8 n \kappa.
\end{eqnarray}
Notice that appearance of the single parameter $\kappa$ instead of
the entire set ${\bm \kappa} = (\kappa_1,\dots,\kappa_p)$ in
Eq.~(\ref{ch-gue}) indicates that correlation functions
$\Pi_{n|p}({\bm \varsigma};{\bm \kappa})$ with different ${\bm
\kappa}$ but with identical traces ${\rm tr}_p\,{\bm \kappa}$
satisfy the very same equation. It is the boundary conditions
\footnote[2]{Indeed, the boundary conditions at infinity,
$$
    \Pi_{n|p}({\bm \varsigma};{\bm \kappa})\Big|_{|\varsigma_\alpha|\rightarrow \infty}
    \sim \prod_{\alpha=1}^p \varsigma_\alpha^{n\kappa_\alpha},
$$
do distinguish between the correlation functions characterised by
different ${\bm \kappa}$'s, see Eq.~(\ref{rpf-gue}).} that pick up
the right solution for the given set ${\bm \kappa} =
(\kappa_1,\dots,\kappa_p)$.
\newline\newline
{\it Remark.}---For $p=1$, the above equation reads:
\begin{eqnarray} \fl \label{ch-0}
\widetilde{{\rm KP}}_1^{\rm G}: \quad
    \left[ \frac{\partial^4}{\partial \varsigma^4}
        + 4 \left[ 2(n-\kappa) - \varsigma^2\right]\frac{\partial^2}{\partial \varsigma^2}
        + 4 \varsigma \frac{\partial}{\partial \varsigma}
    \right] \log \Pi_{n}(\varsigma;\kappa) \nonumber\\
    \qquad \qquad\qquad\qquad
    + 6 \left(
        \frac{\partial^2}{\partial \varsigma^2} \log \Pi_{n}( \varsigma;\kappa)
    \right)^2 = 8 n \kappa.
\end{eqnarray}
This can be recognised as the Chazy I equation (see Appendix
\ref{App-chazy})
\begin{eqnarray}
\label{ek-ch-1}
    \varphi^{\prime\prime\prime} + 6(\varphi^\prime)^2 + 4 \left[ 2(n-\kappa) - \varsigma^2   \right]\varphi^\prime
    + 4 \varsigma \varphi  - 8n\kappa=0,
\end{eqnarray}
where
\begin{eqnarray}
\label{phi-def}
    \varphi (\varsigma) =
    \frac{\partial}{\partial \varsigma} \log \Pi_{n}\left( \varsigma;\kappa\right).
\end{eqnarray}
Equation (\ref{ek-ch-1}) can further be reduced to the fourth
Painlev\'e equation in the Jimbo-Miwa-Okamoto $\sigma$ form (Forrester and Witte 2001, Tracy and Widom 1994):
\begin{eqnarray}
\label{phi-piv}\fl
    P_{\rm IV}: \qquad
        (\varphi^{\prime\prime})^2 - 4 (\varphi - \varsigma \varphi^\prime)^2
        + 4 \varphi^\prime (\varphi^\prime+2n)(\varphi^\prime-2\kappa)=0,
\end{eqnarray}
see Appendix \ref{App-chazy} for more details. The boundary
condition to be imposed at infinity is
\begin{eqnarray}
    \varphi(\varsigma)\Big|_{\varsigma\rightarrow \infty} \sim \frac{n\kappa}{\varsigma}\left(
        1 + {\cal O}(\varsigma^{-1})
    \right).
\end{eqnarray}
Equations (\ref{gue-TL-1}), (\ref{gue-TL-2}), (\ref{ch-gue}) and their one-point reductions Eqs.~(\ref{TL-1-cc}), (\ref{TL-2-cc}), (\ref{phi-def}) and (\ref{phi-piv}) are the main results of this subsection. They will play a central r\^ole in the forthcoming analysis of the replica approach to GUE.

\subsection{Laguerre Unitary Ensemble (LUE)}

The correlation function of characteristic polynomials in LUE
is defined by the formula
\begin{eqnarray}\fl
\label{rpf-Lue}
    \Pi_{n|p}^{\rm L}({\bm {\varsigma}};{\bm \kappa}) =\frac{1}{{\cal N}_n^{\rm L}} \int_{{\mathbb R}_+^{\,n}} \prod_{j=1}^{n}
        \left( d\lambda_j \,  e^{-\lambda_j}\lambda_j^\nu\,\prod_{\alpha=1}^p (\varsigma_\alpha - \lambda_j)^{\kappa_\alpha} \right)
    \cdot
    \Delta_{n}^2({\boldsymbol \lambda}),
\end{eqnarray}
where
\begin{eqnarray}
\label{N-Lue} \fl
    {\cal N}_n^{\rm L} =  \int_{{\mathbb R}_+^{\,n}} \prod_{j=1}^{n}
        \left( d\lambda_j \,  e^{-\lambda_j} \lambda_j^\nu \right)
    \cdot
    \Delta_{n}^2({\boldsymbol \lambda}) = \prod_{j=1}^{n} \Gamma(j+1)\Gamma(j+\nu)
\end{eqnarray}
is the normalisation constant, and it is assumed that $\nu>-1$. The associated $\tau$ function equals
\begin{eqnarray}\fl
\label{tau-Lue}
    \tau_{n}^{\rm L}({\bm {\varsigma}},{\bm \kappa};{\bm t}) =\frac{1}{n!} \int_{{\mathbb R}_+^{\,n}} \prod_{j=1}^{n}
        \left( d\lambda_j \, e^{-\lambda_j + v({\bm t};\lambda_j)} \lambda_j^\nu
        \prod_{\alpha=1}^p (\varsigma_\alpha - \lambda_j)^{\kappa_\alpha} \right)
    \cdot
    \Delta_{n}^2({\boldsymbol \lambda}).
\end{eqnarray}
In the above definitions, the superscript ${\rm L}$ stands for LUE but it will be omitted from now on.

\subsubsection{Virasoro constraints}
\noindent\newline\newline In the notation of Section \ref{Sec-3},
the definition Eq.~(\ref{rpf-Lue}) implies that \footnote[4]{Notice that ${\rm dim}(\bm{c^\prime}) =0$ follows from Eq.~(\ref{c-redef}) in which ${\cal Z}_0 = \{0\}$.
}
\begin{eqnarray}
    f(\lambda)=1 \;\; &\mapsto& \;\;\; a_k=\delta_{k,1},\\
    g(\lambda) = \lambda-\nu\;\;&\mapsto& \;\;\;b_k = -\nu \delta_{k,0} + \delta_{k,1},\\
    {\cal D} = {\mathbb R}_+ \;\; &\mapsto&\;\;\; {\rm dim}(\bm{c^\prime}) =0.
\end{eqnarray}
This brings the following Virasoro constraints Eqs.~(\ref{2-Vir}) --
(\ref{vo}) for the $\tau$ function Eq.~(\ref{tau-Lue}):
\begin{eqnarray}
\label{2-Vir-L}\fl
    \left[
        \hat{\cal L}_{q}({\bm t}) +\nu \frac{\partial}{\partial t_{q}}- \frac{\partial}{\partial t_{q+1}} + \sum_{m=0}^{q}
        \vartheta_m({\bm \varsigma},{\bm \kappa})
              \frac{\partial}{\partial t_{q-m}}
     \right] \tau_n({\bm \varsigma},{\bm \kappa};{\bm t})
    ={\hat {\mathcal B}}_{q} \,\tau_n({\bm \varsigma},{\bm \kappa};{\bm
    t}).
\end{eqnarray}
where ${\hat {\mathcal B}}_q$ is defined by Eq.~(\ref{Bq}) and $\hat{\cal L}_{q}({\bm t})$ is the Virasoro operator given by Eq.~(\ref{vo}).

In what follows, we need the three lowest Virasoro constraints for
$q=0$, $q=1$ and~$q=2$. Written for $\log \tau_n({\bm \varsigma},{\bm
\kappa};{\bm t})$, they read:
\begin{eqnarray}
\label{2-Vir-q=-1L} \fl
    \left(
        \sum_{j=1}^\infty jt_j\frac{\partial}{\partial t_{j}}
         -
        \frac{\partial}{\partial t_{1}} \right)\,\log \tau_n ({\bm \varsigma},{\bm \kappa};{\bm t})
     + n \left(n +\nu +\kappa\right) ={\hat {\mathcal B}}_0 \,\log\tau_n ({\bm \varsigma},{\bm \kappa};{\bm t}),
\end{eqnarray}
\begin{eqnarray}
\label{2-Vir-q=0L} \fl
    \left(
    \sum_{j=1}^\infty jt_j\frac{\partial}{\partial t_{j+1}}
    -
        \frac{\partial}{\partial t_{2}} +
    \left(
        2n + \nu + \kappa \right) \frac{\partial}{\partial t_{1}}
    \right) \log\tau_n ({\bm \varsigma},{\bm \kappa};{\bm t}) \nonumber \\
    \qquad \qquad\qquad\qquad
     +  n \,\vartheta_1({\bm \varsigma},{\bm \kappa})
    ={\hat {\mathcal B}}_1 \,\log\tau_n({\bm \varsigma},{\bm \kappa};{\bm t}),
\end{eqnarray}
\begin{eqnarray}
\label{2-Vir-q=+1L} \fl
    \Bigg( \sum_{j=1}^\infty jt_j\frac{\partial}{\partial t_{j+2}} -
        \frac{\partial}{\partial t_{3}}
        + \left(
        2n + \nu +\kappa \right)
        \frac{\partial}{\partial t_{2}} \nonumber\\
        \qquad\qquad\qquad +
    \,\vartheta_1({\bm \varsigma},{\bm \kappa})
       \frac{\partial}{\partial t_{1}}+
       \frac{\partial^2}{\partial t_{1}^2}
     \Bigg) \log\tau_n ({\bm \varsigma},{\bm \kappa};{\bm t}) \nonumber \\\fl
     \qquad\qquad\qquad+\left(\frac{\partial}{\partial t_{1}}\log\tau_n ({\bm \varsigma},{\bm \kappa};{\bm t})\right)^2+n\,
      \,\vartheta_2({\bm \varsigma},{\bm \kappa})
    ={\hat {\mathcal B}}_2 \,\log\tau_n({\bm \varsigma},{\bm \kappa};{\bm t}).
\end{eqnarray}

\subsubsection{Toda Lattice hierarchy}\noindent
\newline\newline\noindent
To generate the Toda Lattice hierarchy
for the correlation function $\Pi^{\rm L}_{n|p}({\bm
{\varsigma}};{\bm \kappa})$ [Eq.~(\ref{rpf-Lue})] of characteristic
polynomials we apply the projection formula
\begin{eqnarray}
\label{tau-pi-LUE}
    \Pi^{\rm L}_{n|p}({\bm {\varsigma}};{\bm \kappa}) = \frac{n!}{{\cal N}_n^{\rm L}}\,
    \tau^{\rm L}_n({\bm \varsigma},{\bm \kappa};{\bm t})\Big|_{{\bm t}={\rm 0}},
\end{eqnarray}
in which the $\tau$ function is defined by Eq.~(\ref{tau-Lue}), to
the first and second equation of the ${\bm t}$-dependent TL hierarchy:
\begin{eqnarray} \fl\label{tl-1-equiv-LUE}
    {\rm TL}_1:\quad \frac{\partial^2}{\partial t_1^2}\log \tau_n({\bm t})= \frac{\tau_{n+1}({\bm t}) \,
    \tau_{n-1} ({\bm t})}{\tau_{n}^2({\bm t})},
    \\ \fl
    \label{tl-2-equiv-LUE} {\rm TL}_2:\quad \frac{\partial^2}{\partial t_1 \partial t_2}
    \log \tau_n({\bm t})
     =
     \frac{\tau_{n+1}({\bm t}) \,
    \tau_{n-1} ({\bm t})}{\tau_{n}^2({\bm t})}
    \,
        \frac{\partial}{\partial t_1} \log \left(\frac{\tau_{n+1}({\bm t})}{\tau_{n-1}({\bm t})}\right),
\end{eqnarray}
see Eqs.~(\ref{tl-1-equiv}) and (\ref{tl-2-equiv}).
\noindent\newline\newline (i) To derive the first equation of the TL
hierarchy for $\Pi^{\rm L}_{n|p}({\bm {\varsigma}};{\bm \kappa})$
from Eqs.~(\ref{tau-pi-LUE}) and (\ref{tl-1-equiv-LUE}), we have to
determine
$$
    \frac{\partial^2}{\partial t_1^2}\log\tau_n({\bm t})\Big|_{\bm{t}=0}
$$
with the help of the Virasoro constraints. Differentiating Eq.~(\ref{2-Vir-q=-1L}) over $t_1$ and setting ${\bm t}={\bm 0}$ afterwards, we obtain:
\begin{eqnarray}
\label{d-11-LUE}
    \frac{\partial^2}{\partial t_1^2}\log\tau_n({\bm t})\Big|_{\bm{t}=0}
    = ( 1 -
    {\hat {\mathcal B}}_{0}) \,\frac{\partial}{\partial t_1}
    \log\tau_n({\bm t}) \Big|_{\bm{t}=0}.
\end{eqnarray}
Second, we set ${\bm t}={\bm 0}$ in Eq.~(\ref{2-Vir-q=-1L}) to
identify the relation
\begin{eqnarray}
\label{d-1-LUE}
    \frac{\partial}{\partial t_1}\log\tau_n({\bm t})\Big|_{\bm{t}=0}
    = n\left(
        n+\nu+\kappa
    \right) -
    {\hat {\mathcal B}}_{0} \,
    \log\tau_n({\bm 0}).
\end{eqnarray}
Combining Eqs.~(\ref{d-11-LUE}) and (\ref{d-1-LUE}), we conclude that
\begin{eqnarray}
        \frac{\partial^2}{\partial t_1^2}\log\tau_n({\bm t})\Big|_{\bm{t}=0}
        = n\left(
        n+\nu+\kappa
    \right) -
    {\hat {\mathcal B}}_{0} (
        1 - {\hat {\mathcal B}}_{0}
    )\,
    \log\tau_n({\bm 0}).
\end{eqnarray}
Finally, substituting this result back to Eq.~(\ref{tl-1-equiv-LUE}),
and taking into account Eqs.~(\ref{tau-pi-LUE}) and (\ref{N-Lue}), we
end up with the first TL equation
\begin{eqnarray}\label{TL-1-LUE}\fl
\widetilde{{\rm TL}}_1^{\rm L}: \qquad
    {\hat {\mathcal B}}_{0} ({\hat {\mathcal B}}_{0}-1) \,
        \log \Pi_{n|p}({\bm {\varsigma}};{\bm \kappa}) \nonumber\\
        \quad= n\,(n +\nu) \left(
            \frac{\Pi_{n+1|p}({\bm {\varsigma}};{\bm \kappa}) \,
            \Pi_{n-1|p}({\bm {\varsigma}};{\bm \kappa})}{\Pi_{n|p}^2({\bm {\varsigma}};{\bm \kappa})} -1
        \right) -  n\kappa
\end{eqnarray}
written in the space of physical parameters ${\bm \varsigma}$.

Notice that the above equation becomes more symmetric if written for the correlation function
\begin{eqnarray}
\label{tilde-Pi-LUE}
    \tilde{\Pi}_{n|p}({\bm {\varsigma}}) = \Pi_{n|p}({\bm {\varsigma}})\,\prod_{\alpha=1}^p \varsigma_\alpha^{ - n \kappa_\alpha}.
\end{eqnarray}
The corresponding TL equation reads:
\begin{eqnarray}\label{TL-1-LUE-alt}\fl
\widetilde{\widetilde{{\rm TL}}}_1^{\rm L}: \;
    {\hat {\mathcal B}}_{0} ({\hat {\mathcal B}}_{0}-1) \,
        \log \tilde{\Pi}_{n|p}({\bm {\varsigma}};{\bm \kappa})= n\,(n +\nu) \left(
            \frac{\tilde{\Pi}_{n+1|p}({\bm {\varsigma}};{\bm \kappa}) \,
            {\tilde \Pi}_{n-1|p}({\bm {\varsigma}};{\bm \kappa})}{{\tilde \Pi}_{n|p}^2({\bm {\varsigma}};{\bm \kappa})} -1
        \right).
\end{eqnarray}
\noindent\newline\newline (ii) The second equation of the TL
hierarchy for $\Pi_{n|p}({\bm {\varsigma}};{\bm \kappa})$ can be
derived along the same lines. Equation (\ref{tl-2-equiv-LUE}) suggests
that, in addition to the derivative $\partial/\partial t_1
\log\tau_n$ at ${\bm t}={\bm 0}$ given by Eq.~(\ref{d-1-LUE}), one needs
to know the mixed derivative
$$
    \frac{\partial^2}{\partial t_1\partial t_2}\log\tau_n({\bm t})\Big|_{\bm{t}=0}.
$$
It can be calculated by combining Eq.~(\ref{2-Vir-q=-1L})
differentiated over $t_2$ with Eqs.~(\ref{2-Vir-q=0L}) and
(\ref{d-1-LUE}). Straightforward calculations bring
\begin{eqnarray}
        \frac{\partial^2}{\partial t_1\partial t_2}\log\tau_n({\bm t})\Big|_{\bm{t}=0} =
        (2-{\hat {\mathcal B}}_{0})
        \frac{\partial}{\partial t_2}\log\tau_n({\bm t})\Big|_{\bm{t}=0},
\end{eqnarray}
where
\begin{eqnarray} \fl
        \frac{\partial}{\partial t_2}\log\tau_n({\bm t})\Big|_{\bm{t}=0}
        = n
        \left[
                \left(
        2n + \nu +\kappa\right) \left(
        n + \nu +\kappa \right) + \,\vartheta_1({\bm \varsigma},{\bm \kappa})
        \right] \nonumber\\
        \qquad\qquad-
        \left[
            \left(
        2n + \nu +\kappa \right) \, {\hat {\mathcal B}}_{0} + {\hat {\mathcal B}}_{1}
        \right] \log \tau_n({\bm 0}).
\end{eqnarray}
The final result reads:
\begin{eqnarray} \label{TL-2-LUE}\fl
\widetilde{{\rm TL}}_2^{\rm L}: \quad
     ({\hat {\mathcal B}}_{0}-2)\left[
        \left(
            2 n + \nu + \kappa
        \right) {\hat {\mathcal B}}_{0} + {\hat {\mathcal B}}_{1}
     \right]\,
        \log \Pi_{n|p}({\bm {\varsigma}};{\bm \kappa}) \nonumber \\
        \fl \qquad \qquad=
        n(n+\nu)\,
            \frac{\Pi_{n+1|p}({\bm {\varsigma}};{\bm \kappa}) \,
            \Pi_{n-1|p}({\bm {\varsigma}};{\bm \kappa})}{\Pi_{n|p}^2({\bm {\varsigma}};{\bm \kappa})} \nonumber\\
            \times \left[
            2 \left( 2 n +\nu+\kappa \right)
            -\,{\hat {\mathcal B}}_{0}
         \log \left(
            \frac{\Pi_{n+1|p}({\bm {\varsigma}};{\bm \kappa})}{\Pi_{n-1|p}({\bm {\varsigma}};{\bm \kappa})}
        \right)\right] \nonumber\\
        \quad
        - 2n \left( 2 n +\nu+ \kappa \right) \left( n +\nu+ \kappa \right) - n\,\vartheta_1({\bm \varsigma},{\bm \kappa}).
\end{eqnarray}
This equation takes a more compact form if written for the correlation function $\tilde{\Pi}_{n|p}$ defined by Eq.~(\ref{tilde-Pi-LUE}):
\begin{eqnarray} \label{TL-2-LUE-alt}\fl
\widetilde{\widetilde{{\rm TL}}}_2^{\rm L}: \quad
     ({\hat {\mathcal B}}_{0}-2)\left[
        \left(
            2 n + \nu + \kappa
        \right) {\hat {\mathcal B}}_{0} + {\hat {\mathcal B}}_{1}
     \right]\,
        \log \tilde{\Pi}_{n|p}({\bm {\varsigma}};{\bm \kappa}) \nonumber \\
        \fl \qquad \qquad=
        n(n+\nu)\,
            \frac{\tilde{\Pi}_{n+1|p}({\bm {\varsigma}};{\bm \kappa}) \,
            \tilde{\Pi}_{n-1|p}({\bm {\varsigma}};{\bm \kappa})}{\tilde{\Pi}_{n|p}^2({\bm {\varsigma}};{\bm \kappa})} \nonumber\\
            \times \left[
            2 (2 n +\nu)
            -\,{\hat {\mathcal B}}_{0}
         \log \left(
            \frac{{\tilde \Pi}_{n+1|p}({\bm {\varsigma}};{\bm \kappa})}{\tilde{\Pi}_{n-1|p}({\bm {\varsigma}};{\bm \kappa})}
        \right)\right] \nonumber\\
        \qquad
        - 2n (n+\nu) \left( 2 n +\nu+\kappa \right).
\end{eqnarray}

\subsubsection{KP hierarchy and Painlev\'e V equation}
\noindent\newline\newline
The same technology is at work for projecting the KP hirerachy Eq.~(\ref{kph}) onto ${\bm t}={\bm 0}$. Below, only the first KP equation
\begin{eqnarray} \fl \label{kp1-exp-rep-LUE}
{\rm KP}_1:\quad \left(
    \frac{\partial^4}{\partial t_1^4} + 3 \frac{\partial^2}{\partial t_2^2}
    - 4 \frac{\partial^2}{\partial t_1 \partial t_3}
\right)\log\, \tau_n({\bm t}) + 6 \left(
    \frac{\partial^2}{\partial t_1^2} \log\, \tau_n({\bm t})
\right)^2 = 0
\end{eqnarray}
will be treated. Notice that no superscript $(s)$ appears in
Eq.~(\ref{kp1-exp-rep-LUE}) as the LUE confinement potential does not
depend on the matrix size $n$. To make the forthcoming calculation more efficient, it is beneficial to introduce the notation
\begin{eqnarray}
\label{T-not}
    T_{\ell_1 \ell_2 \dots \ell_k} = \left(
    \prod_{j=1}^k \frac{\partial}{\partial t_{\ell_j}}
    \right) \log \tau_n({\bm t})\Bigg|_{{\bm t}={\bm 0}},\;\; T = \log \tau_n({\bm 0}),
\end{eqnarray}
which brings the KP equation Eq.~(\ref{kp1-exp-rep-LUE}) projected onto ${\bm t}={\bm 0}$ to the form
\begin{eqnarray} \label{KPT}
    T_{1111} + 3 T_{22} - 4T_{13} + 6T_{11}^2=0.
\end{eqnarray}
\noindent\newline
(i) First, we observe that $T_{11}$ and $T_{1111}$ can be determined from the following chain of relations, obtained by repeated differentiation of the first Virasoro constraint Eq.~(\ref{2-Vir-q=-1L}) with respect to $t_1$:
\begin{eqnarray}
\cases{
         T_{1\phantom{1}\phantom{1}\phantom{1}} = n\left(
    n+\nu +\kappa
  \right) - {\hat {\mathcal B}}_{0} \, T, &\\
      T_{11\phantom{1}\phantom{1}}  = (1-{\hat {\mathcal B}}_{0}) \, T_1, &\\
     T_{111\phantom{1}} = (2-{\hat {\mathcal B}}_{0}) \, T_{11}, &\\
     T_{1111} = (3-{\hat {\mathcal B}}_{0}) \, T_{111}. &
     }
\end{eqnarray}
Hence,
\begin{eqnarray} \label{T11-LUE}
    T_{11} = n\left(
    n+\nu +\kappa
  \right) - (1-{\hat {\mathcal B}}_{0}) {\hat {\mathcal B}}_{0} \, T
\end{eqnarray}
and
\begin{eqnarray}
 \label{T1111-LUE}
        T_{1111}
        = 3!\, n \left(
            n+\nu+\kappa
        \right) - ( 3 - {\hat {\mathcal B}}_{0} ) ( 2 - {\hat {\mathcal B}}_{0} ) ( 1 - {\hat {\mathcal B}}_{0} )
        {\hat {\mathcal B}}_{0}\,   T.
\end{eqnarray}
\noindent\newline
(ii) Second, to determine $T_{13}$, we differentiate the first Virasoro constraint Eq.~(\ref{2-Vir-q=-1L}) with respect to $t_3$, and make use of the second and third constraints as they stand [Eqs.~(\ref{2-Vir-q=0L}) and (\ref{2-Vir-q=+1L})] to obtain:
\begin{eqnarray}\fl \qquad
\label{eq463}
\cases{
         T_{13} = (3-{\hat {\mathcal B}}_{0})\, T_3, &\\
      T_{3\phantom{1}}  = (2n+\nu+\kappa)\, T_2 + \,\vartheta_1({\bm \varsigma},{\bm \kappa}) T_1 + T_1^2 + T_{11} -
      {\hat {\mathcal B}}_{2}\, T + n \,\vartheta_2({\bm \varsigma},{\bm \kappa}), &\\
     T_{2\phantom{1}} = (2n+\nu+\kappa)\, T_1 -
     {\hat {\mathcal B}}_{1}\, T + n\,\vartheta_1({\bm \varsigma},{\bm \kappa}). &}
\end{eqnarray}
Although easy to derive, an explicit expression for $T_{13}$ is too cumbersome to be explicitly stated here.
\noindent\newline\newline
(iii) To calculate $T_{22}$, the last unknown ingredient of Eq.~(\ref{KPT}), we differentiate the first and second Virasoro constraints
[Eqs.~(\ref{2-Vir-q=-1L}) and (\ref{2-Vir-q=0L})] with respect to $t_2$ to realize that
\begin{eqnarray}
\label{eq464}
\cases{
    T_{22} = 2T_3 + (2n +\nu +\kappa)\, T_{12} - {\hat {\mathcal B}}_{1}\, T_2, & \\
    T_{12} = (2-{\hat {\mathcal B}}_{0})\, T_2, &}
\end{eqnarray}
Combining Eqs.~(\ref{eq463}) and (\ref{eq464}), one readily derives a closed expression for $T_{22}$.
\noindent\newline\newline
Finally, we substitute so determined $T_{1111}$, $T_{11}$, $T_{13}$ and $T_{22}$ into Eq.~(\ref{KPT}) to generate a nonlinear
differential equation for $\log \Pi_{n|p}({\bm \varsigma};{\bm
\kappa})$ in the form
\begin{eqnarray} \fl \label{ch-Lue}
\widetilde{{\rm KP}}_1^{\rm L}: \quad
    \Bigg[
        {\hat {\mathcal B}}_0^4
    -2\hat {\mathcal B}_{0}^3
    -\left[ (\nu+\kappa)^2+4\,\vartheta_1({\bm \varsigma},{\bm \kappa}) -1\right] {\hat {\mathcal B}}_0^2
    + 2 \,\vartheta_1({\bm \varsigma},{\bm \kappa}) {\hat {\mathcal B}}_0
    +3 {\hat {\mathcal B}}_1^2
        \nonumber\\\fl
    \qquad\qquad
    +(2n+\nu+\kappa) {\hat {\mathcal B}}_1 (2{\hat {\mathcal B}}_0-1)
    - 2 {\hat {\mathcal B}}_2 (2{\hat {\mathcal B}}_0+1)
    \Bigg] \, \log \Pi_{n|p}({\bm \varsigma};{\bm \kappa}) \nonumber\\\fl\qquad
    \qquad + 6 \left(
        {\hat {\mathcal B}}_0^2 \log \Pi_{n|p}({\bm \varsigma};{\bm \kappa})
    \right)^2
    -4\left( {\hat {\mathcal B}}_0\log \Pi_{n|p}({\bm \varsigma};{\bm \kappa})\right)\left( {\hat {\mathcal B}}_0^2 \log \Pi_{n|p}({\bm \varsigma};{\bm \kappa})\right)\nonumber\\
    \qquad\qquad\qquad= n\left[ \left(\nu+\kappa\right)\,\vartheta_1({\bm \varsigma},{\bm \kappa})+\vartheta_2({\bm \varsigma},{\bm \kappa}) \right].
\end{eqnarray}
\newline
{\it Remark.}---For $p=1$, the above equation reads:
\begin{eqnarray} \fl
    \Bigg[
        \varsigma^4 \frac{\partial^4}{\partial \varsigma^4}
        + 4 \varsigma^3 \frac{\partial^3}{\partial \varsigma^3}
        + 2 \varsigma^2 (1-\varsigma^2) \frac{\partial^2}{\partial \varsigma^2}
        - (\nu+\kappa)^2 \left( \varsigma^2 \frac{\partial^2}{\partial \varsigma^2} +
             \varsigma \frac{\partial}{\partial \varsigma}
        \right) \nonumber \\
              \fl
              \quad \quad
               + (2n+\nu-\kappa)
               \left(
                2 \varsigma^3 \,\frac{\partial^2}{\partial \varsigma^2}
        + \varsigma^2\,\frac{\partial}{\partial \varsigma} \right)
    \Bigg]  \log \Pi_{n|p}(\varsigma;\kappa)\nonumber\\
    \fl \quad \quad + 2 \varsigma^2
    \left[
        \left(
            \frac{\partial}{\partial \varsigma} + \varsigma \frac{\partial^2}{\partial \varsigma^2}
        \right)\, \log \Pi_{n|p}(\varsigma;\kappa)
    \right]
    \left[
        \left(
            \frac{\partial}{\partial \varsigma} + 3 \varsigma \frac{\partial^2}{\partial \varsigma^2}
        \right)\, \log \Pi_{n|p}(\varsigma;\kappa)
    \right]\nonumber \\
     = n\kappa \varsigma (\nu+\kappa+\varsigma).
\end{eqnarray}
It can further be simplified if written for the function
\begin{eqnarray}
\label{pi-phi-link}
    \varphi(\varsigma) = \varsigma \frac{\partial}{\partial \varsigma} \,\log \Pi_{n}(\varsigma;\kappa) - n\kappa.
\end{eqnarray}
Straightforward calculations yield:
\begin{eqnarray} \fl \label{pv-LUE-chazy}
    \varsigma^2 \varphi^{\prime\prime\prime} + \varsigma \varphi^{\prime\prime} -
    \left[
        \varsigma^2 - 2 (2n+\nu -\kappa) \,\varsigma + 4 \kappa n + (\nu+\kappa)^2
    \right]\, \varphi^\prime \nonumber\\
    + \left[ 2(2\kappa -n) -\nu -2\varsigma\right] \,\varphi + 6\, \varsigma (\varphi^\prime)^2 - 4 \varphi \varphi^\prime = 2n\kappa (n+\nu).
\end{eqnarray}
This can be recognized as the Chazy I form (see Appendix~\ref{App-chazy}) of the fifth Painlev\'e transcendent. Equivalently, $\varphi$ satisfies the Painlev\'e V equation in the
Jimbo-Miwa-Okamoto form (Forrester and Witte 2002, Tracy and Widom 1994):
\begin{eqnarray}
\label{phi-pv}\fl
    P_{\rm V}: \quad
              \left(\varsigma \varphi^{\prime\prime} \right)^2
        - \left[ \varphi- \varsigma \varphi^{\prime}
        + 2 (\varphi^\prime)^2 +
        (2n+\nu-\kappa) \varphi^\prime
        \right]^2 \nonumber \\
        \hspace{2cm}
        + 4 \varphi^\prime
        (\varphi^\prime+n)
        (\varphi^\prime+n+\nu)
        (\varphi^\prime-\kappa) = 0.
\end{eqnarray}
Both equations have to be supplemented by the boundary condition
\begin{eqnarray}
    \varphi(\varsigma)\Big|_{\varsigma\rightarrow \infty} \sim \frac{n(n+\nu)\kappa}{\varsigma}\left(
        1 + {\cal O}(\varsigma^{-1})
    \right)
\end{eqnarray}
following from Eq.~(\ref{pi-phi-link}) and the asymptotic analysis of Eq.~(\ref{rpf-Lue}). Equations (\ref{TL-1-LUE}), (\ref{TL-1-LUE-alt}), (\ref{TL-2-LUE}), (\ref{TL-2-LUE-alt}), (\ref{ch-Lue}) and (\ref{phi-pv}) represent the main results of this subsection.

\subsection{Discussion}
This Section concludes the detailed exposition of integrable theory of correlation functions of RMT characteristic polynomials. Among the main results derived are:
\begin{itemize}
  \item The multivariate first [Eqs. (\ref{gue-TL-1}) and (\ref{TL-1-LUE-alt})] and second [Eqs. (\ref{gue-TL-2}) and (\ref{TL-2-LUE-alt})] equations of the Toda Lattice hierarchy \footnote{See also their single variable reductions Eqs. (\ref{TL-1-cc}) and (\ref{TL-2-cc}) derived for the GUE.} which establish nonlinear differential recurrence relations between ``nearest neighbor'' correlation functions ${\Pi}_{n|p}({\bm {\varsigma}};{\bm \kappa})$ and ${\Pi}_{n\pm 1|p}({\bm {\varsigma}};{\bm \kappa})$, and
  \item The nonlinear multivariate differential equations [Eqs. (\ref{ch-gue}) and (\ref{ch-Lue})] satisfied by ${\Pi}_{n|p}({\bm {\varsigma}};{\bm \kappa})$ alone. These can be considered as multivariate generalisations of the corresponding Painlev\'e equations arising in the one-point setup $p=1$ [Eqs. (\ref{phi-piv}) and (\ref{phi-pv})]
\end{itemize}
Other nonlinear multivariate relations between the correlation functions ${\Pi}_{n|p}({\bm {\varsigma}};{\bm \kappa})$ and ${\Pi}_{n\pm q|p}({\bm {\varsigma}};{\bm \kappa})$ can readily be obtained from the {\it modified} Toda Lattice and Kadomtsev-Petviashvili hierarchies listed in Section \ref{Sec-3-4}.

Finally, let us stress that a similar calculational framework applies to other $\beta=2$ matrix integrals depending on one (Osipov and Kanzieper 2007) or more (Osipov and Kanzieper 2009; Osipov, Sommers and \.Zyczkowski 2010) parameters. The reader is referred to the above papers for further details.

\section{Integrability of Zero-Dimensional Replica Field Theories}
\label{Sec-5}
\subsection{Introduction}
In this Section, the integrable theory of CFCP will be utilised to present a tutorial exposition of the exact
approach to zero dimensional replica field theories
formulated in a series of publications (Kanzieper 2002, Splittorff and Verbaarschot 2003, Osipov and
Kanzieper 2007). Focussing, for definiteness, on the calculation of the finite-$N$ average eigenlevel density in the GUE (whose exact form (Mehta 2004)
\begin{eqnarray}
    \varrho_N(\epsilon) = \frac{1}{2^{N} \Gamma(N) \sqrt{\pi}}
    \,e^{-\epsilon^2} \left[
        H_N^\prime(\epsilon) H_{N-1}(\epsilon) -
        H_N(\epsilon) H_{N-1}^\prime(\epsilon)
    \right]\qquad \nonumber
\end{eqnarray}
has been known for decades), we shall put a special emphasis on a {\it comparative analysis}
of three alternative formulations -- fermionic, bosonic and supersymmetric -- of the replica method. This will allow us to meticulously analyse the {\it fermionic-bosonic factorisation phenomenon} of RMT spectral correlation functions in the {\it fermionic} and {\it bosonic} variations of the replica method, where its existence is not self-evident, to say the least.

\subsection{Density of eigenlevels in finite-$N$ GUE}
To determine the mean density of eigenlevels in the GUE, we define
the average one-point Green function
\begin{eqnarray}
    G(z;N) = \left< {\rm tr} (z - {\boldsymbol {\mathcal H}})^{-1}\right>_{{\boldsymbol {\mathcal H}}\in {\rm GUE}_N}
\end{eqnarray}
that can be restored from the replica partition function ($n \in
{\mathbb R}^+$)
\begin{eqnarray}
\label{rpf-def}
    {\mathcal Z}^{(\pm)}_{n}(z;N) = \left<
        {\rm \det}^{\pm n} (z - {\boldsymbol {\mathcal H}})
    \right>_{\boldsymbol {\mathcal H} \in {\rm GUE}_N}
\end{eqnarray}
through the replica limit
\begin{eqnarray}
    G(z;N) = \pm \lim_{n\rightarrow 0} \frac{1}{n} \frac{\partial}{\partial z} {\mathcal Z}^{(\pm)}_{ n}(z;N).
\end{eqnarray}
Equation (\ref{rpf-def}) can routinely be mapped onto either
fermionic or bosonic replica field theories, the result being (see, e.g., Kanzieper 2010)
\begin{eqnarray}
\label{fer-pf}
    {\mathcal Z}^{(+)}_{n}(z;N) = \frac{1}{c_n}\, \imag^{-nN}
    \int
    ({\cal D}_n{\boldsymbol {\mathcal Q}})\, \,e^{-{\rm tr}_n {\boldsymbol {\mathcal Q}}^2}\,
    {\rm det}{}_n^{N} \left(
        \imag z - {\boldsymbol {\mathcal Q}}
    \right)
\end{eqnarray}
and
\begin{eqnarray}
\label{bos-pf}
    {\mathcal Z}^{(-)}_{n}(z;N) = \frac{1}{c_{n}}\,
    \int
    ({\cal D}_{n}{\boldsymbol {\mathcal Q}})\, \,e^{-{\rm tr}_{n} {\boldsymbol {\mathcal Q}}^2}\,
    {\rm det}{}_{n}^{-N} \left(
        z - {\boldsymbol {\mathcal Q}}
    \right).
\end{eqnarray}
Both integrals run over $n\times n$ Hermitean matrix ${\boldsymbol
{\mathcal Q}}$; the normalisation constant $c_{n}$ equals
\begin{eqnarray}
    c_{n} = \int
    ({\cal D}_{n}{\boldsymbol {\mathcal Q}})\, \,e^{-{\rm tr}_{n} {\boldsymbol {\mathcal Q}}^2}.
\end{eqnarray}
By derivation, the replica parameter $n$ in Eqs. (\ref{fer-pf}) and
(\ref{bos-pf}) is restricted to integers, $n \in {\mathbb Z}^+$.
Notably, Eqs.~(\ref{fer-pf}) and (\ref{bos-pf}) are particular cases
of the correlation function $\Pi_{n|p}({\boldsymbol
\varsigma};{\boldsymbol \kappa})$ studied in previous sections.

\subsubsection{Fermionic replicas}\label{Sec-FR}\noindent\newline\newline
Indeed, comparison of Eq.~(\ref{fer-pf}) with the definition
Eq.~(\ref{rpf-gue}) yields
\begin{eqnarray}
\label{ZPi-F}
    {\mathcal Z}^{(+)}_{n}(z;N) = (-\imag)^{nN} \Pi_{n}(\imag z; N),
\end{eqnarray}
where the shorthand notation $\Pi_{n}(z; N)$ is used to denote
$\Pi_{n|1}^{\rm G}(z; N)$, in accordance with the earlier notation in
Eqs.~(\ref{TL-1-cc}) and (\ref{TL-2-cc}). This observation results in the Painlev\'e IV representation of the
fermionic replica partition function [see Eqs.~(\ref{phi-def}) and
(\ref{phi-piv})]:
\begin{eqnarray}
\label{052}
    \frac{\partial}{\partial z} \log {\mathcal Z}^{(+)}_{n}(z;N) =
    \imag\, \varphi(t;n,N)\Big|_{t=\imag z},
\end{eqnarray}
where $\varphi(t;n,N)$ is the fourth Painlev\'e transcendent
satisfying the equation
\begin{eqnarray}
\label{phi-piv-00}
        (\varphi^{\prime\prime})^2 - 4 (\varphi - t\, \varphi^\prime)^2
        + 4 \varphi^\prime (\varphi^\prime+2n)(\varphi^\prime-2N)=0
\end{eqnarray}
subject to the boundary conditions~\footnote{Equation (\ref{phi-bc-f}) follows from  Eqs.~(\ref{052}), (\ref{ZPi-F}) and the footnote below Eq.~(\ref{ch-gue}).}
\begin{eqnarray}\label{phi-bc-f}
    \varphi(t;n,N) \sim \frac{nN}{t}, \qquad |t| \rightarrow  \infty, \qquad t\in {\mathbb C}.
\end{eqnarray}
Here and above, $n \in {\mathbb Z}_+$.
\newline\newline\noindent
Equations (\ref{052}) and (\ref{phi-piv-00}) open the way for calculating the average Green function $G(z;N)$ via the fermionic replica
limit
\begin{eqnarray}
\label{RL-f}
    G(z;N) = \lim_{n\rightarrow 0} \frac{1}{n} \frac{\partial}{\partial z} {\mathcal Z}_n^{(+)}(z;N)
     =  \imag \lim_{n\rightarrow 0} \frac{1}{n} \, \varphi(t;n,N)\Big|_{t=\imag z}.
\end{eqnarray}
For the prescription Eq.~(\ref{RL-f}) to be operational, the Painlev\'e representation of ${\mathcal Z}^{(+)}_{n}(z;N)$ should hold \footnote{Previous studies (Kanzieper 2002, Osipov and Kanzieper 2007) suggest that this is indeed the case.}
for $n\in {\mathbb R}_+$. Notice that for generic real $n$, the fermionic replica partition function
${\mathcal Z}^{(+)}_{n}(z;N)$ is no longer an analytic function of $z$ and exhibits a discontinuity across the real axis. For this reason,
the Painlev\'e equation Eq.~(\ref{phi-piv-00}) should be solved separately for $\mathfrak{Re}\, t<0$ ($\mathfrak{Im}\, z >0$) and  $\mathfrak{Re}\, t>0$ ($\mathfrak{Im}\, z <0$).
\newline\newline\noindent
{\it Replica limit and the Hamiltonian formalism.}---To implement the replica limit, we employ the Hamiltonian formulation of the Painlev\'e IV (Noumi 2004, Forrester and Witte 2001) which associates
$\varphi(t;n,N)$ with the polynomial Hamiltonian (Okamoto 1980a)
\begin{eqnarray}
\label{phi-H}
    \varphi(t;n,N) \equiv H_{\rm f}\left\{P,Q,t\right\} = (2P + Q + 2t) P Q + 2 n P - N Q
\end{eqnarray}
of a dynamical system $\{Q,P,H_{\rm f}\}$, where $Q=Q(t;n,N)$ and $P=P(t;n,N)$ are canonical coordinate and momentum. For such a system,
Hamilton's equations of motion read:
\begin{eqnarray}
    \dot{Q} &=& + \frac{\partial H_{\rm f}}{\partial P} = Q (Q + 4 P + 2t) + 2n, \\
    \dot{P} &=& -  \frac{\partial H_{\rm f}}{\partial Q} = - P (2Q+2P+2t) +N.
\end{eqnarray}
Since
\begin{eqnarray}
\label{RL-h}
    G(z;N)  =  \imag \lim_{n\rightarrow 0} \frac{1}{n} \, H_{\rm f}\left\{P,Q,t\right\}\Big|_{t=\imag z},
\end{eqnarray}
we need to develop a small-$n$ expansion for the Hamiltonian $H_{\rm f}\left\{P,Q,t\right\}$. Restricting ourselves to the linear in $n$ terms,
\begin{eqnarray}
\label{H-exp}
    H_{\rm f}\left\{P,Q,t\right\} = nH_1^{({\rm f})}(t;N) + {\mathcal O}(n^2)
\end{eqnarray}
and
\begin{eqnarray}
    P(t;n,N) &=& p_0(t;N) + n p_1(t;N) + {\mathcal O}(n^2), \\
    Q(t;n,N) &=& q_0(t;N) + n q_1(t;N) + {\mathcal O}(n^2),
\end{eqnarray}
we conclude that $q_0(t;N)=0$. This derives directly from the expansion Eq.~(\ref{H-exp}) in which absence of the term of order ${\mathcal O}(n^0)$ is guaranteed by the normalisation condition ${\mathcal Z}_0^{(+)}(z;N)=1$. As the result \footnote[1]{We will drop the superscript $({\rm f})$ wherever this does not cause a notational confusion.},
\begin{eqnarray}
    \label{gh1}
    G(z;N) = \imag H_1^{({\rm f})} (\imag z;N),
\end{eqnarray}
where
\begin{eqnarray} \label{h1}
    H_1(t;N) &=& 2 p_0 q_1 (p_0+t) + 2 p_0 -N q_1, \\
    \label{h1-dot}
    \dot{H}_1(t;N) &=& 2 p_0 q_1.
\end{eqnarray}
Here, $p_0=p_0(t;N)$ and $q_1 = q_1(t;N)$ are solutions to the system of coupled first order equations:
\begin{eqnarray}
\label{system}
\left\{
\begin{array}{cll}
    \dot{p}_0 &=& - 2 p_0^2 - 2 p_0 t + N,\\
    \dot{q}_1 &=& 4 p_0 q_1 + 2 q_1 t + 2.
\end{array}\right.
\end{eqnarray}
Since the initial conditions are known for $H_1(t;N)$ rather than for $p_0(t;N)$ and $q_1(t;N)$ separately, below we determine these two functions up to integration constants.

The function $p_0(t;N)$ satisfies the Riccati differential equation whose solution is
\begin{eqnarray}
\label{p0-ans}
    p_0(t;N) = \frac{1}{2} \left[ \frac{\dot{u}_+(t)}{u_+(t)} - t \right],
\end{eqnarray}
where
\begin{eqnarray}
\label{ut}
    u_+(t) = c_1 D_{-N-1}(t\sqrt{2}) + c_2 (-\imag)^{N} D_{N}(it \sqrt{2})
\end{eqnarray}
is, in turn, a solution to the equation of parabolic cylinder
\begin{eqnarray}
    \ddot{u}_+(t) - (2N+1+t^2)u_+(t)=0.
\end{eqnarray}
Two remarks are in order. First, factoring out $(-\imag)^{N}$ in the second term in Eq.~(\ref{ut}) will simplify the formulae to follow. Second, the solution Eq.~(\ref{p0-ans}) for $p_0(t;N)$ actually depends on a {\it single} constant (either $c_1/c_2$ or $c_2/c_1$) as it must be.

To determine $q_1(t;N)$, we substitute Eq.~(\ref{p0-ans}) into the second formula of Eq.~(\ref{system}) to derive:
\begin{eqnarray}
\label{q1-int}
    q_1(t;N) = 2 u_+^2(t) \int \frac{dt}{u_+^2(t)}.
\end{eqnarray}
Making use of the integration formula (see Appendix \ref{App-D-int})
\begin{eqnarray}
    \label{osipov-integral}
    \int \frac{dt}{u_+^2(t)} = \frac{1}{\sqrt{2}}\, \frac{\alpha_1 D_{-N-1}(t\sqrt{2}) + \alpha_2 (-\imag)^N D_{N}(it\sqrt{2})}{u_+(t)},
\end{eqnarray}
where two constants $\alpha_1$ and $\alpha_2$ are subject to the constraint
\begin{eqnarray}
\label{const-a12}
    c_1 \alpha_2 - c_2 \alpha_1 = 1,
\end{eqnarray}
we further reduce Eq.~(\ref{q1-int}) to
\begin{eqnarray}
\label{q1-ans}
    q_1(t;N) = \sqrt{2} u_+(t) \left[
        \alpha_1 D_{-N-1}(t\sqrt{2}) + \alpha_2 (-\imag)^N D_{N}(it\sqrt{2})
    \right].
\end{eqnarray}
Equations (\ref{h1-dot}), (\ref{p0-ans}), (\ref{q1-ans}) and the identity
\begin{eqnarray} \fl
    \dot{u}_+(t) - t\, u_+(t) = -\sqrt{2} \left[
        c_1 D_{-N}(t\sqrt{2}) - c_2 (-\imag)^{N-1} N D_{N-1}(\imag t \sqrt{2})
    \right]
\end{eqnarray}
(obtained from Eq.~(\ref{ut}) with the help of Eqs.~(\ref{d-rec-1}) and (\ref{d-rec-2})) yield $\dot{H}_1(t;N)$ in the form
\begin{eqnarray} \fl
\label{H1-sol}
    \dot{H}_1(t;N) = -2
       \left[
        \alpha_1 D_{-N-1}(t\sqrt{2}) + \alpha_2 (-\imag)^N D_{N}(\imag t\sqrt{2})
    \right] \nonumber \\
    \times
            \left[
            c_1 D_{-N}(t\sqrt{2}) - c_2 (-\imag)^{N-1} N D_{N-1}(\imag t \sqrt{2})
        \right].
\end{eqnarray}
Notice that appearance of
four integration constants ($c_1$, $c_2$, $\alpha_1$ and $\alpha_2$) in Eq.~(\ref{H1-sol}) is somewhat illusive: a little thought shows that
there is a pair of independent constants, either $(c_1/c_2,\alpha_2 c_2)$ or their derivatives.\newline\newline
To determine the unknown constants in Eq.~(\ref{H1-sol}), we make use of the asymptotic formulae for the functions of parabolic cylinder (collected in Appendix \ref{App-D-int}) in an attempt to meet the boundary conditions~\footnote{~Equation (\ref{h1-as}) is straightforward to derive from Eqs.~(\ref{H-exp}), (\ref{phi-H}) and (\ref{phi-bc-f}).}
\begin{eqnarray}\label{h1-as}
    H_1(t;N) \sim \frac{N}{t}, \quad \dot{H}_1(t;N) \sim -\frac{N}{t^2}, \qquad |t| \rightarrow  \infty, \qquad t\in {\mathbb C}.
\end{eqnarray}
Following the discussion next to Eq.~(\ref{RL-f}), the two cases $\mathfrak{Re\,} t<0$ and $\mathfrak{Re\,} t>0$ will be treated separately.

\begin{itemize}
  \item  {\it The case} $\mathfrak{Re\,} t<0$. Asymptotic analysis of Eq.~(\ref{H1-sol}) at $t\rightarrow -\infty$ yields
  \begin{eqnarray}
        \frac{\alpha_2}{\alpha_1} = (-1)^{N-1} \frac{\sqrt{2\pi}}{N!},
    \nonumber
  \end{eqnarray}
  so that
\begin{eqnarray} \fl
\label{H1-sol-step-10}
    \dot{H}_1(t;N) = 2 D_{-N-1}(-t\sqrt{2}) \left[
    \alpha_1 c_1 D_{-N}(-t\sqrt{2}) - \imag ^{N-1} N  D_{N-1}(\imag t\sqrt{2})
    \right].
\end{eqnarray}
Here, we have used Eq.~(\ref{relD}). To determine the remaining constant $\alpha_1 c_1$, we make use of the boundary
condition Eq.~(\ref{h1-as}) for $t \rightarrow \pm \imag \infty - 0$. Straightforward
calculations bring $\alpha_1 c_1 = 0$. We then conclude that
\begin{eqnarray} \fl \label{h1t-neg}
    \dot{H}_1(t;N) = - 2 (-\imag )^{N-1} N\, D_{-N-1}(-t\sqrt{2}) D_{N-1}(-\imag t \sqrt{2}),\qquad \mathfrak{Re\,} t<0.
\end{eqnarray}
\vspace{0.2cm}
  \item  {\it The case} $\mathfrak{Re\,} t>0$.
        Asymptotic analysis of Eq.~(\ref{H1-sol}) at $t\rightarrow +\infty$ yields $\alpha_2=0$ so that
\begin{eqnarray} \fl
\label{H1-sol-step-1}
    \dot{H}_1(t;N) = 2
         D_{-N-1}(t\sqrt{2})
            \left[
            \frac{c_1}{c_2} D_{-N}(t\sqrt{2}) - (-\imag)^{N-1} N D_{N-1}(\imag t \sqrt{2})
        \right].
\end{eqnarray}
To determine the remaining constant $c_1/c_2$, we make use of the boundary condition Eq.~(\ref{h1-as}) for $t \rightarrow \pm \imag \infty +0$. Straightforward
calculations bring $c_1/c_2=0$. We then conclude that
\begin{eqnarray} \fl \label{h1t-pos}
    \dot{H}_1(t;N) = -2 (-\imag)^{N-1} N\, D_{-N-1}(t\sqrt{2}) D_{N-1}(\imag t \sqrt{2}),\qquad \mathfrak{Re\,} t>0.
\end{eqnarray}
\end{itemize}
\noindent\newline
The calculation of $\dot{H}_1(t;N)$ can be summarised in a single formula
\begin{eqnarray}
\label{h1d-answer}
    \dot{H}_1(t;N) = - 2 (-\imag)^{N-1} N D_{-N-1}(\sigma_{\imag t} t \sqrt{2}) D_{N-1}(\imag \sigma_{\imag t} t \sqrt{2}),
\end{eqnarray}
where $\sigma_{\imag t}={\rm sgn}\,\mathfrak{Im}\, (\imag t)={\rm sgn}\,\mathfrak{Re}\, t$ denotes the sign of $\mathfrak{Re}\, t$. In terms of canonical variables $p_0(t;N)$ and $q_1(t;N)$, this result translates to
\begin{eqnarray}
\label{p0-fer}
    p_0(t;N) &=& \frac{\imag N \sigma_{\imag t}}{\sqrt{2}} \frac{D_{N-1}(\imag t \sigma_{\imag t}\sqrt{2})}{D_N(\imag t \sigma_{\imag t}\sqrt{2})}, \\
    \label{q1-fer}
    q_1(t;N) &=& - \sqrt{2} \sigma_{\imag t} (-\imag)^N D_{-N-1}( t \sigma_{\imag t} \sqrt{2}) D_N(\imag t \sigma_{\imag t}  \sqrt{2}).
\end{eqnarray}
Now $H_1(t;N)$ can readily be restored by integrating Eq.~(\ref{h1d-answer}). We proceed in three steps. (i) First, we make use of differential recurrence relations Eqs.~(\ref{d-rec-1}) and (\ref{d-rec-2}) and the Wronskian Eq.~(\ref{wronsk_D}) to prove the identity
\begin{eqnarray} \fl
    \imag N D_{-N-1}(\sigma_{\imag t} t \sqrt{2}) D_{N-1}(\imag  \sigma_{\imag t} t \sqrt{2}) = \imag^N - D_{-N}(\sigma_{\imag t} t \sqrt{2}) D_{N}(\imag \sigma_{\imag t} t \sqrt{2}).
\end{eqnarray}
The latter allows us to write down $\dot{H}_1(t;N)$ as
\begin{eqnarray}
    \label{h1d-equiv}
    \dot{H}_1(t;N) = - 2 + 2 (-\imag)^{N} D_{-N}(\sigma_{\imag t} t \sqrt{2}) D_{N}(\imag \sigma_{\imag t} t \sqrt{2}).
\end{eqnarray}
(ii) Second, it is beneficial to employ the differential equation Eq.~(\ref{deq}) to derive
\begin{eqnarray}
    \frac{d^2}{dt^2} D_{-N}(\sigma_{\imag t} t \sqrt{2}) &=& (2N-1+t^2) D_{-N}(\sigma_{\imag t} t \sqrt{2}), \\
    \frac{d^2}{dt^2} D_{N}(\imag \sigma_{\imag t} t \sqrt{2}) &=& (2N+1+t^2) D_{N}(\imag \sigma_{\imag t} t \sqrt{2}).
\end{eqnarray}
These two relations imply
\begin{eqnarray}  \fl
    2  D_{-N}(\sigma_{\imag t} t \sqrt{2}) D_{N}(\imag \sigma_{\imag t} t \sqrt{2}) = D_{-N}(\sigma_{\imag t} t \sqrt{2}) \frac{d^2}{dt^2}D_{N}(\imag  \sigma_{\imag t} t \sqrt{2}) \nonumber\\
    \qquad \qquad \qquad- D_{N}(\imag \sigma_{\imag t} t \sqrt{2}) \frac{d^2}{dt^2} D_{-N}(\sigma_{\imag t} t \sqrt{2})
\end{eqnarray}
so that
\begin{eqnarray} \fl
    \dot{H}_1(t;N) = - 2 \nonumber\\
    \fl \qquad \qquad + (-\imag)^{N} \left[
        D_{-N}(\sigma_{\imag t} t \sqrt{2}) \frac{d^2}{dt^2}D_{N}(\imag \sigma_{\imag t} t \sqrt{2}) - D_{N}(\imag \sigma_{\imag t} t \sqrt{2}) \frac{d^2}{dt^2} D_{-N}(\sigma_{\imag t} t \sqrt{2})
        \right].\nonumber\\
        {}
\end{eqnarray}
(iii) Third, we integrate the above equation to obtain
\begin{eqnarray}
    H_1(t;N) = - 2 t + (-\imag)^{N} \hat{{\mathcal W}}_t \left[
        D_{-N}(\sigma_{\imag t} t \sqrt{2}), D_{N}(\imag \sigma_{\imag t} t \sqrt{2})
        \right].
\end{eqnarray}
Here, the integration constant was set to zero in order to meet the boundary conditions Eq.~(\ref{h1-as}) at infinities. The notation  $\hat{{\mathcal W}}_t$ stands for the Wronskian
\begin{eqnarray}
\label{wd}
    \hat{{\mathcal W}}_t[f,g] = f \frac{\partial g}{\partial t}  -  \frac{\partial f}{\partial t} g.
\end{eqnarray}
\newline\noindent
{\it Average Green function and eigenlevel density.}---Now, the average one-point Green function readily follows from Eq.~(\ref{gh1}):
\begin{eqnarray} \label{gf-054-fermions}
    G(z;N) = 2 z + (-\imag)^N \hat{{\mathcal W}}_z
    \left[
        D_{-N}(-\imag z \sigma_z \sqrt{2}),\, D_{N}(z \sigma_z \sqrt{2})
    \right].
\end{eqnarray}
Here, $\sigma_z={\rm sgn}\,\mathfrak{Im}\,z$ denotes the sign of imaginary part of $z=-\imag t$.
\newline\noindent\newline
The average density of eigenlevels can be restored from Eq.~(\ref{gf-054-fermions}) and the relation
\begin{eqnarray}
\label{doe-gue-f}
    \varrho_N(\epsilon) = -\frac{\sigma_z}{\pi} \, \mathfrak{Im}\, G(\epsilon+ \imag \sigma_z 0;N).
\end{eqnarray}
Indeed, noticing from Eqs.~(\ref{ir-minus}) and (\ref{gf-054-fermions}) that
\begin{eqnarray} \fl
    \mathfrak{Im} \left[ (-\imag)^N D_{-N}(-\imag\epsilon \sigma_z \sqrt{2})\right] = -\frac{(-\imag)^{N-1}}{2^{N/2+1}\Gamma(N)}
    \, e^{\epsilon^2/2} \int_{\mathbb R} d\tau \, \tau^{N-1} e^{-\tau^2/4 + \imag \epsilon \sigma_z \tau},
\end{eqnarray}
we conclude, with the help of Eq.~(\ref{ir-plus}), that
\begin{eqnarray}
    \mathfrak{Im} \left[ (-\imag)^N D_{-N}(-\imag \epsilon\sigma_z \sqrt{2})\right] =
    - \frac{\sqrt{\pi} \sigma_z^{N-1}}{2^{N/2} \Gamma(N)}\, e^{-\epsilon^2/2}
    H_{N-1}(\epsilon).
\end{eqnarray}
Here, $H_{N-1}(\epsilon)$ is the Hermite polynomial appearing by virtue of the relation
\begin{eqnarray}
    D_N(z\sqrt{2}) = e^{-z^2/2} \frac{H_N(z)}{2^{N/2}}.
\end{eqnarray}
Consequently,
\begin{eqnarray} \fl
    \mathfrak{Im}\, \left\{(-\imag)^N \hat{{\mathcal W}}_z
    \left[
        D_{-N}(-\imag z \sigma_z \sqrt{2}),\, D_{N}(z \sigma_z \sqrt{2})
    \right]\right\}\qquad\qquad \nonumber \\
    \qquad\qquad = - \frac{\sqrt{\pi} \sigma_z}{2^N \Gamma(N)}\,
     \hat{{\mathcal W}}_\epsilon \left[
        e^{-\epsilon^2/2} H_{N-1}(\epsilon), e^{-\epsilon^2/2} H_{N}(\epsilon)
     \right].
\end{eqnarray}
Taken together with Eqs.~(\ref{doe-gue-f}) and (\ref{gf-054-fermions}), this equation yields the finite-$N$ average density of eigenlevels in the GUE:
\begin{eqnarray}
\label{doe-gue-fin} \fl
    \varrho_N(\epsilon) = \frac{1}{2^N \Gamma(N) \sqrt{\pi}} \,
    \hat{{\mathcal W}}_\epsilon \left[
        e^{-\epsilon^2/2} H_{N-1}(\epsilon), e^{-\epsilon^2/2} H_{N}(\epsilon)
     \right] \nonumber\\
     \qquad\qquad = \frac{1}{2^N \Gamma(N) \sqrt{\pi}} \,e^{-\epsilon^2}
    \hat{{\mathcal W}}_\epsilon \left[
        H_{N-1}(\epsilon), H_{N}(\epsilon)
     \right].
\end{eqnarray}
While this result, obtained via the {\it fermionic} replica limit, is seen to coincide with the celebrated finite-$N$ formula (Mehta 2004)
\begin{eqnarray}
    \label{m-res}
    \varrho_N(\epsilon) = \frac{1}{2^{N} \Gamma(N) \sqrt{\pi}}
    \,e^{-\epsilon^2} \left[
        H_N^\prime(\epsilon) H_{N-1}(\epsilon) -
        H_N(\epsilon) H_{N-1}^\prime(\epsilon)
    \right]\qquad
\end{eqnarray}
originally derived within the orthogonal polynomial technique, the factorisation phenomenon (as defined in Section \ref{Sec-1-2}) has not been immediately detected throughout the calculation of either $G(z;N)$ or $\varrho_N(\epsilon)$. We shall return to this point in Section \ref{Sec-5-3}.

\subsubsection{Bosonic replicas}\label{Sec-5-2-2}\noindent\newline\newline
Comparing Eq.~(\ref{bos-pf}) with the definition Eq.~(\ref{rpf-gue}),
we conclude that
\begin{eqnarray}
\label{054}
    {\mathcal Z}^{(-)}_{n}(z;N) = \Pi_{n}(z; -N),
\end{eqnarray}
where $\mathfrak{Im\,}z \neq 0$. The shorthand notation $\Pi_{n}(z; -N)$ is used to denote
$\Pi_{n|1}^{\rm G}(z; -N)$, in accordance with the earlier notation in
Eqs.~(\ref{TL-1-cc}) and (\ref{TL-2-cc}). Consequently, the Painlev\'e IV representation of the bosonic
replica partition function reads [see Eqs.~(\ref{phi-def}) and
(\ref{phi-piv})]:
\begin{eqnarray}
    \frac{\partial}{\partial z} \log {\mathcal Z}^{(-)}_{n}(z;N) =
    \varphi(t;n,-N)\Big|_{t=z},
\end{eqnarray}
where $\psi(t;n,N)=\varphi(t;n,-N)$ is the fourth Painlev\'e transcendent
satisfying the equation
\begin{eqnarray}
\label{phi-piv-00-bos}
        (\psi^{\prime\prime})^2 - 4 (\psi - t\, \psi^\prime)^2
        + 4 \psi^\prime (\psi^\prime+2n)(\psi^\prime+2N)=0
\end{eqnarray}
subject to the boundary conditions
\begin{eqnarray}\label{phi-bc-b}
    \psi(t;n,N) \sim -\frac{nN}{t}, \qquad |t| \rightarrow  \infty, \qquad t\in {\mathbb C}\setminus {\mathbb R}.
\end{eqnarray}
Here and above, $n \in {\mathbb Z}_+$.
\newline\newline\noindent
The average Green function $G(z;N)$ we are aimed at is given by the bosonic replica
limit
\begin{eqnarray}
\label{RL-bos}
    G(z;N) = - \lim_{n\rightarrow 0} \frac{1}{n} \frac{\partial}{\partial z} {\mathcal Z}_n^{(-)}(z;N)
     =  - \lim_{n\rightarrow 0} \frac{1}{n} \, \psi(t;n,N)\Big|_{t=z}.
\end{eqnarray}
To implement it, we assume that the Painlev\'e representation of ${\mathcal Z}^{(-)}_{n}(z;N)$ holds for $n\in {\mathbb R}_+$.
\newline\newline\noindent
{\it Replica limit and the Hamiltonian formalism.}---Similarly to our treatment of the fermionic case, we employ the Hamiltonian formulation of the Painlev\'e IV (Noumi 2004, Forrester and Witte 2001) which associates
$\psi(t;n,N)$ with the polynomial Hamiltonian (Okamoto 1980a)
\begin{eqnarray}
\label{phi-H-bos}
    \psi(t;n,N) \equiv H_{\rm b}\{P,Q,t\} = (2P + Q + 2t) PQ + 2 n P + N Q
\end{eqnarray}
of a dynamical system $\{Q,P,H_{\rm b}\}$, where $Q=Q(t;n,N)$ and $P=P(t;n,N)$ are canonical coordinate and momentum. For such a system,
Hamilton's equations of motion read:
\begin{eqnarray}
    \dot{Q} &=& + \frac{\partial H_{\rm b}}{\partial P} = Q (Q+ 4 P + 2t) + 2n, \\
    \dot{P} &=& -  \frac{\partial H_{\rm b}}{\partial Q} = - P (2Q+2P+2t) - N.
\end{eqnarray}
Owing to Eq.~(\ref{RL-bos}), we need to develop a small-$n$ expansion for the Hamiltonian $H_{\rm b}\{Q,P,t\}$:
\begin{eqnarray}
\label{H-exp-bos}
    H_{\rm b}\{P,Q,t\} = nH_1^{({\rm b})}(t;N) + {\mathcal O}(n^2).
\end{eqnarray}
Being consistent with yet another expansion
\begin{eqnarray}
    P(t;n,N) &=& p_0(t;N) + np_1(t;N) + {\mathcal O}(n^2), \\
    Q(t;n,N) &=& nq_1(t;N) + {\mathcal O}(n^2),
\end{eqnarray}
it results in the relation \footnote[1]{We will drop the superscript $({\rm b})$ wherever this does not cause a notational confusion.}
\begin{eqnarray}
    \label{gh1-bos}
    G(z;N) = - H_1^{({\rm b})} (z;N),
\end{eqnarray}
where
\begin{eqnarray} \label{h1-bos}
    H_1(t;N) &=& 2 p_0 q_1 (p_0+t) + 2 p_0  + N q_1, \\
    \label{h1-dot-bos}
    \dot{H}_1(t;N) &=& 2 p_0 q_1.
\end{eqnarray}
Here, $p_0=p_0(t;N)$ and $q_1 = q_1(t;N)$ are solutions to the system of coupled first order equations:
\begin{eqnarray}
\label{system-bos}
\left\{
\begin{array}{cll}
    \dot{p}_0 &=& - 2 p_0^2 - 2 p_0 t - N,\\
    \dot{q}_1 &=& 4 p_0 q_1 + 2 q_1 t + 2.
\end{array}\right.
\end{eqnarray}
Since the initial conditions are known for $H_1(t;N)$, rather than for $p_0(t;N)$ and $q_1(t;N)$ separately, below we determine these two functions up to integration constants.

The function $p_0(t;N)$ satisfies the Riccati differential equation whose solution is
\begin{eqnarray}
\label{p0-ans-bos}
    p_0(t;N) = \frac{1}{2} \left[ \frac{\dot{u}_-(t)}{u_-(t)} - t \right],
\end{eqnarray}
where
\begin{eqnarray}
\label{ut-bos}
    u_-(t) = c_1 \imag^N D_{N-1}(t\sqrt{2}) + c_2 \imag^N D_{-N}(\imag t \sqrt{2})
\end{eqnarray}
is, in turn, a solution to the equation of parabolic cylinder
\begin{eqnarray}
    \ddot{u}_-(t) + (2N-1-t^2)u_-(t)=0.
\end{eqnarray}
Factoring out $\imag^{N}$ in the second term in Eq.~(\ref{ut-bos}) will simplify the formulae to follow.

To determine $q_1(t;N)$, we substitute Eq.~(\ref{p0-ans-bos}) into the second formula of Eq.~(\ref{system-bos}) to derive:
\begin{eqnarray}
\label{q1-int-bos}
    q_1(t;N) = 2 u^2_-(t) \int \frac{dt}{u_-^2(t)}.
\end{eqnarray}
Making use of the integration formula (see Appendix \ref{App-D-int})
\begin{eqnarray}
    \label{osipov-integral-bos}
    \int \frac{dt}{u^2_-(t)} = \frac{1}{\sqrt{2}}\, \frac{\alpha_1 D_{N-1}(t\sqrt{2}) + \alpha_2  D_{-N}(it\sqrt{2})}{u_-(t)},
\end{eqnarray}
where two constants $\alpha_1$ and $\alpha_2$ are subject to the constraint
\begin{eqnarray}
\label{const-a12-bos}
    c_1 \alpha_2 - c_2 \alpha_1 = 1,
\end{eqnarray}
we further reduce Eq.~(\ref{q1-int-bos}) to
\begin{eqnarray}
\label{q1-ans-bos}
    q_1(t;N) = \sqrt{2} u_-(t) \left[
        \alpha_1 D_{N-1}(t\sqrt{2}) + \alpha_2 D_{-N}(\imag t\sqrt{2})
    \right].
\end{eqnarray}
Equations (\ref{h1-dot-bos}), (\ref{p0-ans-bos}), (\ref{q1-ans-bos}) and the identity
\begin{eqnarray} \fl
    \dot{u}_-(t) - t\, u_-(t) = -\sqrt{2} \left[
        c_1 \imag^N D_{N}(t\sqrt{2}) + c_2 \imag^{N+1} N D_{-N-1}(\imag t \sqrt{2})
    \right]
\end{eqnarray}
(obtained from Eq.~(\ref{ut-bos}) with the help of Eqs.~(\ref{d-rec-1}) and (\ref{d-rec-2})) yield $\dot{H}_1(t;N)$ in the form
\begin{eqnarray} \fl
\label{H1-sol-dot-bos}
    \dot{H}_1(t;N) = -2
       \left[
       \alpha_1 D_{N-1}(t\sqrt{2}) + \alpha_2 D_{-N}(\imag t\sqrt{2})
    \right] \nonumber \\
    \times
            \left[
        c_1 \imag^N D_{N}(t\sqrt{2}) + c_2 \imag^{N+1} N D_{-N-1}(\imag t \sqrt{2})
    \right].
\end{eqnarray}
To determine the unknown constants in Eq.~(\ref{H1-sol-dot-bos}), we make use of the asymptotic formulae for the functions of parabolic cylinder (collected in Appendix \ref{App-D-int}) to satisfy the boundary conditions [see Eq.~(\ref{phi-bc-b})]
\begin{eqnarray}\label{h1-as-bos} \fl
   \qquad \qquad H_1(t;N) \sim - \frac{N}{t}, \quad \dot{H}_1(t;N) \sim \frac{N}{t^2}, \qquad |t| \rightarrow  \infty, \qquad t\in {\mathbb C} \setminus {\mathbb R}.
\end{eqnarray}
The two cases $\mathfrak{Im\,} t<0$ and $\mathfrak{Im\,} t>0$ should be treated separately.

\begin{itemize}
  \item  {\it The case} $\mathfrak{Im\,} t<0$. Asymptotic analysis of Eq.~(\ref{H1-sol-dot-bos}) at $t\rightarrow -\imag \infty$ yields $c_1 =0$,
  so that
\begin{eqnarray} \fl
\label{H1-sol-step-10-bos}
    \dot{H}_1(t;N) = 2 \imag^{N+1} N D_{-N-1}(\imag t\sqrt{2}) \left[
    D_{N-1}(t\sqrt{2}) - \alpha_2 c_2 D_{-N}(\imag t\sqrt{2})
    \right].
\end{eqnarray}
To determine the remaining constant $\alpha_2 c_2$, we make use of the boundary
condition Eq.~(\ref{h1-as-bos}) for $t \rightarrow \pm \infty - \imag 0$. Straightforward
calculations bring $\alpha_2 c_2 = 0$. We then conclude that
\begin{eqnarray} \fl \label{h1t-neg-bos}
    \dot{H}_1(t;N) = 2\, \imag^{N+1} N\, D_{-N-1}(\imag t\sqrt{2}) D_{N-1}(t \sqrt{2}),\qquad \mathfrak{Im\,} t<0.
\end{eqnarray}
\vspace{0.2cm}
  \item  {\it The case} $\mathfrak{Im\,} t>0$.
        Asymptotic analysis of Eq.~(\ref{H1-sol-dot-bos}) at $t\rightarrow + \imag \infty$ yields
        \begin{eqnarray} \label{cond-1-bos}
          \frac{c_1}{c_2} = - (-\imag)^{N-1}\frac{\sqrt{2\pi}}{(N-1)!}
        \end{eqnarray}
        so that
\begin{eqnarray}
\label{H1-sol-step-1-bos} \fl
    \dot{H}_1(t;N) = -N! \sqrt{\frac{2}{\pi}}
         D_{-N-1}(-\imag t\sqrt{2}) \nonumber\\ \fl
         \qquad\qquad \times
            \left[
            D_{-N}(\imag t\sqrt{2}) + \alpha_1 c_1 (-\imag)^{N-1} \frac{(N-1)!}{\sqrt{2\pi}} D_{-N}(-\imag t \sqrt{2})
        \right].
\end{eqnarray}
To determine the remaining constant $\alpha_1 c_1$, we make use of the boundary condition Eq.~(\ref{h1-as-bos}) for $t \rightarrow \pm  \infty + \imag 0$. Straightforward
calculations bring
\begin{eqnarray}
    \alpha_1 c_1 = (-\imag)^{N-1} \frac{\sqrt{2\pi}}{(N-1)!}.
\end{eqnarray}
We then conclude that
\begin{eqnarray} \fl \label{h1t-pos-bos}
    \dot{H}_1(t;N) = 2\, \imag^{N+1} N\, D_{-N-1}(-\imag t\sqrt{2}) D_{N-1}(- t \sqrt{2}),\qquad \mathfrak{Im\,} t>0.
\end{eqnarray}
\end{itemize}
\noindent\newline
The calculation of $\dot{H}_1(t;N)$ can be summarised in a single formula
\begin{eqnarray}
\label{h1d-answer-bos}
    \dot{H}_1(t;N) = - 2\, \imag^{N-1} N\, D_{-N-1}(-\imag t \sigma_t\sqrt{2}) D_{N-1}(- t \sigma_t \sqrt{2}),
\end{eqnarray}
where $\sigma_t={\rm sgn}\,\mathfrak{Im}\, t$ denotes the sign of $\mathfrak{Im}\, t$. In terms of canonical variables $p_0(t;N)$ and $q_1(t;N)$, this result translates to
\begin{eqnarray}
\label{p0-bos}
    p_0(t;N) &=& \frac{\imag N \sigma_t}{\sqrt{2}} \frac{D_{-N-1}(-\imag t \sigma_t \sqrt{2})}{D_{-N}(-\imag t \sigma_t\sqrt{2})}, \\
\label{q1-bos}
    q_1(t;N) &=&  \sqrt{2} \sigma_t \imag^N D_{-N}(- \imag t \sigma_t \sqrt{2}) D_{N-1}(-t \sigma_t \sqrt{2}).
\end{eqnarray}
In view of Eq.~(\ref{gh1-bos}), the latter result is equivalent to the statement
\begin{eqnarray} \fl
    \frac{\partial}{\partial z} G(z;N)
    = -\dot{H}_1^{({\rm b})}(t;N)\Big|_{t=z} = 2 \, \imag^{N-1} N D_{-N-1}(-\imag z\sigma_z \sqrt{2}) D_{N-1}(-z\sigma_z \sqrt{2}).
\end{eqnarray}

This expression, obtained within the {\it bosonic} replicas, must be compared with its counterpart derived via the {\it fermionic} replicas [Eqs.~(\ref{gh1}) and (\ref{h1d-answer})]:
\begin{eqnarray} \fl
    \frac{\partial}{\partial z} G(z;N)
    = -\dot{H}_1^{({\rm f})}(t;N)\Big|_{t=i z} = 2 \, (-\imag)^{N-1} N D_{-N-1}(-\imag z\sigma_z \sqrt{2}) D_{N-1}(z\sigma_z \sqrt{2}).
\end{eqnarray}
As the two expressions coincide, we are led to conclude that the bosonic version of the replica limit reproduces correct finite-$N$
results for the average Green function and the average density of eigenlevels as given by Eqs.~(\ref{gf-054-fermions}) and (\ref{m-res}), respectively. Again, as is the case of a fermionic calculation carried out in Section \ref{Sec-FR}, the factorisation property did not show up explicitly in the above bosonic calculation. We defer discussing this point
until Section \ref{Sec-5-3}.

\subsubsection{Supersymmetric replicas}\noindent\newline\newline
The very same integrable theory of characteristic polynomials is at
work for a ``supersymmetric'' variation of replicas invented by
Splittorff and Verbaarschot (2003). These authors suggested that the
fermionic and bosonic replica partition functions (satisfying the
fermionic and bosonic Toda Lattice equations \footnote[1]{Notice that
Splittorff and Verbaarschot (2003) use the term ``Toda Lattice
equation'' for the first equation of the TL hierarchy.},
respectively) can be seen as two different branches of a single,
{\it graded} Toda Lattice equation. Below we show that the above
statement, considered in the context of GUE, is also valid {\it beyond}
the first equation of the Toda Lattice
hierarchy.\newline\newline\noindent {\it First (graded) TL
equation.}---Equations (\ref{ZPi-F}) and (\ref{TL-1-cc}) imply that
the fermionic replica partition function ${\mathcal
Z}^{(+)}_{n}(z;N)$ satisfies the first TL equation in the form:
\begin{eqnarray}
\label{TL-1-fer}
    \frac{\partial^2}{\partial z^2}\,
        \log {\mathcal Z}^{(+)}_{n}(z;N)  = - 2 n\,\left(
            \frac{{\mathcal
Z}^{(+)}_{n-1}(z;N) \,{\mathcal
Z}^{(+)}_{n+1}(z;N) }{{\mathcal
Z}^{(+)\, 2}_{n}(z;N)}-1\right).
\end{eqnarray}
Together with the initial conditions ${\mathcal Z}^{(+)}_{0}(z;N)=1$
and
\begin{eqnarray}  \label{Z-1-fer} \fl
    {\mathcal Z}^{(+)}_{1}(z;N) =(-\imag)^N \Pi_{1|1}^{\rm G}(iz;N) &=& \frac{1}{\sqrt{\pi}}
    \int_{\mathbb R} d\lambda\, e^{-\lambda^2} (z+\imag\lambda)^N \nonumber\\
    &=& 2^{-N/2} e^{z^2/2} D_N(z\sqrt{2}),
\end{eqnarray}
this equation uniquely determines fermionic replica partition
functions of any order ($n \ge 2$). Here, $D_N$ is the function of parabolic cylinder of positive order (see Appendix~\ref{App-D-int}).

The first TL equation for the bosonic replica partition function
${\mathcal Z}^{(-)}_{n}(z;N)$ follows from Eqs.~(\ref{054}) and
(\ref{TL-1-cc}),
\begin{eqnarray}
\label{TL-1-bos}
    \frac{\partial^2}{\partial z^2}\,
        \log {\mathcal Z}^{(-)}_{n}(z;N)  = + 2 n\,\left(
            \frac{{\mathcal
Z}^{(-)}_{n-1}(z;N) \,{\mathcal
Z}^{(-)}_{n+1}(z;N) }{{\mathcal
Z}^{(-)\, 2}_{n}(z;N)}-1\right).
\end{eqnarray}
Together with the initial conditions ${\mathcal Z}^{(-)}_{0}(z;N)=1$
and
\begin{eqnarray}  \label{Z-1-bos} \fl
    {\mathcal Z}^{(-)}_{1}(z;N) = \Pi_{1|1}^{\rm G}(z;-N) &=& \frac{1}{\sqrt{\pi}}
    \int_{\mathbb R} d\lambda\, e^{-\lambda^2} (z- \lambda)^{-N} \nonumber\\
    &=& (-\imag \sigma_z)^N 2^{N/2} e^{-z^2/2} \, D_{-N}(-\imag z \sigma_z \sqrt{2}),
    \end{eqnarray}
where $\sigma_z={\rm sgn}\, \mathfrak{Im}\,z$ denotes the sign of $\mathfrak{Im}\,z$, this equation uniquely determines bosonic replica partition
functions of any order ($n \ge 2$). Here, $D_{-N}$ is the function of parabolic cylinder of negative order (see Appendix~\ref{App-D-int}).

Further, defining the {\it graded} replica partition function as
\begin{eqnarray}
\label{graded-pf-1}
    {\mathcal Z}_{n}(z;N) = \cases{
        {\mathcal Z}^{(-)}_{|n|}(z;N),  & $n \in {\mathbb Z}^-$\\
        1, & $n = 0$ \\
        {\mathcal Z}^{(+)}_{|n|}(z;N),  & $n \in {\mathbb Z}^+$,
    }
\end{eqnarray}
we spot from Eqs.~(\ref{TL-1-fer}) and (\ref{TL-1-bos}) that it satisfies the single (graded) TL equation
\begin{eqnarray}
\label{TL-1-graded}
    \frac{\partial^2}{\partial z^2}\,
        \log {\mathcal Z}_{n}(z;N)  = - 2 n\,\left(
            \frac{{\mathcal
Z}_{n-1}(z;N) \,{\mathcal
Z}_{n+1}(z;N) }{{\mathcal
Z}^{2}_{n}(z;N)}-1\right).
\end{eqnarray}
Here, the index $n$ is an arbitrary integer, be it positive or negative. The first graded TL equation must be supplemented by {\it two} initial conditions given by Eqs.~(\ref{Z-1-fer}) and (\ref{Z-1-bos}).
\newline\newline\noindent {\it Second (graded) TL
equation.}---Equations (\ref{ZPi-F}), (\ref{054}) and (\ref{TL-2-cc}) imply that
both fermionic and bosonic replica partition functions ${\mathcal
Z}^{(\pm)}_{n}(z;N)$ additionally satisfy the second TL equation
\begin{eqnarray}
\label{TL-2-fer-bos} \fl
\left(1- z \frac{\partial}{\partial z} \right) \frac{\partial}{\partial z}\,
        \log {\mathcal Z}^{(\pm)}_{n}(z;N)
         =
        n\,
            \frac{{\mathcal Z}^{(\pm)}_{n+1}(z;N) \,
            {\mathcal Z}^{(\pm)}_{n-1}(z;N)}{{\mathcal Z}^{(\pm)\, 2}_{n}(z;N)}
            \,\frac{\partial}{\partial z}
         \log \left(
            \frac{{\mathcal Z}^{(\pm)}_{n+1}(z;N)}{{\mathcal Z}^{(\pm)}_{n-1}(z;N)}
        \right). \nonumber \\
        {}
\end{eqnarray}
Consequently, the graded replica partition function ${\mathcal Z}_{n}(z;N)$ defined by Eq.~(\ref{graded-pf-1}) is determined by the second {\it graded} TL equation
\begin{eqnarray}
\label{TL-2-graded} \fl
\left(1- z \frac{\partial}{\partial z} \right) \frac{\partial}{\partial z}\,
        \log {\mathcal Z}_{n}(z;N)
         =
        n\,
            \frac{{\mathcal Z}_{n+1}(z;N) \,
            {\mathcal Z}_{n-1}(z;N)}{{\mathcal Z}^{2}_{n}(z;N)}
            \,\frac{\partial}{\partial z}
         \log \left(
            \frac{{\mathcal Z}_{n+1}(z;N)}{{\mathcal Z}_{n-1}(z;N)}
        \right) \nonumber \\
        {}
\end{eqnarray}
supplemented by two initial conditions Eqs.~(\ref{Z-1-fer}) and (\ref{Z-1-bos}).
\newline\newline\noindent {\it Replica limit of graded TL
equations.}---To determine the one-point Green function $G(z;N)$ within the framework of supersymmetric replicas, one has to consider the replica limit
\begin{eqnarray}
    G(z;N) = \lim_{n\rightarrow 0} \frac{1}{n} \frac{\partial}{\partial z} \, {\mathcal Z}_n(z;N).
\end{eqnarray}
The first and second graded TL equations bring
\begin{eqnarray}
\label{g-prime}
   G^\prime(z;N) = 2 - 2\, {\mathcal Z}_{-1}(z;N) \,{\mathcal Z}_{1}(z;N),
\end{eqnarray}
and
\begin{eqnarray} \label{G-deq} \fl
    G(z;N)= z G^\prime (z;N) + {\mathcal Z}_{-1}(z;N) {\mathcal Z}_{1}^\prime(z;N)
    - {\mathcal Z}_{-1}^\prime(z;N) {\mathcal Z}_{1}(z;N),
\end{eqnarray}
respectively. Combining the two equations, we derive
\begin{eqnarray}
  \label{g-factor}
    G(z;N) = 2 z - 2 z\, {\mathcal Z}_{-1} \,{\mathcal Z}_{1} +
    {\hat{\mathcal W}}_z\left[
        {\mathcal Z}_{-1}, \,{\mathcal Z}_{1}
    \right],
\end{eqnarray}
where ${\hat{\mathcal W}}_z$  is the Wronskian defined in Eq.~(\ref{wd}); the prime ${}^\prime$ stands for the derivative $\partial/\partial z$. Interestingly, the second graded TL equation has allowed us to integrate Eq.~(\ref{G-deq}) at once!

The resulting Eq.~(\ref{g-factor}) is remarkable: it shows that the average Green function can solely be expressed in terms of bosonic ${\mathcal Z}_{-1}(z;N)$ and fermionic ${\mathcal Z}_{1}(z;N)$ replica partition functions with only one flavor. This structural phenomenon known as the {\it `factorisation property'} was first observed by Splittorff and Verbaarschot (2003) in the context of the GUE density-density correlation function. Striking at first sight, the factorisation property appears to be very natural after
recognising that fermionic and bosonic replica partition functions are the members of a single graded TL hierarchy.

To make Eq.~(\ref{g-factor}) explicit, we utilise Eqs.~(\ref{Z-1-fer}) and (\ref{Z-1-bos}) to observe the identity
\begin{eqnarray} \fl
        {\hat{\mathcal W}}_z\left[
        {\mathcal Z}_{-1}, \,{\mathcal Z}_{1}
    \right] = 2z\, {\mathcal Z}_{-1} \,{\mathcal Z}_{1} + (-\imag)^N W_z
    \left[
        D_{-N}(-\imag z \sigma_z \sqrt{2}),\, D_{N}(z \sigma_z \sqrt{2})
    \right].
\end{eqnarray}
Consequently,
\begin{eqnarray} \label{gf-054}
    G(z;N) = 2 z + (-\imag)^N {\hat{\mathcal W}}_z
    \left[
        D_{-N}(-\imag z \sigma_z \sqrt{2}),\, D_{N}(z \sigma_z \sqrt{2})
    \right].
\end{eqnarray}
This expression for the average Green function, derived within the framework of {\it supersymmetric replicas}, coincides with the one obtained {\it separately} by means of fermionic and bosonic replicas (see, e.g., Eq.~(\ref{gf-054-fermions})). Hence, the result Eq.~(\ref{m-res}) for the finite-$N$ average density of eigenlevels readily follows.

\subsection{Factorisation property in fermionic and bosonic replicas}\label{Sec-5-3}
The factorisation property naturally appearing in the supersymmetric variation of the replica method suggests that a generic correlation function should contain both {\it compact} (fermionic) and {\it non-compact} (bosonic) contributions. Below, the fermionic-bosonic factorisation property will separately be discussed in the context of fermionic and bosonic replicas where its presence is by far less obvious even though the enlightened reader might have anticipated the factorisation property from Eqs. (\ref{gh1}) and (\ref{h1-dot}) for fermionic replicas and from Eqs. (\ref{gh1-bos}) and (\ref{h1-dot-bos}) for bosonic replicas.\noindent\newline\newline
{\it Fermionic replicas.}---The Hamiltonian formulation of the fourth Painlev\'e transcendent employed in Section \ref{Sec-FR} is the key. It follows from Eqs.~(\ref{gh1}) and (\ref{h1-dot}) that the average Green function $G(z;N)$ is expressed in terms of canonical variables $p_0$ and $q_1$ as
\begin{eqnarray}
    \frac{\partial}{\partial z} \, G(z;N) = - 2p_0(\imag z)\, q_1(\imag z),
\end{eqnarray}
where
\begin{eqnarray}
    p_0(\imag z) = -\frac{\imag}{2}\left[ z +
    \frac{\partial}{\partial z} \log D_N(z \sqrt{2})
    \right]
\end{eqnarray}
and
\begin{eqnarray} \fl
 q_1(\imag z) = -\frac{\imag}{N} (-\imag \sigma_z)^N e^{z^2/2} D_N(z\sqrt{2}) \frac{\partial}{\partial z} \left[
        e^{-z^2/2} D_{-N}(-\imag z\sigma_z \sqrt{2})
    \right].
\end{eqnarray}
To derive the last two equations, we have used Eqs.~(\ref{p0-fer}) and (\ref{q1-fer}) in conjunction with Eqs.~(\ref{d-rec-1}) and (\ref{d-rec-2}).

With the help of Eq.~(\ref{Z-1-fer}), the canonical ``momentum'' $p_0$ can be related to the {\it fermionic} partition function for one flavor,
\begin{eqnarray}
\label{p0-fer-Z}
    p_0(\imag z) = -\frac{\imag}{2} \frac{\partial}{\partial z} \log {\mathcal Z}^{(+)}_{1}(z;N).
\end{eqnarray}
This is a {\it compact contribution} to the average Green function. A {\it non-compact contribution} is encoded in the canonical ``coordinate'' $q_1$
which can be related to the {\it bosonic} partition function via Eq.~(\ref{Z-1-bos}):
\begin{eqnarray}
\label{q1-fer-Z}
    q_1(\imag z) = -\frac{\imag}{N} \, {\mathcal Z}^{(+)}_{1}(z;N) \,
    \frac{\partial}{\partial z} {\mathcal Z}^{(-)}_{1}(z;N),
\end{eqnarray}
so that
\begin{eqnarray}
\label{g-prime-fact-alt}
    \frac{\partial}{\partial z} \, G(z;N) = \frac{1}{N} \frac{\partial}{\partial z} {\mathcal Z}^{(+)}_{1}(z;N)\frac{\partial}{\partial z} {\mathcal Z}^{(-)}_{1}(z;N).
\end{eqnarray}
This is yet another factorised representation for $G^\prime(z;N)$ [compare to Eq.~(\ref{g-prime})].
\noindent\newline\newline
{\it Bosonic replicas.}---To identify both compact and non-compact contributions to the average Green function, we turn to Eqs.~(\ref{gh1-bos}) and (\ref{h1-dot-bos}) to represent
the derivative of the average Green function $G(z;N)$ in terms of canonical variables $p_0$ and $q_1$ as
\begin{eqnarray}
\label{dpG}
    \frac{\partial}{\partial z} \, G(z;N) = - 2p_0(z)\, q_1(z),
\end{eqnarray}
where
\begin{eqnarray}
    p_0(z) = -\frac{1}{2}\left[ z -
    \frac{\partial}{\partial z} \log D_{-N}(- \imag z \sigma_z \sqrt{2})
    \right]
\end{eqnarray}
and
\begin{eqnarray} \fl
 q_1(z) = -\frac{1}{N} (-\imag \sigma_z)^N e^{-z^2/2} D_{-N}(-\imag z \sigma_z \sqrt{2}) \frac{\partial}{\partial z} \left[
        e^{z^2/2} D_{N}(z \sqrt{2})
    \right].
\end{eqnarray}
To derive the last two equations, we have used Eqs.~(\ref{p0-bos}) and (\ref{q1-bos}) in conjunction with Eqs.~(\ref{d-rec-1}) and (\ref{d-rec-2}).

With the help of Eq.~(\ref{Z-1-bos}), the canonical ``momentum'' $p_0$ can be related to the {\it bosonic} partition function for one flavor,
\begin{eqnarray}
\label{p0-bos-Z}
    p_0(z) = \frac{1}{2} \frac{\partial}{\partial z} \log {\mathcal Z}^{(-)}_{1}(z;N).
\end{eqnarray}
This is a {\it non-compact contribution} to the average Green function. A {\it compact contribution} comes from the canonical ``coordinate'' $q_1$
which can be related to the {\it fermionic} partition function via Eq.~(\ref{Z-1-fer}):
\begin{eqnarray}
\label{q1-bos-Z}
    q_1(z) = -\frac{1}{N} \, {\mathcal Z}^{(-)}_{1}(z;N) \,
    \frac{\partial}{\partial z} {\mathcal Z}^{(+)}_{1}(z;N),
\end{eqnarray}
so that
\begin{eqnarray}
    \frac{\partial}{\partial z} \, G(z;N) = \frac{1}{N} \frac{\partial}{\partial z} {\mathcal Z}^{(+)}_{1}(z;N)\frac{\partial}{\partial z} {\mathcal Z}^{(-)}_{1}(z;N),
\end{eqnarray}
agreeing with the earlier result Eq.~(\ref{g-prime-fact-alt}).
\noindent\newline\newline
{\it Brief summary.}---The detailed analysis of fermionic and bosonic replica limits performed in the context of the GUE averaged one-point Green function $G(z;N)$
has convincingly demonstrated that the Hamiltonian formulation of the fourth Painlev\'e transcendent provides a natural and, perhaps, most adequate language to identify the factorisation phenomenon. In particular, we have managed to show that the derivative $G^\prime (z;N)$ of the one-point Green function factorises into a product of canonical ``momentum''
\begin{eqnarray}
\label{p0-limit}
p_0(t;N) = \lim_{n\rightarrow 0} P(t;n,N)
\end{eqnarray}
and canonical ``coordinate''
\begin{eqnarray}
\label{q1-limit}
q_1(t;N) = \lim_{n\rightarrow 0} \frac{1}{n} Q(t;n,N).
\end{eqnarray}
As suggested by Eqs.~(\ref{p0-fer-Z}) and (\ref{p0-bos-Z}), the momentum contribution $p_0$ to the average Green function is a regular one; it corresponds to a compact contribution in fermionic replicas and to a non-compact contribution in bosonic replicas:
\begin{eqnarray}
    p_0 \sim  \frac{\partial}{\partial z} \log
    \cases{
        {\mathcal Z}^{(+)}_{1}(z;N)  & (fermionic)\\
        {\mathcal Z}^{(-)}_{1}(z;N)  & (bosonic)
    }
\end{eqnarray}
On the contrary, the coordinate contribution $q_1$ is of a complementary nature: defined by a replica-like limit [Eq.~(\ref{q1-limit})] it brings in a noncompact contribution in fermionic replicas [Eq.~(\ref{q1-fer-Z})] and a compact contribution in bosonic replicas [Eq.~(\ref{q1-bos-Z})]:
\begin{eqnarray}
    q_1 \sim \exp\left(\int p_0\, dz\right) \times \frac{\partial}{\partial z}
    \cases{
        {\mathcal Z}^{(-)}_{1}(z;N)  & (fermionic)\\
        {\mathcal Z}^{(+)}_{1}(z;N) & (bosonic)
    }
\end{eqnarray}

We close this section by noting that the very same calculational framework should be equally effective in performing the replica limit for other random-matrix ensembles and/or spectral observables.

\section{Conclusions}
\label{Sec-6}
In this paper, we have used the ideas of integrability to formulate a theory of the correlation function
\begin{eqnarray}
    \Pi_{n|p} ({\boldsymbol \varsigma}; {\boldsymbol \kappa}) =
    \int d\mu_n({\boldsymbol {\cal H}})\,
        \prod_{\alpha=1}^{p} {\rm det}_n^{\kappa_\alpha}(\varsigma_\alpha-{\boldsymbol{\cal H}}) \nonumber
\end{eqnarray}
of characteristic polynomials for invariant non-Gaussian ensembles of Hermitean random matrices characterised by the probability measure $d\mu_n({\boldsymbol {\cal H}})$ which may well depend on the matrix dimensionality $n$. Contrary to other approaches based on various duality relations, our theory does not assume ``integerness'' of replica parameters ${\boldsymbol \kappa} = (\kappa_1,\cdots,\kappa_p)$ which are allowed to span ${\boldsymbol \kappa}\in{\mathbb R}^p$. One of the consequences of lifting the restriction ${\boldsymbol \kappa}\in {\mathbb Z}^p$ is that we were unable to represent the CFCP {\it explicitly} in a closed determinant form; instead, we have shown that the correlation function $\Pi_{n|p} ({\boldsymbol \varsigma}; {\boldsymbol \kappa})$ satisfies an infinite set of {\it nonlinear differential hierarchically structured} relations. While such a description is, to a large extent, {\it implicit}, it does provide a useful nonperturbative characterisation of $\Pi_{n|p} ({\boldsymbol \varsigma}; {\boldsymbol \kappa})$ which turns out to be much beneficial for an in-depth analysis of the mathematical foundations of zero-dimensional replica field theories.

With certainty, the replicas is not the only application of a nonperturbative approach to CFCP developed in this paper. With minor modifications, its potential readily extends to various problems of charge transfer through quantum chaotic structures (Osipov and Kanzieper 2008, Osipov and Kanzieper 2009), stochastic theory of density matrices (Osipov, Sommers and \.Zyczkowski 2010), random matrix theory approach to QCD physics (Verbaarschot 2010), to name a few. An extension of the above formalism to the CFCP of unitary matrices may bring a useful calculational tool for generating conjectures on behaviour of the Riemann zeta function at the critical line (Keating and Snaith 2000a, 2000b).

Finally, we wish to mention that an integrable theory of CFCP for $\beta=1$ and $\beta=4$ Dyson symmetry classes is very much called for. Building on the insightful work by Adler and van Moerbeke (2001), a solution of this challenging problem seems to be within the reach.

\section*{Acknowledgements}
The authors thank Nicholas Witte for providing references to the early works by Uvarov (1959, 1969). This work was supported by Deutsche Forschungsgemeinschaft SFB/Tr
12, and by the Israel Science Foundation through the grants No
286/04 and No 414/08.
\smallskip\smallskip\smallskip

\newpage

\renewcommand{\appendixpagename}{\large{Appendices}}
\addappheadtotoc
\appendixpage
\setcounter{section}{0}
\renewcommand{\thesection}{\Alph{section}}
\renewcommand{\theequation}{\thesection.\arabic{equation}}
\section{Bilinear Identity in Hirota Form: An Alternative Derivation}\label{App-bi}
In Section \ref{bi-hirota}, the bilinear identity in Hirota form [Eq.~(\ref{bi-hf}) or, equivalently, Eq.~(\ref{bid-2})]
was derived from the one in the integral form [Eq.~(\ref{bi-id-rep})]. As will be shown below, the latter is not actually necessary.

An alternative proof of the bilinear identity in Hirota form [Eq.~(\ref{bid-2})] starts with the very same Eq.~(\ref{fancy}),
\begin{eqnarray} \fl \label{BI-fancy}
    \int_{\cal D} d\lambda \, \Gamma_n(\lambda)\, e^{v({\bm t};\lambda)}
    e^{(a-1)v({\bm t}-{\bm t}^\prime;\lambda)} P_\ell^{(n)}({\bm t};\lambda)
    P_m^{(n)}({\bm t}^\prime;\lambda) \nonumber \\=
    \int_{\cal D} d\lambda \, \Gamma_n(\lambda)\, e^{v({\bm t}^\prime;\lambda)}
    e^{a\,v({\bm t}-{\bm t}^\prime;\lambda)} P_\ell^{(n)}({\bm t};\lambda)
    P_m^{(n)}({\bm t}^\prime;\lambda).
\end{eqnarray}
However, instead of employing Cauchy representations [see Eqs. (\ref{lhs-cauchy}) and (\ref{rhs-cauchy}) in the original proof], we assume existence of Taylor expansions
\begin{eqnarray}
\label{te-1}
    e^{(a-1)v({\bm t}-{\bm t}^\prime;\lambda)}P_m^{(n)}({\bm t}^\prime;\lambda)
    = \sum_{\sigma=0}^\infty A_\sigma\lambda^\sigma
\end{eqnarray}
and
\begin{eqnarray}
\label{te-2}
    e^{  a\,v({\bm t}-{\bm t}^\prime;\lambda)}P_\ell^{(n)}({\bm t};\lambda)
    = \sum_{\sigma=0}^\infty B_\sigma \lambda^\sigma
\end{eqnarray}
to rewrite Eq.~(\ref{BI-fancy}) in the form
\begin{eqnarray} \label{bi-01}\fl
    \sum_{\sigma=0}^\infty A_\sigma \left<
        \lambda^\sigma \Big| P_\ell^{(n)}({\bm t};\lambda)\right>_{\Gamma_n \,e^{v({\bm t})}}
        =
        \sum_{\sigma=0}^\infty B_\sigma \left<
        \lambda^\sigma \Big| P_m^{(n)}({\bm t^\prime};\lambda)\right>_{\Gamma_n \,e^{v({\bm t^\prime})}}.
\end{eqnarray}
Here, we have used the scalar product notation defined in Eq.~(\ref{sc-prod}). The bilinear identity is obtained from Eq.~(\ref{bi-01}) after expressing all its ingredients in terms of the $\tau$ function Eq.~(\ref{tau-f-c1}).
\newline\newline\noindent
{\it (i) The coefficients $A_\sigma$ and $B_\sigma$}. To determine $A_\sigma$, we employ two identities,
\begin{eqnarray}
\label{id-apb-01}
    e^{(a-1)\,v({\bm t}-{\bm t}^\prime;\lambda)}=\sum_{k=0}^\infty
s_k\left((a-1)({\bm t}-{\bm t}^\prime)\right)\lambda^k
\end{eqnarray}
and
\begin{eqnarray} \fl
\label{id-apb-02}
    P_m^{(n)}({\bm t^\prime};\lambda) = \frac{1}{\tau_m^{(m-n)}({\bm t^\prime})}\left(
    \sum_{k=0}^m\lambda^k s_{m-k}\left(-[\bm {\partial_{t^\prime}}]\right) \right)\tau_m^{(m-n)}({\bm t^\prime}),
\end{eqnarray}
where $[\bm {\partial_{t^\prime}}]$ denotes
\begin{eqnarray}
[\bm
{\partial_{t^\prime}}]=\left(\frac{\partial}{\partial {t_1^\prime}},\frac{1}{2}\frac{\partial}{\partial {t_2^\prime}},\cdots,\frac{1}{k}\frac{\partial}{\partial {t_k^\prime}},\cdots\right). \nonumber
\end{eqnarray}
Equation (\ref{id-apb-01}) is a consequence of Eq.~(\ref{vt-def}) and the definition of Schur polynomials (see Table \ref{schur-table}). Equation (\ref{id-apb-02}) follows from the Heine formula
\begin{eqnarray} \label{bheine}\fl
  P_m^{(n)}({\bm t^\prime};\lambda) = \frac{1}{m!\,\tau_m^{(m-n)}({\bm t^\prime})}
    \int_{{\mathcal D}^{m}} \prod_{j=1}^{m}
    \left(
        d\lambda_j\,  (\lambda-\lambda_j)\, \Gamma_{n}(\lambda_j)\,
        e^{v({\boldsymbol t^\prime};\lambda_j)}\right) \cdot
     \Delta_{m}^2({\boldsymbol \lambda})
\end{eqnarray}
and the identity
\begin{eqnarray} \label{BI-prod}
    \prod_{j=1}^m(\lambda-\lambda_j)=\sum_{k=0}^m\lambda^k s_{m-k}\left(-{\bm p}_m(\bm\lambda)\right)
\end{eqnarray}
where ${\bm p}_m(\bm\lambda)$ is an infinite dimensional vector
\begin{eqnarray} \label{b9}
    {\bm p}_m(\bm\lambda)=\left(
    {\rm tr}_m{\bm \lambda}, \frac{1}{2} {\rm tr}_m{\bm \lambda}^2,\cdots,
    \frac{1}{k} {\rm tr}_m{\bm \lambda}^k,\cdots
    \right).
\end{eqnarray}
Indeed, substituting Eq.~(\ref{BI-prod}) into Eq.~(\ref{bheine}), we derive:
\begin{eqnarray} \fl
\label{BI-der}
    P_m^{(n)}({\bm t^\prime};\lambda) = \frac{1}{m!\,\tau_m^{(m-n)}({\bm t^\prime})}
    \int_{{\mathcal D}^{m}} \prod_{j=1}^{m}
    \left(
        d\lambda_j\,  (\lambda-\lambda_j)\, \Gamma_{n}(\lambda_j)\,
        e^{v({\bm t^\prime};\lambda_j)}\right) \cdot
     \Delta_{m}^2({\bm \lambda})\nonumber\\ \fl
     \qquad =\frac{\lambda^m}{m!\,\tau_m^{(m-n)}({\bm t^\prime})}
     \sum_{k=0}^m\frac{1}{\lambda^k}   \int_{{\mathcal D}^{m}}
     \prod_{j=1}^{m}
    \left(
        d\lambda_j\,  \Gamma_{n}(\lambda_j)\,
        e^{v({\boldsymbol t^\prime};\lambda_j)}\right) \cdot s_{k}\left(-{\bm p}_m(\bm\lambda)\right)\,
     \Delta_{m}^2({\boldsymbol \lambda})\nonumber\\
     =\frac{\lambda^m}{\tau_m^{(m-n)}({\bm t^\prime})}\sum_{k=0}^m \frac{1}{\lambda^k}\, s_{k}\left(-[\bm {\partial_{t^\prime}}]\right)\tau_m^{(m-n)}({\bm t^\prime}).
\end{eqnarray}
In the last step, we have used the obvious formula
\begin{eqnarray}
\label{id-717}
    s_k\left(-[\bm {\partial_{t^\prime}}]\right)
    \prod_{j=1}^{m}
    e^{v({\boldsymbol t^\prime};\lambda_j)}
    =
    s_k\left(-{\bm p}_m(\bm\lambda)\right)\,  \prod_{j=1}^{m}
    e^{v({\boldsymbol t^\prime};\lambda_j)}.
\end{eqnarray}
Having established Eq.~(\ref{id-apb-02}), we substitute it and Eq.~(\ref{id-apb-01}) into
Eq.~(\ref{te-1}) to obtain
\begin{eqnarray} \label{Aj}\fl
    A_\sigma = \frac{1}{\tau_m^{(m-n)}({\bm t^\prime})}
    \sum_{k=\max(0,\sigma-m)}^{\sigma} s_{k}\left((a-1)({\bm t}-{\bm
    t}^\prime)\right)\, s_{m-\sigma+k}\left(-[\bm {\partial}_{\bm
    t^\prime}]\right)\tau_m^{(m-n)}({\bm t}^\prime).
\end{eqnarray}
Similar in spirit calculation yields
\begin{eqnarray} \label{Bj}\fl
    B_\sigma = \frac{1}{\tau_\ell^{(\ell-n)}({\bm t})}
    \sum_{k=\max(0,\sigma-\ell)}^{\sigma} s_{k}\left(a({\bm t}-{\bm
    t}^\prime)\right)\, s_{\ell-\sigma+k}\left(-[\bm {\partial}_{\bm
    t}]\right)\tau_\ell^{(\ell-n)}({\bm t}).
\end{eqnarray}
\newline\newline\noindent
{\it (ii) Scalar products}. Now we are ready to express the scalar products in Eq.~(\ref{bi-01}) in terms
of the $\tau$ function Eq.~(\ref{tau-f-c1}). Having in mind the Heine formula Eq.~(\ref{bheine}), we rewrite the scalar product in the l.h.s. of Eq.~(\ref{bi-01}) as
\begin{eqnarray} \fl
\left<
        \lambda^\sigma \Big| P_\ell^{(n)}({\bm t};\lambda)\right>_{\Gamma_n \,e^{v({\bm t})}}
         &=& \frac{1}{\ell!\,\tau_\ell^{(\ell-n)}({\bm t})}
    \int_{{\mathcal D}^{\ell+1}} \prod_{j=1}^{\ell+1}
    \left(
        d\lambda_j\,  \Gamma_{n}(\lambda_j)\,
        e^{v({\bm t};\lambda_j)}\right) \cdot
     \Delta_{\ell}^2({\boldsymbol \lambda}) \nonumber\\
     &\times& \lambda_{\ell+1}^\sigma \prod_{j=1}^\ell  (\lambda_{\ell+1}-\lambda_j).
\end{eqnarray}
Due to the identity
\begin{eqnarray}
    \Delta_\ell({\bm \lambda}) \prod_{j=1}^\ell  (\lambda_{\ell+1}-\lambda_j) =
    \Delta_{\ell+1}({\bm \lambda}),
\end{eqnarray}
the scalar product is further reduced to
\begin{eqnarray} \fl
\frac{1}{\ell!\,\tau_\ell^{(\ell-n)}({\bm t})}
    \int_{{\mathcal D}^{\ell+1}} \prod_{j=1}^{\ell+1}
    \left(
        d\lambda_j\,  \Gamma_{n}(\lambda_j)\,
        e^{v({\bm t};\lambda_j)}\right) \cdot
     \Delta_{\ell+1}^2({\boldsymbol \lambda}) \, \frac{\lambda_{\ell+1}^\sigma}{\prod_{j=1}^\ell  (\lambda_{\ell+1}-\lambda_j)}.
\end{eqnarray}
Symmetrising the integrand,
\begin{eqnarray}
     \frac{\lambda_{\ell+1}^\sigma}{\prod_{j=1}^\ell  (\lambda_{\ell+1}-\lambda_j)}
     \mapsto \frac{1}{\ell+1} \sum_{\alpha=1}^{\ell+1}
     \frac{\lambda_{\alpha}^\sigma}{\prod_{j=1,\,j\neq \alpha}^{\ell+1}
     (\lambda_{\alpha}-\lambda_j)}
\end{eqnarray}
and employing the remarkable relation (taken at $n=\ell+1$) \footnote[3]{Equation (\ref{eq-remark}) is essentially Eq.~(\ref{id-2}) written in terms of the Schur polynomials.}
\begin{eqnarray}
\label{eq-remark}
    \sum_{\alpha=1}^n \left(
    \lambda_\alpha^{n-1+\sigma} \prod_{j=1,\,j\neq \alpha}^{n}
     \frac{1}{\lambda_{\alpha}-\lambda_j}
     \right) = \cases{0, & $\sigma <0$;\\
     s_\sigma\left( {\bm p}_n({\bm \lambda}) \right),& $\sigma \ge 0$,}
\end{eqnarray}
we deduce:
\begin{eqnarray} \fl
\left<
        \lambda^\sigma \Big| P_\ell^{(n)}({\bm t};\lambda)\right>_{\Gamma_n \,e^{v({\bm t})}}
         &=& \frac{1}{(\ell+1)!\,\tau_\ell^{(\ell-n)}({\bm t})}
    \int_{{\mathcal D}^{\ell+1}} \prod_{j=1}^{\ell+1}
    \left(
        d\lambda_j\,  \Gamma_{n}(\lambda_j)\,
        e^{v({\bm t};\lambda_j)}\right) \cdot
     \Delta_{\ell+1}^2({\boldsymbol \lambda}) \nonumber\\
     &\times&
     \cases{0, & $\sigma <\ell$;\\
     s_{\sigma-\ell}\left( {\bm p}_{\ell+1}({\bm \lambda}) \right),& $\sigma \ge \ell$.}
\end{eqnarray}
For $\sigma \ge \ell$ (which is the only nontrivial case), this scalar product reduces to
\begin{eqnarray} \label{pr-1}\fl
\left<
        \lambda^\sigma \Big| P_\ell^{(n)}({\bm t};\lambda)\right>_{\Gamma_n \,e^{v({\bm t})}}
         &=& \frac{1}{\tau_\ell^{(\ell-n)}({\bm t})}\,
         s_{\sigma-\ell}\left([\bm{\partial_t}] \right)\,
         \tau_{\ell+1}^{(\ell+1-n)}({\bm t}),\;\;\; \sigma \ge \ell.
\end{eqnarray}
By the same token,
\begin{eqnarray} \label{pr-2}\fl
\left<
        \lambda^\sigma \Big| P_m^{(n)}({\bm t^\prime};\lambda)\right>_{\Gamma_n \,e^{v({\bm t^\prime})}}
         &=& \frac{1}{\tau_m^{(m-n)}({\bm t^\prime})}\,
         s_{\sigma-m}\left([\bm{\partial_{t^\prime}}] \right)\,
         \tau_{m+1}^{(m+1-n)}({\bm t^\prime}),\;\;\; \sigma \ge m.
\end{eqnarray}
\newline\newline\noindent
{\it (iii) The bilinear identity}. Now we are in position to derive the bilinear identity. To this end we substitute Eqs.~(\ref{Aj}), (\ref{Bj}), (\ref{pr-1}) and (\ref{pr-2}) into Eq.~(\ref{bi-01}). Up to the prefactor
\begin{eqnarray}
    \frac{1}{\tau_m^{(m-n)}({\bm t^\prime}) \tau_\ell^{(\ell-n)}({\bm t})}, \nonumber
\end{eqnarray}
its l.h.s. reads:
\begin{eqnarray} \label{bi-01-lhs}\fl
    \sum_{\sigma=\ell}^\infty A_\sigma \left<
        \lambda^\sigma \Big| P_\ell^{(n)}({\bm t};\lambda)\right>_{\Gamma_n \,e^{v({\bm t})}}
        \mapsto \nonumber\\ \fl
        \sum_{\sigma=\ell}^\infty
    \sum_{k=\max(0,\sigma-m)}^{\sigma} s_{k}\left((a-1)({\bm t}-{\bm
    t}^\prime)\right)\, s_{m-\sigma+k}\left(-[\bm {\partial}_{\bm
    t^\prime}]\right)\,
         s_{\sigma-\ell}\left([\bm{\partial_t}] \right)\,
         \tau_{\ell+1}^{(\ell+1-n)}({\bm t}) \tau_m^{(m-n)}({\bm t}^\prime). \nonumber
\end{eqnarray}
Owing to the identity \footnote[4]{
Equation (\ref{id-123}) follows from Eq.~(\ref{id-717}) and the expansion
\begin{eqnarray}
\prod_{j=1}^m (1-x \lambda_j) =\exp\left[
    \sum_{\alpha=1}^\infty \frac{x^\alpha}{\alpha} \left(
        \sum_{j=1}^m \lambda_j^\alpha
    \right)
\right] = \sum_{k=0}^\infty x^k s_k \left(
    -\bm{p}_m({\bm \lambda})
\right). \nonumber
\end{eqnarray}
Indeed, since the l.h.s. is the polynomial in $x$ of the degree $m$, the Schur polynomials
$s_k \left(
    -\bm{p}_m({\bm \lambda})
\right)$ in the r.h.s. must nullify for $k>m$.
}
\begin{eqnarray} \label{id-123}
  s_k\left(-[\bm {\partial_{t^\prime}}]\right) \tau_m^{(m-n)}({\bm t^\prime}) = 0\;\;\;{\rm for}\;\;\; k>m,
\end{eqnarray}
the latter reduces to
\begin{eqnarray} \fl
    \sum_{\sigma=\ell}^\infty
    \sum_{k=\max(0,\sigma-m)}^{\infty} s_{k}\left((a-1)({\bm t}-{\bm
    t}^\prime)\right)\, s_{m-\sigma+k}\left(-[\bm {\partial}_{\bm
    t^\prime}]\right)\,
         s_{\sigma-\ell}\left([\bm{\partial_t}] \right)\,
         \tau_{\ell+1}^{(\ell+1-n)}({\bm t}) \tau_m^{(m-n)}({\bm t}^\prime). \nonumber
\end{eqnarray}
Interchange of the two series brings
\begin{eqnarray}
\fl
     \sum_{k=\max(0,\ell-m)}^{\infty}
    s_{k}\left((a-1)({\bm t}-{\bm
    t}^\prime)\right)
      \sum_{\sigma=0}^{k+m-\ell}
      s_{k+m-\ell-\sigma}\left(-[\bm {\partial}_{\bm
    t^\prime}]\right)\,
         s_{\sigma}\left([\bm{\partial_t}] \right)\,
         \tau_{\ell+1}^{(\ell+1-n)}({\bm t}) \tau_m^{(m-n)}({\bm t}^\prime). \nonumber
\end{eqnarray}
Making use of yet another identity \footnote[2]{It readily follows from the equality
\begin{eqnarray*}
   e^{\sum_{j=0}^\infty (t_j-t^\prime_j)x^j}=e^{\sum_{j=0}^\infty t_j x^j}e^{-\sum_{j=0}^\infty t^\prime_j x^j}
\end{eqnarray*}
Taylor-expanded around $x=0$ and the definition Eq.~(\ref{SCHUR}) of the Schur polynomials.
}
\begin{eqnarray}
    s_k ({\bm t}-{\bm t^\prime}) = \sum_{\sigma=0}^k \, s_{k-\sigma}(-{\bm t^\prime})\, s_\sigma({\bm t}),
\end{eqnarray}
we derive:
\begin{eqnarray} \label{as-01}
\fl
 \sum_{\sigma=\ell}^\infty A_\sigma \left<
        \lambda^\sigma \Big| P_\ell^{(n)}({\bm t};\lambda)\right>_{\Gamma_n \,e^{v({\bm t})}} \mapsto\nonumber
        \\
        \fl
        \sum_{k=\max(0,\ell-m)}^{\infty}
    s_{k}\left((a-1)({\bm t}-{\bm
    t}^\prime)\right)
      s_{k+m-\ell}\left(
    [\bm {\partial}_{\bm t}] -[\bm {\partial}_{\bm t^\prime}]\right)\,
         \tau_{\ell+1}^{(\ell+1-n)}({\bm t}) \tau_m^{(m-n)}({\bm t}^\prime).
\end{eqnarray}
Setting the vectors ${\bm t}$ and ${\bm
t}^\prime$ to
\begin{eqnarray} \label{ttprime-2}
    ({\bm t},{\bm t}^\prime) \mapsto ({\bm t}+{\bm x},{\bm t}-{\bm x}), \nonumber
\end{eqnarray}
(see Eq.~(\ref{ttprime})), and applying the Property 3 of Appendix \ref{App-hi} [Eq.~(\ref{prop-3})], we
rewrite Eq.~(\ref{as-01}) in terms of Hirota operators:
\begin{eqnarray}
\fl e^{
    \bm{(x}\cdot\bm{D)}}
     \sum_{k=\max(0,\ell-m)}^{\infty}
    s_{k}\left(2(a-1)({\bm x})\right)
      s_{k+m-\ell}\left(
    [\bm D]\right)\,
         \tau_{\ell+1}^{(\ell+1-n)}({\bm t}) \circ \tau_m^{(m-n)}({\bm t}).
\end{eqnarray}
Finally, replacing $n$ with $n=m-s$, we reproduce the l.h.s. of Eq.~(\ref{bid-2}). The r.h.s. of
Eq.~(\ref{bid-2}) can be reproduced along the same lines starting with the r.h.s. of Eq.~(\ref{bi-01}). This ends the alternative proof of the bilinear identity in Hirota form.

\section{Hirota Operators}
\label{App-hi}
{\bf Definition}. Let $f({\bm t})=f(t_1, t_2,\cdots)$ and $g({\bm t})=g(t_1, t_2,\cdots)$ be differentiable multivariate functions. The Hirota derivative is a bilinear differential operator, $D_k f({\bm t})\circ g({\bm t})$, defined by
\begin{eqnarray}\label{hd-def} \fl
D_k\; f({\bm t})\!\circ g({\bm t}) &\equiv& \frac{\partial}{\partial
x_k}\,f({\bm t}+{\bm x})\, g({\bm t}-{\bm x})\Big|_{{\bm x}={\bm 0}}
        =\left(\frac{\partial}{\partial t_k}-\frac{\partial}{\partial t_k^\prime}\right)\,f({\bm t})\,g({\bm t^\prime})\Big|_{{\bm t^\prime}={\bm t}} \nonumber
        \\&=&g({\bm t})\frac{\partial f({\bm t})}{\partial t_k}-f({\bm t})\frac{\partial g({\bm t})}{\partial t_k}.
\end{eqnarray}
\newline\noindent
{\bf Property 1}. Let ${\cal P}(\bm{D})$ be a multivariate function
defined on an infinite set of differential operators
$\bm{D} = (D_1, D_2,\cdots)$ such that
\begin{eqnarray} \label{b2}
    {\cal P}({\bm D})\, f({\bm t}) \circ g({\bm t}) =
    {\cal P}(\bm{\partial_\xi})\, f({\bm t + \bm \xi})  g({\bm t}-\bm \xi)\Big|_{\bm \xi = \bm 0}
\end{eqnarray}
with $\bm{\partial_\xi}=(\partial/\partial \xi_1,\partial/\partial \xi_2,\cdots)$. Then it holds:
\begin{eqnarray} \label{prop-1}
    {\cal P}({\bm D})\, f({\bm t}) \circ g({\bm t}) = {\cal P}
    (-{\bm D})\, g({\bm t}) \circ f({\bm t}).
\end{eqnarray}
\newline\noindent
{\bf Property 2a}. Setting $g({\bm t})=f({\bm t})$ in Eq. (\ref{prop-1}), one observes:
\begin{eqnarray} \label{prop-2}
    \prod_{k} D_k^{\alpha_k}\, f({\bm t}) \circ f({\bm t}) = 0\;\;\; \mbox{for}\;\;\;
    \sum_k \alpha_k = \mbox{odd}.
\end{eqnarray}
\newline\noindent
{\bf Property 2b}. It follows from Eq. (\ref{hd-def}) that
\begin{eqnarray}
    D_k D_\ell f(\bm t)\!\circ f(\bm t) = 2f^2({\bm t})\,
    \frac{\partial^2}{\partial t_k\partial t_\ell}\, \log f(\bm t)
\end{eqnarray}
and
\begin{eqnarray} \fl
    D_k D_\ell D_m D_n f(\bm t)\!\circ f(\bm t) = 2f^2({\bm t})\,
    \frac{\partial^4}{\partial t_k\partial t_\ell \partial t_m \partial t_n}\, \log f(\bm t) \nonumber \\
    \fl \qquad\qquad\qquad + 4f^2({\bm t})
    \Bigg[
            \left(\frac{\partial^2}{\partial t_k\partial t_\ell}\, \log f(\bm t)\right)
            \left(\frac{\partial^2}{\partial t_m\partial t_n}\, \log f(\bm t)\right) \nonumber\\
            \fl   \qquad\qquad\qquad\qquad
            +
            \left(\frac{\partial^2}{\partial t_k\partial t_m}\, \log f(\bm t)\right)
            \left(\frac{\partial^2}{\partial t_\ell\partial t_n}\, \log f(\bm t)\right) \nonumber\\
            \fl   \qquad\qquad\qquad\qquad
            +
            \left(\frac{\partial^2}{\partial t_k\partial t_n}\, \log f(\bm t)\right)
            \left(\frac{\partial^2}{\partial t_\ell\partial t_m}\, \log f(\bm t)\right)
    \Bigg].
\end{eqnarray}
\newline\noindent
{\bf Property 3}. Let $(\bm x \cdot \bm{D})$ denote the scalar product $(\bm x \cdot \bm{D})=
\sum_{k=1}^\infty x_k D_k$. The exponential identity holds:
\begin{equation}
\label{prop-3}
e^{(\bm x \cdot \bm{D})}\; f(\bm t)\!\circ g(\bm t)=f(\bm t+\bm x)\,g(\bm
t-\bm x).
\end{equation}
\noindent\newline For a more exhaustive list of the properties of Hirota differential operators, the reader is referred to Hirota's book (Hirota 2004).

\section{Jacobi Unitary Ensemble (JUE)}\label{App-JUE}
The correlation function of characteristic polynomials in JUE
is defined by the formula
\begin{eqnarray}\label{rpf-jue}\fl
    \Pi_{n|p}^{\rm J}({\bm {\varsigma}};{\bm \kappa}) =\frac{1}{{\cal N}_n^{\rm J}} \int_{(-1,1)^n} \prod_{j=1}^{n}\left( d\lambda_j \, (1-\lambda_j)^\mu (1+\lambda_j)^\nu \prod_{\alpha=1}^p (\varsigma_\alpha - \lambda_j)^{\kappa_\alpha} \right)
    \cdot
    \Delta_{n}^2({\boldsymbol \lambda})
\end{eqnarray}
where
\begin{eqnarray}\label{N-jue}
    {\cal N}_n^{\rm J} = 2^{n^2+n(\mu+\nu)}\prod_{j=1}^n \frac{\Gamma(j+1)\Gamma(j+\mu)\Gamma(j+\nu)}{\Gamma(j+n+\mu+\nu)}
\end{eqnarray}
is the normalisation constant. It is assumed that both $\mu > -1$ and $\nu >-1$. The associated $\tau$ function equals
\begin{eqnarray}\label{tau-jue}\fl
    \tau_{n}^{\rm J}({\bm {\varsigma}},{\bm \kappa};{\bm t}) =\frac{1}{n!} \int_{(-1,1)^n} \prod_{j=1}^{n}\left( d\lambda_j \, (1-\lambda_j)^\mu (1+\lambda_j)^\nu\, e^{v({\bm t};\lambda_j)}
        \prod_{\alpha=1}^p (\varsigma_\alpha - \lambda_j)^{\kappa_\alpha} \right)
    \cdot
    \Delta_{n}^2({\boldsymbol \lambda}).\nonumber\\{}
\end{eqnarray}
The superscript ${\rm J}$ standing for JUE will be omitted from now on.

Although the very same technology is at work for a nonperturbative calculation of the JUE correlation function Eq.~(\ref{rpf-jue}), its treatment becomes significantly more
cumbersome. For this reason, only the final results of the calculations will be presented below.

\subsection{Virasoro constraints}
In the notation of Section \ref{Sec-3},
the definition Eq.~(\ref{rpf-jue}) implies that
\footnote[4]{Notice that ${\rm dim}(\bm{c^\prime}) =0$ follows from Eq.~(\ref{c-redef}) in which ${\cal Z}_0 = \{-1,+1\}$.
}
\begin{eqnarray}\fl
    \qquad f(\lambda)=1-\lambda^2 \;\; &\mapsto& \;\;\; a_k=\delta_{k,0} - \delta_{k,2},\\
    \fl
    \qquad g(\lambda) = (\mu-\nu) +(\mu+\nu)\,\lambda\;\;&\mapsto& \;\;\;b_k = (\mu-\nu) \delta_{k,0} + (\mu+\nu)\delta_{k,1},\\
    \fl
    \qquad {\cal D} = (-1,+1) \;\; &\mapsto&\;\;\; {\rm dim}(\bm{c^\prime}) =0.
\end{eqnarray}
This brings the Virasoro constraints Eqs.~(\ref{2-Vir}) -- (\ref{vo}) for the associated $\tau$ function Eq.~(\ref{tau-jue}) in the form
\begin{eqnarray}
\label{2-Vir-J} \fl
    \Bigg[ \hat{\cal L}_{q}({\bm t})- \hat{\cal L}_{q+2}({\bm t})
    + \sum_{m=0}^{q}
    \vartheta_m({\bm \varsigma}, {\bm \kappa})
       \frac{\partial}{\partial t_{q-m}}
 -\sum_{m=0}^{q+2}
    \vartheta_m({\bm \varsigma}, {\bm \kappa})
       \frac{\partial}{\partial t_{q+2-m}}
    \nonumber\\
    -(\mu-\nu)\frac{\partial}{\partial t_{q+1}}
    -(\mu+\nu)\frac{\partial}{\partial t_{q+2}}
    \Bigg] \tau_n({\bm \varsigma},{\bm \kappa};{\bm t})={\hat {\mathcal D}}_{q}\,\tau_n({\bm \varsigma},{\bm \kappa};{\bm t}),
\end{eqnarray}
where the operator ${\hat {\mathcal D}}_q$ is defined through operators Eq.~(\ref{Bq}) as
\begin{eqnarray}\label{Dq}
 {\hat {\mathcal D}}_q={\hat {\mathcal B}}_{q}-{\hat {\mathcal B}}_{q+2}= \sum_{\alpha=1}^p (1-\varsigma^2)\, \varsigma_\alpha^{q+1}\frac{\partial}{\partial \varsigma_\alpha},
\end{eqnarray}
and $\hat{\cal L}_{q}({\bm t})$ is the Virasoro operator given by Eq.~(\ref{vo}).
\newline\newline\noindent
The three lowest Virasoro constraints for
$q=-1$, $q=0$, and $q=+1$ read:
\begin{eqnarray}
\label{J-Vir-q=-1} \fl
    \Bigg(
    \sum_{j=2}^\infty jt_j\frac{\partial}{\partial t_{j-1}}
    -\sum_{j=1}^\infty jt_j\frac{\partial}{\partial t_{j+1}}
    -\left( 2n+\mu+\nu+\kappa\right) \frac{\partial}{\partial t_1}
    \Bigg) \log \tau_n ({\bm \varsigma},{\bm \kappa};{\bm t})
    \nonumber\\\qquad\quad + nt_1
    -n\left[ \mu-\nu+\,\vartheta_1({\bm \varsigma},{\bm \kappa})\right]
    ={\hat {\mathcal D}}_{-1}\,\log \tau_n ({\bm \varsigma},{\bm \kappa};{\bm t}),
\end{eqnarray}
\begin{eqnarray}
\label{J-Vir-q=0} \fl
    \Bigg(
    \sum_{j=1}^\infty jt_j\frac{\partial}{\partial t_j}
    -\sum_{j=1}^\infty jt_j\frac{\partial}{\partial t_{j+2}}
    -\left[ \mu-\nu+\,\vartheta_1({\bm \varsigma},{\bm \kappa})\right]\frac{\partial}{\partial t_1}
    -\frac{\partial^2}{\partial t_1^2}
     \nonumber\\\qquad\qquad\quad
    -\left( 2n+\mu+\nu+\kappa\right) \frac{\partial}{\partial t_2}
    \Bigg) \log \tau_n({\bm \varsigma},{\bm \kappa};{\bm t})
    \nonumber\\\fl
    \qquad\qquad-\left( \frac{\partial}{\partial t_1}\log \tau_n({\bm \varsigma},{\bm \kappa};{\bm t})\right)^2
    +n\left[n+ \kappa-\,\vartheta_2({\bm \varsigma},{\bm \kappa})\right]
    ={\hat {\mathcal D}}_0\,\log \tau_n({\bm \varsigma},{\bm \kappa};{\bm t}),
\end{eqnarray}
and
\begin{eqnarray}
\label{J-Vir-q=1} \fl
    \Bigg(
    \sum_{j=1}^\infty jt_j\frac{\partial}{\partial t_{j+1}}
    -\sum_{j=1}^\infty jt_j\frac{\partial}{\partial t_{j+3}}
    +\left[ 2 n +\kappa-\,\vartheta_2({\bm \varsigma},{\bm \kappa})\right]\frac{\partial}{\partial t_1}
        -2 \frac{\partial^2}{\partial t_1 \partial t_2}
     \nonumber\\ \fl \qquad
    -\left[\mu-\nu+\,\vartheta_1({\bm \varsigma},{\bm \kappa})\right]\frac{\partial}{\partial t_2} -\left( 2n+\mu+\nu+\kappa\right) \frac{\partial}{\partial t_3}
    \Bigg) \log \tau_n({\bm \varsigma},{\bm \kappa};{\bm t})\nonumber\\ \fl  \qquad
     +n\left[\,\vartheta_1({\bm \varsigma},{\bm \kappa})-\,\vartheta_3({\bm \varsigma},{\bm \kappa})\right]
    \nonumber\\\fl
    \qquad -2 \left( \frac{\partial}{\partial t_1}\log \tau_n({\bm \varsigma},{\bm \kappa};{\bm t})\right)
    \left( \frac{\partial}{\partial t_2}\log \tau_n({\bm \varsigma},{\bm \kappa};{\bm t})\right)
    ={\hat {\mathcal D}}_1\,\log \tau_n({\bm \varsigma},{\bm \kappa};{\bm t}).
\end{eqnarray}

\subsection{Toda Lattice equation}
Projecting the first Toda Lattice equation Eq.~(\ref{tl-1-equiv}) onto the hyperplane ${\bm t}={\bm 0}$ with the help of the first [Eq.~(\ref{J-Vir-q=-1})] and second [Eq.~(\ref{J-Vir-q=0})] Virasoro constraints, one derives:
\begin{eqnarray}\label{TL-JUE}\fl
\widetilde{{\rm TL}}_1^{\rm J}:\nonumber\\
    \fl \quad\left[
        (2n+\mu+\nu+\kappa)^2 (
            {\hat {\mathcal D}}_{-1}^2 + {\hat {\mathcal D}}_{0}
        ) - (\mu+\nu+\kappa)\left[\mu-\nu+ \,\vartheta_1({\bm \varsigma},{\bm \kappa})\right]\, {\hat {\mathcal D}}_{-1}
    \right]\, \log \Pi_{n|p}({\bm {\varsigma}};{\bm \kappa})
            \nonumber\\ \fl \quad
            + \left({\hat {\mathcal D}}_{-1}\, \log \Pi_{n|p}({\bm {\varsigma}};{\bm \kappa})\right)^2
            \nonumber\\ \fl \qquad
            +  n(n+\mu+\nu+\kappa)
    \left[
        (2n+\mu+\nu+\kappa)^2 - \left[\mu-\nu+\,\vartheta_1({\bm \varsigma},{\bm \kappa})\right]^2
    \right]
     \nonumber\\
    \fl \quad\qquad = \frac{(2n+\mu+\nu+\kappa)^2}{(2n+\mu+\nu)^2}\frac{\left[(2n+\mu+\nu+\kappa)^2-1\right]}{\left[(2n+\mu+\nu)^2-1\right]}
    \nonumber \\
    \fl \quad\qquad \times  \, 4n (n+\mu)(n+\nu)(n+\mu+\nu)\,    \frac{\Pi_{n+1|p}({\bm {\varsigma}};{\bm \kappa}) \,
            \Pi_{n-1|p}({\bm {\varsigma}};{\bm \kappa})}{\Pi_{n|p}^2({\bm {\varsigma}};{\bm \kappa})}.
\end{eqnarray}
Notice the structural similarity between $\widetilde{{\rm TL}}_1^{\rm J}$ and the first Toda Lattice equation for the CyUE [Eq.~(\ref{TL-CyUE})].

\subsection{KP equation and Painlev\'e VI}
Projecting Eq.~(\ref{kp1-exp-rep-LUE}) onto ${\bm t}={\bm 0}$ with the help of all three Virasoro constraints Eqs.~(\ref{J-Vir-q=-1}) -- (\ref{J-Vir-q=1}), one derives:
\begin{eqnarray} \fl
    \label{KP-1-J} \widetilde{{\rm KP}}_1^{\rm J}:\quad
    \Bigg[
        {\hat {\mathcal D}}_{-1}^4 + 4 {\hat {\mathcal D}}_{0} {\hat {\mathcal D}}_{-1}^2 +
        \alpha_{[1,-1]} {\hat {\mathcal D}}_{1} {\hat {\mathcal D}}_{-1} + \alpha_{[0,0]} {\hat {\mathcal D}}_{0}^2
        + \alpha_{[0,-1]} {\hat {\mathcal D}}_{0} {\hat {\mathcal D}}_{-1}\nonumber\\
         \fl \qquad \quad + \alpha_{[-1,-1]} {\hat {\mathcal D}}_{-1}^2
        + \beta_{-1} {\hat {\mathcal D}}_{-1} + \beta_0 {\hat {\mathcal D}}_{0} +  \beta_1 {\hat {\mathcal D}}_{1}
        + \beta_2 {\hat {\mathcal D}}_{2}
            \Bigg] \,
    \log\Pi_{n|p}({\bm {\varsigma}};{\bm \kappa})\nonumber\\
    \fl \qquad \quad+
    \left( {\hat {\mathcal D}}_{-1}^2 \log\Pi_{n|p}({\bm {\varsigma}};{\bm \kappa}) \right)
    \left[
     6 {\hat {\mathcal D}}_{-1}^2 + 4 {\hat {\mathcal D}}_{0}
    \right]\, \log\Pi_{n|p}({\bm {\varsigma}};{\bm \kappa})
    \nonumber\\
    \fl \qquad \quad+ \left( {\hat {\mathcal D}}_{-1}
     \log\Pi_{n|p}({\bm {\varsigma}};{\bm \kappa}) \right) \left[
        4 {\hat {\mathcal D}}_{0} {\hat {\mathcal D}}_{-1} + 2 {\hat {\mathcal D}}_{1} - 2 {\hat {\mathcal D}}_{-1}
     \right] \, \log\Pi_{n|p}({\bm {\varsigma}};{\bm \kappa}) = \gamma.
\end{eqnarray}
Here, $\alpha_{[j,k]}$, $\beta_j$ and $\gamma$ are the short-hand notation for the following functions:
\begin{eqnarray}
\fl
\qquad
    \alpha_{[-1,-1]} &=& 4n^2 + 2  - \left[\mu-\nu+\,\vartheta_1({\bm \varsigma},{\bm \kappa})\right]^2 -  4
    (\mu +\nu + \kappa)\left[\kappa-\,\vartheta_2({\bm \varsigma},{\bm \kappa})-n\right], \nonumber\\
        \fl
\qquad
\alpha_{[0,-1]} &=& - 2(\mu +\nu + \kappa)\left[\mu-\nu+\,\vartheta_1({\bm \varsigma},{\bm \kappa})\right], \nonumber \\
 \fl
\qquad
    \alpha_{[0,0]} &=& 3 \left(
          2 n + \mu +\nu + \kappa
        \right)^2, \nonumber \\
\fl \qquad    \alpha_{[1,-1]} &=& 2\left[
        1 - 2 \left(
          2 n + \mu +\nu + \kappa
        \right)^2
    \right], \nonumber \\
\fl \qquad
    \beta_{-1} &=&  (\mu+\nu+\kappa) \left[ \mu-\nu+\,\vartheta_3({\bm \varsigma},{\bm \kappa})\right]+\left[\mu-\nu+\,\vartheta_1({\bm \varsigma},{\bm \kappa})\right]\left[\kappa-\,\vartheta_2({\bm \varsigma},{\bm \kappa})\right], \nonumber\\
    \fl \qquad
    \beta_{0} &=& -2 \left[ (2n+\mu+\nu+\kappa)^2 + (\mu+\nu+\kappa)\left[\kappa-\,\vartheta_2({\bm \varsigma},{\bm \kappa})\right]
    \right], \nonumber\\
    \fl \qquad
    \beta_{1} &=& - \left[\mu-\nu+\,\vartheta_1({\bm \varsigma},{\bm \kappa})\right] (\mu+\nu+\kappa), \nonumber\\
    \fl \qquad
    \beta_{2} &=& 2 \left(
          2 n + \mu +\nu + \kappa
        \right)^2, \nonumber\\
\fl \qquad
   \gamma &=& 2n (n+\mu+\nu+\kappa)\Big[
    \left[\kappa-\,\vartheta_2({\bm \varsigma},{\bm \kappa})\right]^2 + (\mu+\nu+\kappa)\left[\kappa-\,\vartheta_2({\bm \varsigma},{\bm \kappa})\right] \nonumber\\
    &+&
    \left[\mu-\nu+\,\vartheta_1({\bm \varsigma},{\bm \kappa})\right]\left[\,\vartheta_1({\bm \varsigma},{\bm \kappa})-\,\vartheta_3({\bm \varsigma},{\bm \kappa})\right]
   \Big]. \nonumber
\end{eqnarray}
{\it Remark.} For $p=1$, the equation $\widetilde{{\rm KP}}_1^{\rm J}$ simplifies. Indeed, introducing the function
\begin{eqnarray}
\label{phi-defJ}
    \varphi (\varsigma) = (\varsigma^2-1)
    \frac{\partial}{\partial \varsigma} \log \Pi_{n}\left( \varsigma ;\kappa\right)
    + \varsigma \sum_{2 \le i<j \le 4} b_i b_j - b_1 \sum_{j=2}^4 b_j,
\end{eqnarray}
where $b_j$'s are given by
\begin{eqnarray}
    b_1 &=& +\frac{1}{2}(\mu-\nu),\nonumber\\
    b_2 &=& -\frac{1}{2}(\mu+\nu),\nonumber\\
    b_3 &=& +\frac{1}{2}(\mu+\nu+2n),\nonumber\\
    b_4 &=& -\frac{1}{2}(\mu+\nu+2n+2\kappa),
\end{eqnarray}
one observes that Eq.~(\ref{KP-1-J}) transforms to
\begin{eqnarray}\label{Ch-Jue}
   (\varsigma^2-1)^2\varphi^{\prime\prime\prime} + 2\varsigma(\varsigma^2-1)\, \varphi^{\prime\prime}
    &+& 6(\varsigma^2-1) \,( \varphi^{\prime})^2 - 8\varsigma\,\varphi \varphi^\prime  +2\varphi^2 \nonumber\\
    &-& 4 \nu_1\, \varphi^\prime
    - 4 \nu_4 \varsigma -2 \nu_2 = 0
\end{eqnarray}
with the parameters
\begin{eqnarray}
\fl \qquad
    \nu_1=\sum_{j=1}^4 b_j^2,\quad \nu_2 = \sum_{i<j} b_i^2 b_j^2,\quad \nu_3 = \sum_{i<j<k} b_i^2 b_j^2 b_k^2,\quad \nu_4 = b_1b_2b_3b_4.
\end{eqnarray}
Equation (\ref{Ch-Jue}) can equivalently be written as
\begin{eqnarray}
\label{p6-jue}
\left[ (\varsigma^2-1) \varphi^{\prime\prime}
        \right]^2 &-& 4 (\varphi^\prime)^3 +
        4\varphi^\prime(\varsigma \varphi^\prime-\varphi)^2 \nonumber \\
         &-&  4\nu_1 \left(\varphi^\prime\right)^2 - 8\nu_4\left[\varsigma \varphi^\prime-\varphi\right] - 4\nu_2\varphi^\prime -4 \nu_3 = 0.
\end{eqnarray}
The boundary condition at infinity reads:
\begin{eqnarray}
    \varphi(\varsigma)\Big|_{\varsigma\rightarrow \infty} \sim -\frac{1}{4}(\mu +\nu +2 n)^2 \varsigma\left(
        1 + {\cal O}(\varsigma^{-1})
    \right).
\end{eqnarray}
It is easy to convince yourself that Eq.~(\ref{p6-jue}) can be reduced to the sixth Painlev\'e transcendent.
Introducing the new function
\begin{eqnarray}
    h(t) = \frac{1}{2} \varphi (\varsigma)\Big|_{\varsigma=2t-1},
\end{eqnarray}
one straightforwardly arrives at the $\sigma$ form of the sixth Painlev\'e transcendent (Forrester and Witte 2004):
\begin{eqnarray}\label{Jac-P6}\fl
    h^\prime \left[
    t(t-1)\, h^{\prime\prime}
    \right]^2 + \left[
        h^\prime \left(
            2h - (2t-1)h^\prime
        \right) + b_1 b_2 b_3 b_4
    \right]^2 = \prod_{j=1}^4 (h^\prime +b_j^2 ),
\end{eqnarray}
see Appendix \ref{App-chazy} for more details.

\section{Cauchy Unitary Ensemble (CyUE)}\label{App-CyUE}
For the Cauchy Unitary Ensemble (CyUE), the correlation function of characteristic polynomials
is defined by the formula
\begin{eqnarray}\label{rpf-Cue}\fl
    \Pi_{n|p}^{\rm Cy}({\bm {\varsigma}};{\bm \kappa}) =\frac{1}{{\cal N}_n^{\rm{Cy}}} \int_{\mathbb R^n} \prod_{j=1}^{n}\left( \frac{d\lambda_j}{(1+\lambda_j^2)^{\nu}} \prod_{\alpha=1}^p (\varsigma_\alpha - \lambda_j)^{\kappa_\alpha} \right)
    \cdot
    \Delta_{n}^2({\boldsymbol \lambda})
\end{eqnarray}
where
\begin{eqnarray}\label{N-Cue}
    {\cal N}_n^{\rm Cy} = \frac{(2\pi)^n}{2^{n(2\nu-n)}}\prod_{j=1}^{n} \frac{\Gamma(j+1)\Gamma(2\nu+1-n-j)}{\Gamma^2(\nu+1-j)}
\end{eqnarray}
is the normalisation constant. It is assumed that the parameter $\nu$ is real and $\nu>n+ (\kappa-1)/2$, where $\kappa=\tr_p{\bm\kappa}$. The associated $\tau$ function equals
\begin{eqnarray}\label{tau-Cue}\fl
    \tau_{n}^{\rm Cy}({\bm {\varsigma}},{\bm \kappa};{\bm t}) =\frac{1}{n!} \int_{\mathbb R^n} \prod_{j=1}^{n}\left( \frac{d\lambda_j\,e^{v({\bm t};\lambda_j)}}{(1+\lambda_j^2)^{\nu}}
        \prod_{\alpha=1}^p (\varsigma_\alpha - \lambda_j)^{\kappa_\alpha} \right)
    \cdot
    \Delta_{n}^2({\boldsymbol \lambda}).
\end{eqnarray}
The superscript ${\rm Cy}$ standing for CyUE will further be omitted. Similarly to the JUE case, involved character of the calculations makes us present only the main results.

\subsection{Virasoro constraints}
In the notation of Section \ref{Sec-3},
the definition Eq.~(\ref{rpf-Cue}) implies that
\begin{eqnarray}\fl
    \qquad f(\lambda)=1+\lambda^2 \;\; &\mapsto& \;\;\; a_k=\delta_{k,0} + \delta_{k,2},\\
    \fl
    \qquad g(\lambda) = 2\nu\,\lambda\;\;&\mapsto& \;\;\;b_k = 2\nu\delta_{k,1},\\
    \fl
    \qquad {\cal D} = {\mathbb R} \;\; &\mapsto&\;\;\; {\rm dim}(\bm{c^\prime}) =0.
\end{eqnarray}
This brings the Virasoro constraints Eqs.~(\ref{2-Vir}) -- (\ref{vo}) for the associated $\tau$ function Eq.~(\ref{tau-Cue}) in the form
\begin{eqnarray}
\label{2-Vir-C} \fl
    \Bigg[ \hat{\cal L}_{q}({\bm t})+ \hat{\cal L}_{q+2}({\bm t})
    + \sum_{m=0}^{q}
    \,\vartheta_m({\bm \varsigma},{\bm \kappa})
           \frac{\partial}{\partial t_{q-m}}
 +\sum_{m=0}^{q+2}
    \,\vartheta_m({\bm \varsigma},{\bm \kappa})
       \frac{\partial}{\partial t_{q+2-m}}
    \nonumber\\\qquad\qquad\qquad
    -2\nu\frac{\partial}{\partial t_{q+2}}
    \Bigg] \tau_n({\bm \varsigma},{\bm \kappa};{\bm t})={\hat {\mathcal D}}_{q}\,\tau_n({\bm \varsigma},{\bm \kappa};{\bm t}),
\end{eqnarray}
where ${\hat {\mathcal D}}_q$ is defined through operators Eq.~(\ref{Bq}) as
\begin{eqnarray}\label{DqC}
 {\hat {\mathcal D}}_q={\hat {\mathcal B}}_{q}+{\hat {\mathcal B}}_{q+2} = \sum_{\alpha=1}^p (1+\varsigma^2)\, \varsigma_\alpha^{q+1}\frac{\partial}{\partial \varsigma_\alpha}
\end{eqnarray}
(not to be confused with the operator ${\hat {\mathcal D}}_q$ defined by Eq.~(\ref{Dq}) of the previous subsection) and $\hat{\cal L}_{q}({\bm t})$ is the Virasoro operator given by Eq.~(\ref{vo}).
\newline\newline\noindent
The three lowest Virasoro constraints for $q=-1$, $q=0$, and $q=+1$ read:
\begin{eqnarray}
\label{C-Vir-q=-1} \fl
    \Bigg(
    \sum_{j=2}^\infty jt_j\frac{\partial}{\partial t_{j-1}}
    +\sum_{j=1}^\infty jt_j\frac{\partial}{\partial t_{j+1}}
    -\left(2\nu-2n- \kappa\right) \frac{\partial}{\partial t_1}
    \Bigg) \log \tau_n ({\bm \varsigma},{\bm \kappa};{\bm t})
    \nonumber\\\qquad\qquad\qquad  + nt_1
    +n\,\vartheta_1({\bm \varsigma},{\bm \kappa})
    ={\hat {\mathcal D}}_{-1}\,\log \tau_n ({\bm \varsigma},{\bm \kappa};{\bm t}),
\end{eqnarray}
\begin{eqnarray}
\label{C-Vir-q=0} \fl
    \Bigg(
    \sum_{j=1}^\infty jt_j\frac{\partial}{\partial t_j}
    +\sum_{j=1}^\infty jt_j\frac{\partial}{\partial t_{j+2}}
    +\,\vartheta_1({\bm \varsigma},{\bm \kappa}) \frac{\partial}{\partial t_1}
    +\frac{\partial^2}{\partial t_1^2}
     \nonumber\\\fl\qquad\qquad
    -\left( 2\nu-2n-\kappa\right) \frac{\partial}{\partial t_2}
    \Bigg) \log \tau_n({\bm \varsigma},{\bm \kappa};{\bm t})
    +\left( \frac{\partial}{\partial t_1}\log \tau_n({\bm \varsigma},{\bm \kappa};{\bm t})\right)^2
    \nonumber\\\qquad
    + n^2
    +n\left[ \kappa +\,\vartheta_2({\bm \varsigma},{\bm \kappa})\right]
    ={\hat {\mathcal D}}_0\,\log \tau_n({\bm \varsigma},{\bm \kappa};{\bm t}),
\end{eqnarray}
\begin{eqnarray}
\label{C-Vir-q=1} \fl
    \Bigg(
    \sum_{j=1}^\infty jt_j\frac{\partial}{\partial t_{j+1}}
    +\sum_{j=1}^\infty jt_j\frac{\partial}{\partial t_{j+3}}
    +\left[ 2 n +\kappa+\,\vartheta_2({\bm \varsigma},{\bm \kappa})\right]\frac{\partial}{\partial t_1}
     \nonumber\\ \fl \qquad
    + \,\vartheta_1({\bm \varsigma},{\bm \kappa})\frac{\partial}{\partial t_2} +\left( \kappa +2n - 2 \nu\right) \frac{\partial}{\partial t_3}
       +2 \frac{\partial^2}{\partial t_1 \partial t_2}
    \Bigg) \log \tau_n({\bm \varsigma},{\bm \kappa};{\bm t})\nonumber\\ \fl \qquad  +n\left[\,\vartheta_1({\bm \varsigma},{\bm \kappa})+\,\vartheta_3({\bm \varsigma},{\bm \kappa})\right]
    \nonumber\\\fl
    \qquad +2 \left( \frac{\partial}{\partial t_1}\log \tau_n({\bm \varsigma},{\bm \kappa};{\bm t})\right)
    \left( \frac{\partial}{\partial t_2}\log \tau_n({\bm \varsigma},{\bm \kappa};{\bm t})\right)
    ={\hat {\mathcal D}}_1\,\log \tau_n({\bm \varsigma},{\bm \kappa};{\bm t}).
\end{eqnarray}

\subsection{Toda Lattice equation}
Projecting the first Toda Lattice equation Eq.~(\ref{tl-1-equiv}) onto the hyperplane ${\bm t}={\bm 0}$ with the help of the first [Eq.~(\ref{C-Vir-q=-1})] and second [Eq.~(\ref{C-Vir-q=0})] Virasoro constraints, one derives:
\begin{eqnarray}\label{TL-CyUE}\fl
\widetilde{{\rm TL}}_1^{\rm Cy}:\nonumber\\
    \fl \quad\left[
        (2\nu-2n-\kappa)^2 (
            {\hat {\mathcal D}}_{-1}^2 - {\hat {\mathcal D}}_{0}
        ) - (2\nu-\kappa)\,\vartheta_1({\bm \varsigma},{\bm \kappa})\, {\hat {\mathcal D}}_{-1}
    \right]\, \log \Pi_{n|p}({\bm {\varsigma}};{\bm \kappa})
            \nonumber\\ \fl \quad
            + \left({\hat {\mathcal D}}_{-1}\, \log \Pi_{n|p}({\bm {\varsigma}};{\bm \kappa})\right)^2
            +  n(2\nu-n-\kappa)
    \left[
        (2\nu-2n-\kappa)^2 + \,\vartheta_1({\bm \varsigma},{\bm \kappa})^2
    \right]
     \nonumber\\
    \fl \quad\qquad = (2\nu-2n-\kappa)^2[(2\nu-2n-\kappa)^2-1]\nonumber\\
    \fl \quad\qquad\qquad\qquad\qquad\qquad\times \frac{n(2\nu-n)}{4(n-\nu)^2-1}
            \frac{\Pi_{n-1|p}({\bm {\varsigma}};{\bm \kappa})\Pi_{n+1|p}({\bm {\varsigma}};{\bm \kappa})}{\Pi_{n|p}^2({\bm {\varsigma}};{\bm \kappa})}.
\end{eqnarray}
Notice the structural similarity between $\widetilde{{\rm TL}}_1^{\rm Cy}$ and the first Toda Lattice equation for the JUE [Eq.~(\ref{TL-JUE})].

\subsection{KP equation and Painlev\'e VI}
Projecting Eq.~(\ref{kp1-exp-rep-LUE}) onto ${\bm t}={\bm 0}$ with the help of all three Virasoro constraints Eqs.~(\ref{C-Vir-q=-1}) -- (\ref{C-Vir-q=1}), one derives:
\begin{eqnarray} \fl
    \label{KP-1-C} \widetilde{{\rm KP}}_1^{\rm Cy}:\quad
    \Bigg[
        {\hat {\mathcal D}}_{-1}^4 - 4 {\hat {\mathcal D}}_{0} {\hat {\mathcal D}}_{-1}^2 +
        \alpha_{[1,-1]} {\hat {\mathcal D}}_{1} {\hat {\mathcal D}}_{-1} + \alpha_{[0,0]} {\hat {\mathcal D}}_{0}^2
        + \alpha_{[0,-1]} {\hat {\mathcal D}}_{0} {\hat {\mathcal D}}_{-1}\nonumber\\
         \fl \qquad \quad + \alpha_{[-1,-1]} {\hat {\mathcal D}}_{-1}^2
        + \beta_{-1} {\hat {\mathcal D}}_{-1} + \beta_0 {\hat {\mathcal D}}_{0} +  \beta_1 {\hat {\mathcal D}}_{1}
        + \beta_2 {\hat {\mathcal D}}_{2}
            \Bigg] \,
    \log\Pi_{n|p}({\bm {\varsigma}};{\bm \kappa})\nonumber\\
    \fl \qquad \quad+
    \left( {\hat {\mathcal D}}_{-1}^2 \log\Pi_{n|p}({\bm {\varsigma}};{\bm \kappa}) \right)
    \left[
     6 {\hat {\mathcal D}}_{-1}^2 - 4 {\hat {\mathcal D}}_{0}
    \right]\, \log\Pi_{n|p}({\bm {\varsigma}};{\bm \kappa})
    \nonumber\\
    \fl \qquad \quad - \left( {\hat {\mathcal D}}_{-1}
     \log\Pi_{n|p}({\bm {\varsigma}};{\bm \kappa}) \right) \left[
        4 {\hat {\mathcal D}}_{0} {\hat {\mathcal D}}_{-1} - 2 {\hat {\mathcal D}}_{1} - 2 {\hat {\mathcal D}}_{-1}
     \right] \, \log\Pi_{n|p}({\bm {\varsigma}};{\bm \kappa}) = \gamma.
\end{eqnarray}
Here, $\alpha_{[j,k]}$, $\beta_j$ and $\gamma$ are the short-hand notation for the following functions:
\begin{eqnarray}
\fl
\qquad
    \alpha_{[-1,-1]} &=& -4n^2 - 2  - \,\vartheta_1({\bm \varsigma},{\bm \kappa})^2 -  4  (2\nu - \kappa)\left[\kappa + \,\vartheta_2({\bm \varsigma},{\bm \kappa})-n\right], \nonumber\\
        \fl
\qquad
\alpha_{[0,-1]} &=& 2(2 \nu - \kappa)\, \,\vartheta_1({\bm \varsigma},{\bm \kappa}), \nonumber \\
 \fl
\qquad
    \alpha_{[0,0]} &=& 3 \left(
          2 \nu -2n - \kappa
        \right)^2, \nonumber \\
\fl \qquad    \alpha_{[1,-1]} &=& 2\left[
        1 - 2 \left(
          2 \nu -2n - \kappa
        \right)^2
    \right], \nonumber \\
\fl \qquad
    \beta_{-1} &=&  (2\nu-\kappa) \,\vartheta_3({\bm \varsigma},{\bm \kappa}) +\,\vartheta_1({\bm \varsigma},{\bm \kappa})
    \left[\kappa+\,\vartheta_2({\bm \varsigma},{\bm \kappa})\right], \nonumber\\
    \fl \qquad
    \beta_{0} &=& -2 \left[ (2\nu-2n - \kappa)^2 - (2 \nu-\kappa)\left[\kappa+\,\vartheta_2({\bm \varsigma},{\bm \kappa})\right]
    \right], \nonumber\\
    \fl \qquad
    \beta_{1} &=& - \,\vartheta_1({\bm \varsigma},{\bm \kappa}) (2\nu-\kappa), \nonumber\\
    \fl \qquad
    \beta_{2} &=& -2 \left(
          2 \nu -2n - \kappa
        \right)^2, \nonumber\\
\fl \qquad
   \gamma &=& 2n (2\nu-n-\kappa) \Big[
    -\left[\kappa+\,\vartheta_2({\bm \varsigma},{\bm \kappa})\right]^2 + (2 \nu-\kappa)\left[\kappa+\,\vartheta_2({\bm \varsigma},{\bm \kappa})\right]
    \nonumber\\
    &+&
    \,\vartheta_1({\bm \varsigma},{\bm \kappa}) \left[\,\vartheta_1({\bm \varsigma},{\bm \kappa})+\,\vartheta_3({\bm \varsigma},{\bm \kappa})\right]
   \Big]. \nonumber
\end{eqnarray}
\newline\noindent
{\it Remark}. For $p=1$, the equation $\widetilde{{\rm KP}}_1^{\rm Cy}$ can be simplified. Indeed, introducing the function
\begin{eqnarray}
\label{phi-defC}
    \varphi (\varsigma) = (1+\varsigma^2)
    \frac{\partial}{\partial \varsigma} \log \Pi_{n}\left( \varsigma ;\kappa\right)
    + \varsigma \sum_{2 \le i<j \le 4} b_i b_j - b_1 \sum_{j=2}^4 b_j,
\end{eqnarray}
where $b_j$'s are given by
\begin{eqnarray}
    b_1 &=& 0,\nonumber\\
    b_2 &=& \nu-n,\nonumber\\
    b_3 &=& -\nu,\nonumber\\
    b_4 &=& \kappa-\nu+n,
\end{eqnarray}
one observes that Eq.~(\ref{KP-1-C}) takes the form of the Chazy I equation
\begin{eqnarray}\label{Ch-Cyue} \fl \quad
   (1+\varsigma^2)^2\varphi^{\prime\prime\prime} + 2\varsigma(1+\varsigma^2)\, \varphi^{\prime\prime}
    + 6(1+\varsigma^2) \,( \varphi^{\prime})^2 - 8\varsigma\,\varphi \varphi^\prime  +2\varphi^2 \nonumber\\
    \qquad \qquad \qquad \qquad  \qquad \qquad + 4 \nu_1\, \varphi^\prime
    + 2 \nu_2 = 0
\end{eqnarray}
with the parameters
\begin{eqnarray}
\fl \qquad
    \nu_1=\sum_{j=1}^4 b_j^2,\quad \nu_2 = \sum_{i<j} b_i^2 b_j^2,\quad \nu_3 = \sum_{i<j<k} b_i^2 b_j^2 b_k^2.
\end{eqnarray}
Equation (\ref{Ch-Cyue}) can equivalently be written as
\begin{eqnarray}\fl
\left[ (1+\varsigma^2) \varphi^{\prime\prime}
        \right]^2 +
        4(\varphi^\prime)^3 + 4\varphi^\prime (\varsigma \varphi^\prime-\varphi)^2
        +  4\nu_1 \left(\varphi^\prime\right)^2 + 4\nu_2\varphi^\prime +4\nu_3 = 0.
\end{eqnarray}
The boundary condition at infinity reads:
\begin{eqnarray}
    \varphi(\varsigma)\Big|_{\varsigma\rightarrow \infty} \sim -(n-\nu)^2 \varsigma\left(
        1 + {\cal O}(\varsigma^{-1})
    \right).
\end{eqnarray}
Introducing the new function
\begin{eqnarray}
    h(t) = \frac{1}{2 i} \varphi (\varsigma)\Big|_{\varsigma=i(2t-1)},
\end{eqnarray}
one straightforwardly verifies that it satisfies the equation
\begin{eqnarray}
    \left[
    t(t-1)\, h^{\prime\prime}
    \right]^2 + h^\prime \left[
            2h - (2t-1)h^\prime
    \right]^2 = \prod_{j=2}^4 (h^\prime +b_j^2 ).
\end{eqnarray}
This coincides with the $\sigma$ form of the sixth Painlev\'e transcendent [Eq.~(\ref{Jac-P6})] with $b_1=0$.

\section{Chazy I Equation as a Master Painlev\'e Equation}
\label{App-chazy}
This Appendix, based on Chapter 6 of Forrester (2010) and Section 4.3 of Adler and van Moerbeke (2001), collects very basic facts on six Painlev\'e transcendents and a closely related differential equation belonging to Chazy I class (Cosgrove and Scoufis 1993).
\subsection{Painlev\'e property and six Painlev\'e transcendents}

Painlev\'e transcendents are second order differential equations
of the form
\begin{eqnarray}
    \label{Cpainleve-eqs}
    y^{\prime\prime} = F(t, y, y^\prime),
\end{eqnarray}
for which all movable singularities of $y(t)$ are limited to
poles, given $F$ is a rational function in all its arguments. Note
that this requirement, known as the {\it Painlev\'e property},
does not rule out the existence of immovable (i.e. fixed)
essential singularities. We remind that a singularity is called
movable if its location depends on one or more integration
constants.

Painlev\'e (1900, 1902) and Gambier (1910) have
shown that the requirement that all movable singularities are
restricted to poles leads to 50 types of equations, six of which
cannot further be reduced to either (i) linear second order
differential equations or (ii) the differential equation for the
Weierstrass ${\cal P}$-function,
\begin{eqnarray}
    \label{CW-eq}
    (y^\prime)^2 = 4 y^3 - g_2 y - g_3,
\end{eqnarray}
($g_2,g_3$ are constants) or (iii) the Riccati equation
\begin{eqnarray}
    \label{CR-eq}
    y^\prime = a(t) \, y^2 + b(t) \, y + c(t),
\end{eqnarray}
where $a(t)$, $b(t)$ and $c(t)$ are analytic functions of $t$. (The two later equations
represent an irreducible class of first order differential
equations of the form $P(t,y,y^\prime)=0$ with $P$ being a
polynomial in $y^\prime,\,y$ with coefficients meromorphic in $t$,
such that $y(t)$ is free from movable essential singularities.)
\newline\newline\noindent
Labeled $P_{\rm I}$ to $P_{\rm VI}$, these six equations are
explicitly given by
\begin{eqnarray}
    \label{CP1-6}\fl
      P_{\rm I} &\rightsquigarrow\quad & y^{\prime\prime} = 6 y^2 + t, \\\fl
      P_{\rm II} &\rightsquigarrow\quad  & y^{\prime\prime} = 2 y^3 + t\,y + \alpha, \\\fl
      P_{\rm III} &\rightsquigarrow\quad & y^{\prime\prime} = \frac{1}{y} \, \left( y^\prime\right)^2
      - \frac{1}{t} \, y^\prime + \alpha y^3 + \frac{1}{t} \, \left(
      \beta y^2+\gamma\right) + \frac{\delta}{y},\\\fl
      P_{\rm IV:} &\rightsquigarrow\quad & y^{\prime\prime} = \frac{1}{2y} \, \left(
      y^\prime\right)^2 + \frac{3}{2} \, y^3 + 4 t\, y^2 +
      2 (t^2 - \alpha)y + \frac{\beta}{y}, \\\fl
      P_{\rm V} &\rightsquigarrow\quad & y^{\prime\prime} = \left(
        \frac{1}{2y} + \frac{1}{1-y}
      \right) \left( y^\prime \right)^2 - \frac{1}{t}\, y^\prime
      + \frac{(y-1)^2}{t^2} \left(
        \alpha y + \frac{\beta}{y}
      \right) \nonumber \\ \fl
      & & \hspace{6cm}  + \frac{\gamma y}{t} + \frac{\delta y (y+1)}{y-1},\\\fl
      P_{\rm VI} &\rightsquigarrow\quad & y^{\prime\prime} = \frac{1}{2}
        \left(
            \frac{1}{y} + \frac{1}{y-1} + \frac{1}{y-t}
        \right) \left( y^\prime \right)^2
        -
        \left(
            \frac{1}{t} + \frac{1}{t-1} + \frac{1}{y-t}
        \right) y^\prime \nonumber \\\fl
        & & \hspace{1cm} + \frac{y(y-1)(y-t)}{t^2 (t-1)^2}
        \left(
            \alpha + \frac{\beta t}{y^2} + \frac{\gamma (t-1)}{(y-1)^2}
            + \frac{\delta t (t-1)}{(y-t)^2}
        \right).
\end{eqnarray}
Here, $\alpha$, $\beta$, $\gamma$ and $\delta$ denote complex constants. The equations $P_{\rm I}$, $P_{\rm II}$, and $P_{\rm IV}$ have essential singularities at the point $\infty$, the equations $P_{\rm III}$ and $P_{\rm V}$ have critical points $0$ and $\infty$, whilst the $P_{\rm VI}$ equation has critical points at $0$, $1$, and $\infty$.

\subsection{Hamiltonian formulation of of Painlev\'e transcendents and their \\ Jimbo-Miwa-Okamoto $\sigma$ forms }
Each $P_{\rm J}$ of $P_{\rm I}$ -- $P_{\rm VI}$ is equivalent to a Hamiltonian system $\{Q,P,H_{\rm J}\}$
\begin{eqnarray}\label{HJ}
\left\{
     \begin{array}{c}
       \dot{Q} = + \displaystyle\frac{\partial H_{\rm J}}{\partial P}, \\
       {}\\
       \dot{P} = - \displaystyle\frac{\partial H_{\rm J}}{\partial Q},
     \end{array}
     \right.
\end{eqnarray}
where $\dot{Q}$ and $\dot{P}$ denote derivatives with respect to $t$, and the Hamiltonian $H_{\rm J}=H_{\rm J}\{P,Q,t\}$ is a polynomial or rational function. The equation in canonical coordinate $Q$, obtained from Eq.~(\ref{HJ}) after eliminating canonical momentum $P$, is appropriate $P_{\rm J}$ equation. Existence of the Hamiltonian formulation of Painlev\'e equations can be traced back to absence of movable branch points in $P_{\rm J}$ (Malmquist 1922).

Explicit forms of the Hamiltonians $H_{\rm J}$, as well as the differential equations satisfied by them, can be found in original papers of Okamoto (1980a, 1980b), see also Noumi (2004) and Chapter 6 of Forrester (2010). In the random-matrix-theory literature, the differential equations for $H_{\rm J}$ more often appear in the so-called Jimbo-Miwa-Okamoto
$\sigma$-form:
\begin{eqnarray}
    \label{Csigma1-1}\fl
      \sigma P_{\rm I} & \rightsquigarrow\quad&  \left( \sigma_{\rm I}^{\prime\prime}\right)^2
      + 4 \left( \sigma_{\rm I}^{\prime} \right)^3 - 2 (\sigma_{\rm I} -
      t \sigma_{\rm I}^\prime)=0, \\\fl
      \label{Csigma1-2}
      \sigma P_{\rm II}  &\rightsquigarrow\quad & \left(\sigma^{\prime\prime}_{\rm II}\right)^2
      + 4 \sigma^\prime_{\rm II} \left[
        \left( \sigma^\prime_{\rm II} \right)^2  - t \sigma^\prime_{\rm II}
        + \sigma_{\rm II}
      \right]- b^2 = 0, \\\fl
      \label{Csigma1-3}
      \sigma P_{\rm III} & \rightsquigarrow\quad &
        \left( t \sigma^{\prime\prime}_{\rm III} \right)^2
         + \sigma_{\rm III}^\prime (4 \sigma^\prime_{\rm III}-1)
        (\sigma_{\rm III} - t \sigma^\prime_{\rm III}) \nonumber \\
        \fl
        & & \hspace{3cm}+ \left(
        \sigma^{\prime}_{\rm III}
        \right)^2 \prod_{j=1}^2 b_j
              -
        \frac{1}{4^3}\,\Big(\sum_{j=1}^2 b_j\Big)^2=0,\\\fl
        \label{Csigma1-4}
      \sigma P_{\rm IV}  & \rightsquigarrow\quad &
      \left(\sigma^{\prime\prime}_{\rm IV} \right)^2
        - 4 (\sigma_{\rm IV} - t \sigma^\prime_{\rm IV})^2
        + 4 \sigma^{\prime}_{\rm IV}
        \prod_{j=1}^2\left(
        \sigma^{\prime}_{\rm IV} + b_j
        \right)= 0, \\
        \label{Csigma1-5}\fl
      \sigma P_{\rm V}  & \rightsquigarrow\quad &
      \left( t \sigma^{\prime\prime}_{\rm V} \right)^2
        - \Big[ \sigma_{\rm V} - t \sigma^{\prime}_{\rm V}
        + 2 (\sigma^\prime_{\rm V})^2 +
         \sigma^\prime_{\rm V} \sum_{j=1}^4 b_j
        \Big]^2
        + 4 \prod_{j=1}^4(\sigma^\prime_{\rm V}+b_j) = 0,\label{Csigma1-5}\\
        \label{Csigma1-6}\fl
      \sigma P_{\rm VI}  &\rightsquigarrow\quad   &
        \sigma_{\rm VI}^\prime \left[
    t(t-1)\, \sigma_{\rm VI}^{\prime\prime}
    \right]^2 + \Big[
        \sigma_{\rm VI}^\prime \left(
            2\sigma_{\rm VI} - (2t-1)\sigma_{\rm VI}^\prime
        \right) + \prod_{j=1}^4 b_j
    \Big]^2 \nonumber \\ \fl
    & & \hspace{6cm}- \prod_{j=1}^4 (\sigma_{\rm VI}^\prime +b_j^2 )=0.
\end{eqnarray}
These follow from equations for the polynomial Hamiltonians after appropriate linear transformations of $H_{\rm J}$ (Forrester 2010).

\subsection{Chazy classes and a master Painlev\'e equation}

Differential equation belonging to Chazy classes (Chazy 1911) is the third order differential equation of the form
\begin{eqnarray}
    \label{Cchazy-1}
    y^{\prime\prime\prime} = F(t,y,y^\prime,y^{\prime\prime})
\end{eqnarray}
with $F$ being a rational function in $y,y^\prime$, and
$y^{\prime\prime}$ and locally analytic in $t$, whose general
solution is free of movable branch points. In his classification, Chazy discovered 13 irreducible classes (reviewed in Section 6 of Cosgrove (2000)), of
which the class Chazy I is given by
\begin{eqnarray}
    \label{Cchazy-class-1}\fl
    y^{\prime\prime\prime} + \frac{P^\prime}{P}\, y^{\prime\prime}
    + \frac{6}{P} \, y^{\prime\;2} - \frac{4 P^\prime}{P^2}\,
    y y^\prime + \frac{P^{\prime\prime}}{P^2}\, y^2
    + \frac{4 Q}{P^2} \, y^\prime - \frac{2 Q^\prime}{P^2}\, y
    + \frac{2 R}{P^2} = 0
\end{eqnarray}
with arbitrary polynomials $P(t)$, $Q(t)$ and $R(t)$ of third, second and first degree, respectively.

Cosgrove and Scoufis (1993) have proven that Chazy I equation (\ref{Cchazy-class-1}) admits the first integral, which is of second order in $y$ and quadratic in
$y^{\prime\prime}$,
\begin{eqnarray}
    \label{Cchazy-integral}\fl
    (y^{\prime \prime})^2 + \frac{4}{P^2}\, \Bigg[
    \left(
        P y^{\prime \, 2} + Q y^\prime + R
    \right) y^\prime
    -
    \left(
        P^\prime y^{\prime \, 2} + Q^\prime y^\prime + R^\prime
    \right) y
    \nonumber\\\qquad\qquad\qquad+
    \frac{1}{2}
    \left(
    P^{\prime\prime} y^\prime + Q^{\prime\prime}
    \right) y^2
    -
    \frac{1}{6} \, P^{\prime\prime\prime} y^3 + c
    \Bigg] = 0,
\end{eqnarray}
$c$ is the integration constant. They also show that this equation is the `master Painlev\'e equation' because it
unifies all of the six Painlev\'e transcendents into a single
equation.

Making use of Eqs.~(\ref{Cchazy-class-1}) and (\ref{Cchazy-integral}) in conjunction with Eqs.~(\ref{Csigma1-1}) -- (\ref{Csigma1-6}), one readily
verifies that $\sigma P_{\rm J}$ can be brought to the
Chazy I form ($CP_{\rm J}$):
\begin{eqnarray}
    \label{CCP-form1}\fl
      CP_{\rm I} &\;\rightsquigarrow\quad & \sigma_{\rm I}^{\prime\prime\prime} + 6 (\sigma_{\rm I}^\prime)^2  +
      t = 0, \\\fl
      CP_{\rm II} &\;\rightsquigarrow\quad & \sigma_{\rm II}^{\prime\prime\prime} + 6 (\sigma_{\rm II}^\prime)^2
      - 4 t \sigma_{\rm II}^\prime + 2 \sigma_{\rm II} = 0, \\\fl
      CP_{\rm III} &\;\rightsquigarrow\quad &
        t^2 \sigma_{\rm III}^{\prime\prime\prime} + t
        \sigma_{\rm III}^{\prime\prime} - 6 t
        \, (\sigma_{\rm III}^\prime)^2 +
        4 \sigma_{\rm III} \sigma_{\rm III}^\prime
        + \Big[
            t+ \prod_{j=1}^2 b_j
        \Big]\, \sigma_{\rm III}^\prime
        - \frac{1}{2}\, \sigma_{\rm III} = 0,\\\fl
      CP_{\rm IV} &\;\rightsquigarrow\quad &
        \sigma_{\rm IV}^{\prime\prime\prime} + 6 (\sigma_{\rm IV}^\prime)^2
        + 4 \Big[ \sum_{j=1}^2 b_j-t^2 \Big] \sigma_{\rm IV}^\prime
        + 4 t \sigma_{\rm IV} + 2 \prod_{j=1}^2 b_j =0, \\\fl
      CP_{\rm V} &\;\rightsquigarrow\quad &
      t^2\sigma_{\rm V}^{\prime\prime\prime} + t\, \sigma_{\rm V}^{\prime\prime}
    + 6t \,( \sigma_{\rm V}^{\prime})^2 - 4\,
    \sigma_{\rm V} \sigma_{\rm V}^\prime \\
    \fl
    & &
    + \Bigg[\Big(t -\sum_{j=1}^4 b_j\Big)^2 - 4 \sum_{1\le i < j \le 4} b_i b_j\Bigg]  \, \sigma_{\rm V}^\prime + \Big[t - \sum_{j=1}^4 b_j\Big]\sigma_{\rm V}
    = 0,\\
    \fl
      CP_{\rm VI} &\;\rightsquigarrow\quad &
      [t(t-1)]^2\sigma_{\rm VI}^{\prime\prime\prime} + t(t-1)(2t-1)\, \sigma_{\rm VI}^{\prime\prime}
    + 6t(t-1) \,( \sigma_{\rm VI}^{\prime})^2  \nonumber \\\fl
    & & \qquad \qquad- 4(2t-1)\,\sigma_{\rm VI} \sigma_{\rm VI}^\prime
    + \sigma_{\rm VI}^\prime \sum_{j=1}^4 b_j^2 +2\sigma_{\rm VI}^2 \nonumber\\
    \fl & &
     \qquad \qquad \qquad+ \frac{1}{2}\left[
        t \sum_{1\le i < j \le 4} b_i^2 b_j^2 + \sum_{1\le i < j <k\le 4} b_i^2 b_j^2 b_k^2
    \right] = 0.
\end{eqnarray}

\section{Functions of Parabolic Cylinder and Two Related Integrals}
\label{App-D-int}
In this Appendix, we collect some useful formulae related to the functions of parabolic cylinder and also treat two related integrals Eqs.~(\ref{osipov-integral}) and (\ref{osipov-integral-bos}) encountered in Section \ref{Sec-5}.\newline\newline
\noindent
{\it (i)~Integral representations, differential equation, and a Wronskian.}---The functions of parabolic cylinder $D_p(w)$ and $D_{-p-1}(iw)$ admit integral representations
\begin{eqnarray}\fl
\label{ir-plus}
    D_p(w) =\frac{1}{\sqrt{\pi}} 2^{-p-1/2} e^{-ip\pi/2} e^{w^2/4}
    \int_{\mathbb R} dx \, x^p e^{-x^2/2 + i x w},\qquad \mathfrak{Re} p > -1, \\
    \fl \label{ir-minus}
    D_{-p-1}(iw) = \frac{1}{\Gamma(p+1)} e^{w^2/4}
    \int_{{\mathbb R}_+} dx \, x^{p} e^{-x^2/2 - i x w},\qquad \mathfrak{Re} p > -1,
\end{eqnarray}
and are general, linear independent solutions to the equation
\begin{eqnarray}
\label{deq}
    \frac{d^2 U}{dw^2} + \left(
        p+\frac{1}{2} -\frac{w^2}{4}
    \right) U = 0.
\end{eqnarray}
The corresponding Wronskian is given by
\begin{eqnarray}\label{wronsk_D}
    {\hat {\mathcal W}}_w[D_{-p-1}(w), D_{p}(iw)]=i^{p}.
\end{eqnarray}
\newline\newline
\noindent
{\it (ii)~Functional and recurrence identitites.}---The following functional identity holds:
\begin{eqnarray}
\label{relD} \fl
D_{p}(iw) = \frac{\Gamma(p+1)}{\sqrt{2\pi}} \left[
     (-i)^{p} D_{-p-1}(w) + i^{p} D_{-p-1}(-w)
\right].
\end{eqnarray}
In addition to the three term recurrence relation,
\begin{eqnarray}
    D_{p+1}(w) - w D_p(w) + p D_{p-1}(w)=0,
\end{eqnarray}
there exist two differential recurrence relations:
\begin{eqnarray}
\label{d-rec-1}
    D_p^\prime(w) +\frac{w}{2} D_p(w) - pD_{p-1}(w) &=& 0, \\
\label{d-rec-2}
    D_p^\prime(w) -\frac{w}{2} D_p(w) + D_{p+1}(w) &=& 0.
\end{eqnarray}
\newline\newline
\noindent
{\it (iii)~Asymptotic expansions.}---Throughout the paper we make use of the large-$w$ asymptotic expansions:\newline
\begin{itemize}
        \item For $|{\arg w}| < 3\pi/4$:
        \begin{eqnarray}
        D_p(w) \sim w^p e^{-w^2/4}\left[
                    1+ {\mathcal O}\left(
                        \frac{1}{w^2}
                        \right)
                        \right].
    \end{eqnarray}
    \item For $\pi/4 < \arg w < 5\pi/4$:
    \begin{eqnarray} \fl
                D_p(w) \sim w^p e^{-w^2/4} \left[
                1+ {\mathcal O}\left(
                \frac{1}{w^2}
                \right)
                \right] - \frac{\sqrt{2\pi}}{\Gamma(-p)} e^{i\pi p} w^{-p-1} e^{w^2/4} \left[
                1+ {\mathcal O}\left(
                \frac{1}{w^2}
                \right)
                \right].\nonumber \\
    {}
\end{eqnarray}
    \item For $ -5\pi/4 < \arg w < -\pi/4$:
\begin{eqnarray}\fl
    D_p(w) \sim w^p e^{-w^2/4} \left[
        1+ {\mathcal O}\left(
            \frac{1}{w^2}
        \right)
    \right] - \frac{\sqrt{2\pi}}{\Gamma(-p)} e^{-i\pi p} w^{-p-1} e^{w^2/4} \left[
        1+ {\mathcal O}\left(
            \frac{1}{w^2}
        \right)
    \right]. \nonumber \\ {}
\end{eqnarray}
\newline
\end{itemize}
\noindent
{\it (iv)~Integrals Eq.~(\ref{osipov-integral}) and Eq.~(\ref{osipov-integral-bos}).}---To calculate the integrals Eq.~(\ref{osipov-integral}) and Eq.~(\ref{osipov-integral-bos}) encountered in Section \ref{Sec-5}, we first formulate the Lemma.\newline\newline\noindent {\bf Lemma.} Let $u_1(t)$ and $u_2(t)$ be linearly independent functions whose Wronskian is a constant:
\begin{eqnarray}
    {\hat {\mathcal W}}_t[u_1,u_2] = u_1 u_2^\prime - u_1^\prime u_2 = w_0 \neq 0.
\end{eqnarray}
Then,
\begin{eqnarray}\label{Oi_gen}
    \int \frac{dt}{(c_1 u_1(t) + c_2 u_2(t))^2} = \frac{\alpha_1 u_1(t) + \alpha_2 u_2(t)}{c_1 u_1(t) + c_2 u_2(t)},
\end{eqnarray}
where $\alpha_1$ and $\alpha_2$ are the constants satisfying the relation
\begin{eqnarray}
\label{const}
    c_1 \alpha_2 - c_2 \alpha_1 = \frac{1}{w_0}.
\end{eqnarray}
\newline{\it Proof.} Differentiate both sides of Eq.~(\ref{Oi_gen}) and make use of Eq.~(\ref{const}).
\newline\newline\noindent
{\bf Remark 1.} The integral Eq.~(\ref{osipov-integral}) follows from the Lemma upon the choice
\begin{eqnarray}
    u_1(t) = D_{-N-1}(t), \;\;\; u_2(t) = D_{N}(it).
\end{eqnarray}
Since $w_0= i^{N}$, we conclude that
\begin{eqnarray} \label{ad-01} \fl
    \int \frac{dt}{\left(c_1 D_{-N-1}(t) + c_2 D_{N}(it)\right)^2} =
    \,\frac{\alpha_1 D_{-N-1}(t) + \alpha_2 D_{N}(it)}{c_1 D_{-N-1}(t) + c_2 D_{N}(it)},
\end{eqnarray}
where
\begin{eqnarray}
\label{const-2}
    c_1 \alpha_2 - c_2 \alpha_1 = (-i)^N.
\end{eqnarray}
\newline\newline\noindent
{\bf Remark 2.} The integral Eq.~(\ref{osipov-integral-bos}) follows from the Lemma upon the choice
\begin{eqnarray}
    u_1(t) = D_{N-1}(t), \;\;\; u_2(t) = D_{-N}(it).
\end{eqnarray}
Since $w_0= (-i)^{N}$, we conclude that
\begin{eqnarray} \label{ad-01-bos} \fl
    \int \frac{dt}{\left(c_1 D_{N-1}(t) + c_2 D_{-N}(it)\right)^2} =
    \,\frac{\alpha_1 D_{N-1}(t) + \alpha_2 D_{-N}(it)}{c_1 D_{N-1}(t) + c_2 D_{-N}(it)},
\end{eqnarray}
where
\begin{eqnarray}
\label{const-2-bos}
    c_1 \alpha_2 - c_2 \alpha_1 = i^N.
\end{eqnarray}

\newpage
\section*{References}
\fancyhead{} \fancyhead[RE,LO]{References}
\fancyhead[LE,RO]{\thepage}
\addcontentsline{toc}{section}{\protect\enlargethispage*{100pt}References}
\begin{harvard}

\item[] Adler M, Shiota T and van Moerbeke P 1995
        Random matrices, vertex operators and the Virasoro algebra
        \PL {\it A} {\bf 208} 67

\item[] Adler M and van Moerbeke P 2001
        Hermitian, symmetric and symplectic random ensembles: PDEs for the
        distribution of the spectrum
        {\it Ann. Math.} {\bf 153} 149

\item[] Andreev A V and Simons B D 1995
        Correlators of spectral determinants in quantum chaos
        {\it Phys. Rev. Lett.} {\bf 75} 2304

\item[] Andreev A V, Agam O, Simons B D and Altshuler B L 1996
        Quantum chaos, irreversible classical dynamics, and random matrix theory
        {\it Phys. Rev. Lett.} {\bf 76} 3947

\item[] Andr\'eief C 1883
        Note sur une relation les int\'egrales d\'efinies des produits des fonctions
        {\it M\'em. de la Soc. Sci.} {\bf 2} 1

\item[] Baik J, Deift P and Strahov E 2003
        Products and ratios of characteristic polynomials of random
        Hermitian matrices
        {\it J. Math. Phys.} {\bf 44} 3657

\item[] Basor E L, Chen Y and Widom H 2001
        Determinants of Hankel Matrices
        {\it J. Fun. Analysis} {\bf 179} 214

\item[] Beenakker C W J 1997
        Random Matrix Theory of quantum transport
        {\it Rev. Mod. Phys.} {\bf 69} 731

\item[] Bohigas O, Giannoni M J and Schmit C 1984
        {\it Phys. Rev. Lett.} {\bf 52} 1

\item[] Borodin A and Strahov E 2005
        Averages of characteristic polynomials in random matrix
        theory
        {\it Commun. Pure Appl. Math.} {\bf 59} 161

\item[] Borodin A, Olshanski G and Strahov E 2006
        Giambelli compatible point processes
        {\it Adv. Appl. Math.} {\bf 37} 209

\item[] Br\'ezin E and Hikami S 2000
        Characteristic polynomials of random matrices
        {\it Commun. Math. Phys.} {\bf 214} 111

\item[] Br\'ezin E and Hikami S 2008
        Intersection theory from duality and replica
        {\it Commun. Math. Phys.} {\bf 283} 507

\item[] Chazy J 1911
        Sur les \'equations diff\'erentielles du troisi\`eme ordre et d'ordre sup\'erieur dont
        l'int\'egrale g\'en\'erale a ses points critiques fixes
        {\it Acta Math.} {\bf 34} 317

\item[] Cosgrove C M and  Scoufis G 1993
        Painlev\'e classification of a class of differential equations
        of the second order and second degree
        {\it Stud. Appl. Math.} {\bf 88} 25

\item[] Cosgrove C M 2000
        Chazy classes IX--XI of third-order differential equations
        {\it Stud. Appl. Math.} {\bf 104} 171

\item[] de~Bruijn N G 1955
        On some multiple integrals involving determinants
        {\it J. Indian Math. Soc.} {\bf 19} 133

\item[] Desrosiers P 2009
        Duality in random matrix ensembles for all $\beta$
        {\it Nucl. Phys. B} {\bf 817[PM]} 224

\item[] Dyson F J 1962
        The threefold way. Algebraic structure of symmetry groups and ensembles in quantum mechanics
        {\it J. Math. Phys.} {\bf 3} 1199

\item[] Edwards S F and Anderson P W 1975
        Theory of spin glasses
        \JPF {\bf 5} 965

\item[] Efetov K B 1983
        Supersymmetry and theory of disordered metals
        {\it Adv. Phys.} {\bf 32} 53

\item[] Forrester P J and Witte N S 2001
        Application of the $\tau$-function theory of Painlev\'e equations to random matrices: PIV, PII and the GUE
        {\it Commun. Math. Phys.} {\bf 219} 357

\item[] Forrester P J and Witte N S 2002
        Application of the $\tau$-function theory of Painlev\'e equations to random matrices: PV, PIII, the LUE, JUE, and CUE
        {\it Commun. Pure Appl. Math.} {\bf LV} 0679

\item[] Forrester P J and Witte N S 2004
        Application of the $\tau$-function theory of Painlev\'e equations to random matrices: PVI, the JUE, CyUE, cJUE and scaled limits
        {\it Nagoya Math. J.} {\bf 174} 29

\item[] Forrester P J 2010
        {\it Log-Gases and Random Matrices} (Princeton: Princeton University
       Press) to appear

\item[] Fyodorov Y V and Strahov E 2003
        An exact formula for general spectral correlation function
        of random Hermitean matrices
        \JPA {\bf 36} 3203

\item[] Fyodorov Y V 2004
        Complexity of random energy landscapes, glass transition and absolute value of spectral determinant of random matrices
        \PRL {\bf 92} 240601 [Erratum: \PRL {\bf 93} 149901]

\item[] Gambier B 1910
        Sur les \'equations du second ordre
        et du premier degr\'e dont l'int\'egrale g\'en\'erale est \`{a}
        points critique fix\'es
        {\it Acta Math.} {\bf 33} 1

\item[] Garoni T M 2005
        On the asymptotics of some large Hankel determinants generated by Fisher-Hartwig symbols defined on the real line
        {\it J. Math. Phys.} {\bf 46} 043516

\item[] Gr\"onqvist J, Guhr T and Kohler H 2004
        The $k$-point random matrix kernels obtained from one-point supermatrix models
        \JPA {\bf 37} 2331

\item[] Guhr T 1991
        Dyson's correlation functions and graded symmetry
        \JMP {\bf 32} 336

\item[] Guhr T 2006
        Arbitrary unitarily invariant random matrix ensembles and
        supersymmetry
        \JPA {\bf 39} 13191

\item[] Hardy G H, Littlewood J E and P\'olya G 1934
       {\it Inequalities} (Cambridge: Cambridge University
       Press)

\item[] Heine E 1878
        {\it Handbuch der Kugelfunctionen} vol 1 (Berlin) p 288

\item[] Hirota R 2004
        {\it The Direct Method in Soliton Theory} (Cambridge: Cambridge University
       Press)

\item[] Hughes C P, Keating J P and O'Connell N 2000
        Random matrix theory and the derivative of the Riemann zeta function
        {\it Proc. R. Soc. Lond. A} {\bf 456} 2611

\item[] Its A and Krasovsky I 2008
        Hankel determinant and orthogonal polynomials for the Gaussian weight with
        a jump
        {\it Integrable Systems and Random Matrices: In Honor of Percy Deift} ed J Baik, T Kriecherbauer, L-C Li, K McLaughlin, and C Tomei
        (AMS: Contemporary Mathematics, vol 458) p 215

\item[] Kanzieper E 2002
        Replica field theories, Painlev\'e transcendents, and exact
        correlation functions
        \PRL {\bf 89} 250201

\item[] Kanzieper E 2005
        Exact replica treatment of non-Hermitean
        complex random matrices
        {\it Frontiers in Field Theory} ed O Kovras (New York: Nova Science
        Publishers) p 23

\item[] Kanzieper E 2010
        Replica approach in Random Matrix Theory
        {\it The Oxford Handbook of Random Matrix Theory} ed G Akemann, J Baik and P Di Francesco (Oxford: Oxford
        University Press) to be published

\item[] Keating J P and Snaith N C 2000a
        Random matrix theory and $\xi(1/2+\imag t)$
        {\it Commun. Math. Phys.} {\bf 214} 57

\item[] Keating J P and Snaith N C 2000b
        Random matrix theory and $L$-functions at $s=1/2$
        {\it Commun. Math. Phys.} {\bf 214} 91

\item[] Krasovsky I V 2007
        Correlations of the characteristic polynomials in the Gaussian unitary ensemble or a singular Hankel determinant
        {\it Duke Math. J.} {\bf 139} 581

\item[] Macdonald I G 1998
       {\it Symmetric Functions and Hall Polynomials} (Oxford: Oxford University
       Press)

\item[] Malmquist J 1922
        Sur les \'equations diff\'erentielles du second ordre dont
        l'int\'egral g\'en\'eral a ses points critiques fixes
        {\it Ark. Mat. Astr. Fys.} {\bf 17} 1

\item[] Mehta M L 2004
        {\it Random Matrices}
        (Amsterdam: Elsevier)

\item[] Mehta M L and Normand J-M 2001
        Moments of the characteristic polynomial in the three ensembles of random matrices
        \JPA {\bf 34} 4627

\item[] Morozov A 1994
        Integrability and matrix models
        {\it Physics--Uspekhi} {\bf 37} 1

\item[] M\"uller S, Heusler S, Braun P, Haake F and Altland A 2004
        Semiclassical foundation of universality in quantum chaos
        {\it Phys. Rev. Lett.} {\bf 93} 014103

\item[] Noumi M 2004
       {\it Painlev\'e Equations Through Symmetry} (Providence: American Mathematical Society)

\item[] Okamoto K 1980a
        Polynomial Hamiltonians associated with Painlev\'e equations, I
        {\it Proc. Japan Acad. Ser. A} {\bf 56} 264

\item[] Okamoto K 1980b
        Polynomial Hamiltonians associated with Painlev\'e equations, II: Differential equations satisfied by polynomial Hamiltonians
        {\it Proc. Japan Acad. Ser. A} {\bf 56} 264

\item[] Osipov V Al and Kanzieper E 2007
        Are bosonic replicas faulty?
        \PRL {\bf 99} 050602

\item[] Osipov V Al and Kanzieper E 2008
        Integrable theory of quantum transport in chaotic cavities
        \PRL {\bf 101} 176804

\item[] Osipov V Al and Kanzieper E 2009
        Statistics of thermal to shot noise crossover in chaotic cavities
        {\it J. Phys. A: Math. Theor.} {\bf 42} 475101

\item[] Osipov V Al, Sommers H-J, and \.Zyczkowski K 2010
        Random Bures mixed states and the distribution of their purity
        {\it J. Phys. A: Math. Theor.} {\bf 43} 055302

\item[] Painlev\'e P 1900
        M\'emoire sur les \'equations diff\'erentielles dont
        l'int\'egrale g\'en\'erale est uniforme
        {\it Bull. Soci\'et\'e Math\'ematique de France} {\bf 28} 201

\item[] Painlev\'e P 1902
        Sur les \'equations diff\'erentielles du second ordre et
        d'ordre sup\'erieur dont l'int\'egrale g\'en\'erale est uniforme
        {\it Acta Math.} {\bf 25} 1

\item[] Splittorff K and Verbaarschot J J M 2003
        Replica limit of the Toda Lattice equation
        \PRL {\bf 90} 041601

\item[] Splittorff K and Verbaarschot J J M 2004
        Factorization of correlation functions and the replica limit of the Toda lattice equation
        {\it Nucl. Phys. B} {\bf 683[FS]} 467

\item[] Strahov E and Fyodorov YV 2003
        Universal results for correlations of characteristic
        polynomials
        {\it Commun. Math. Phys.} {\bf 241} 343

\item[] Szeg\"o G 1939
        {\it Orthogonal Polynomials} (New York: American Mathematical Society)

\item[] Tracy C A and Widom H 1994
        Fredholm determinants, differential equations and matrix models
        {\it Commun. Math. Phys.} {\bf 163} 33

\item[] Tu M H, Shaw J C and Yen H C 1996
        A note on integrability in matrix models
        {\it Chinese J. Phys.} {\bf 34} 1211

\item[] Uvarov V B 1959
        Relation between polynomials orthogonal with different weights
        {\it Dokl. Akad. Nauk SSSR} {\bf 126} 33

\item[] Uvarov V B 1969
        The connection between systems of polynomials that are orthogonal with respect to different distribution functions
        {\it USSR Comput. Math. Math. Phys.} {\bf 9} 25

\item[] Verbaarschot J J M, Weidenm\"uller H A and Zirnbauer M R 1985
        Grassmann integration in stochastic quantum physics: The case of compound-nucleus scattering
        {\it Phys. Reports} {\bf 129} 367

\item[] Verbaarschot J J M and Zirnbauer M 1985
        Critique of the replica trick
        \JPA {\bf 17} 1093

\item[] Verbaarschot J J M 2010
        Quantum Chromodynamics
        {\it The Oxford Handbook of Random Matrix Theory} ed G Akemann, J Baik and P Di Francesco (Oxford: Oxford
        University Press) to be published

\smallskip
\end{harvard}

\end{document}